\documentclass[a4paper,fleqn]{cas-dc}

\usepackage[authoryear]{natbib}

\usepackage{graphicx}
\usepackage{amsmath}
\usepackage{booktabs}
\usepackage{listings}
\usepackage{xcolor}
\usepackage{enumitem}
\usepackage{url}          
\definecolor{codegreen}{rgb}{0,0.6,0}
\definecolor{codegray}{rgb}{0.5,0.5,0.5}
\definecolor{codepurple}{rgb}{0.58,0,0.82}
\definecolor{backcolour}{rgb}{0.95,0.95,0.92}

\lstdefinestyle{mystyle}{
    backgroundcolor=\color{backcolour},
    commentstyle=\color{codegreen},
    keywordstyle=\color{magenta},
    numberstyle=\tiny\color{codegray},
    stringstyle=\color{codepurple},
    basicstyle=\ttfamily\footnotesize,
    breakatwhitespace=false,
    breaklines=true,
    captionpos=b,
    keepspaces=true,
    numbers=left,
    numbersep=5pt,
    showspaces=false,
    showstringspaces=false,
    showtabs=false,
    tabsize=2
}
\lstdefinelanguage{json}{
    basicstyle=\normalfont\ttfamily,
    morestring=[b]",
    morecomment=[l]{//},
    morecomment=[s]{/*}{*/},
    literate=
        *{:}{{{\color{codepurple}{:}}}}{1}
         {,}{{{\color{codepurple}{,}}}}{1}
         {\{}{{{\color{codepurple}{\{}}}}{1}
         {\}}{{{\color{codepurple}{\}}}}}{1}
         {[}{{{\color{codepurple}{[}}}}{1}
         {]}{{{\color{codepurple}{]}}}}{1},
}

\newcommand{\gpuphot}{\textsc{GPUPHOT}}
\newcommand{\micron}{$\mu$m}

\begin{document}
\let\WriteBookmarks\relax
\def\floatpagepagefraction{1}
\def\textpagefraction{.001}

\shorttitle{GPUPHOT: GPU-Accelerated Photometry Framework}

\shortauthors{S. Lemes-Perera et al.}

\title [mode = title]{GPUPHOT: A Python Framework for High-Performance GPU-Accelerated Photometry and Distributed Astronomical Data Reduction}

\author[1,2,3]{Samuel Lemes-Perera}[orcid=0000-0003-1044-154X]
\cormark[1] 
\ead{samuel@lightbridges.es}
\credit{Conceptualization, Methodology, Software, Validation, Formal analysis, Investigation, Data Curation, Writing - Original Draft, Visualization}

\author[1,4,5]{Miguel R. Alarcon}[orcid=0000-0002-8134-2592]
\credit{Methodology, Formal analysis, Visualization, Validation, Writing - Review \& Editing}

\author[1,4,5]{Miquel Serra-Ricart}[orcid=0000-0002-2394-0711]
\credit{Supervision, Resources, Validation, Writing - Review \& Editing}

\author[2]{Pino Caballero-Gil}[orcid=0000-0002-0859-5876]
\credit{Conceptualization, Supervision, Writing - Review \& Editing}

\cortext[cor1]{Corresponding author}

\affiliation[1]{organization={Light Bridges S.L.},
    addressline={Observatorio Astronómico del Teide, Carretera del Observatorio del Teide, s/n},
    city={Santa Cruz de Tenerife},
    postcode={E-38500},
    country={Spain}}

\affiliation[2]{organization={Departamento de Ingeniería Informática y de Sistemas, Universidad de La Laguna (ULL)},
    addressline={Calle Padre Herrera s/n},
    city={San Cristóbal de La Laguna},
    postcode={E-38206},
    country={Spain}}

\affiliation[3]{organization={Instituto Tecnológico y de Energías Renovables (ITER)},
    addressline={Polígono Industrial de Granadilla s/n},
    city={Granadilla de Abona},
    postcode={E-38600},
    country={Spain}}

\affiliation[4]{organization={Instituto de Astrofísica de Canarias (IAC)},
    addressline={C/ Vía Láctea s/n},
    city={San Cristóbal de La Laguna},
    postcode={E-38205},
    country={Spain}}

\affiliation[5]{organization={Departamento de Astrofísica, Universidad de La Laguna (ULL)},
    addressline={Av. Astrofísico Francisco Sánchez s/n},
    city={San Cristóbal de La Laguna},
    postcode={E-38206},
    country={Spain}}

\begin{abstract}
We present \gpuphot, an open-source Python framework for GPU-accelerated
real-time photometry and astrometry of astronomical CCD and scientific CMOS
images.
Four of its seven stages run entirely on the GPU through CuPy (background
estimation, source detection, point spread function modeling and aperture
photometry); the catalog crossmatch selects a CPU or GPU backend by problem
size, and the zero-point and astrometric calibrations run on the CPU, within
a containerized distributed system.
Checkpoint-based GPU memory monitoring and an opt-in spatial-binning fallback
let a single container image run on GPUs from a 4\,GB laptop card to 80\,GB
datacenter accelerators.
We benchmark the framework on nineteen observations from two robotic
facilities, the Two-meter Twin Telescope and the Transient Survey Telescope,
spanning 4.2 to 151.2\,megapixels and 228 to 134{,}206 cataloged sources, on
eight NVIDIA GPU platforms with one process per image.
The H100 completes the densest 151.2-megapixel field end-to-end in 31.9\,s,
and GPU source detection on the A100 is 6--15$\times$ faster than the CPU
library \texttt{sep} on the same frames.
Once three sources of non-determinism are pinned, catalogs and zero points
agree to the fourth decimal across GPU models and host CPUs.
In this one-process-per-image regime the CuPy~14 / cuML stack showed a
7\,\% lower peak memory and a median 38\,\% higher per-image latency than
CuPy~12, both due to the RAPIDS allocator displacing the CuPy memory pool at
import; restoring the pool removes the memory difference and cuts the median
overhead on datacenter GPUs to 11\,\%, with identical output.
Long-lived workers, the production mode, run the two stacks near parity.
We recommend the CuPy~14 stack for its reproducibility and maintained
libraries, not for memory or speed.
\gpuphot\ and its benchmark scripts are released for autonomous facilities
that require end-to-end photometric reduction at survey cadence.
\end{abstract}

\begin{highlights}
\item GPU photometry pipeline in production at two robotic facilities since 2023
\item 151-megapixel frames reduced in 32 s on an H100; detection 6--15x faster than sep
\item Identical catalogs and zero points across GPU models and hosts, three sources pinned
\item CuPy 14/cuML latency and memory differences traced to a single allocator defect
\item Benchmarks on eight GPUs, from a 4 GB laptop to 80 GB datacenter cards, released
\end{highlights}


\maketitle

\section{Introduction}
\label{sec:introduction}

Modern robotic telescope facilities demand autonomous, low-latency photometric reduction at rates that CPU-only pipelines cannot sustain at full image resolution. Projects such as the Vera C. Rubin Observatory's Legacy Survey of Space and Time \citep{ivezicLSSTScienceDrivers2019} and the Zwicky Transient Facility \citep{bellmZwickyTransientFacility2019} exemplify this pressure at survey scale. The same challenge applies to agile facilities. The Two-meter Twin Telescope (TTT)\footnote{\url{https://ttt.iac.es}} and the Transient Survey Telescope (TST)\footnote{\url{https://tst.iac.es}} at Teide Observatory (Instituto de Astrof\'isica de Canarias, Tenerife) produce thousands of frames per night from instruments spanning 4.2 to 151.2 megapixels (MP), and every frame must be reduced in real time for transient detection and time-domain follow-up. We present \gpuphot, an open-source Python framework that addresses this demand by executing a complete photometric and astrometric pipeline on NVIDIA GPUs. The framework has been the real-time reduction pipeline of the TTT and TST since 2023 and of ATLAS-Teide, the Teide unit of the Asteroid Terrestrial-impact Last Alert System \citep{tonryATLASHighcadenceAllsky2018, licandroATLASTEIDENextGenerations2023}, since 2025; by September 2026 it had reduced 3.74 million frames, 297 trillion pixels or 1.19\,PB as single-precision images, according to the results database of these facilities.

\subsection{State of the Art and Limitations}
Several established software packages have been the standard for astronomical photometry for decades. \texttt{DAOPHOT} \citep{stetsonDAOPHOTComputerProgram1987}, one of the earliest packages for CCD stellar photometry and still the reference for point spread function (PSF) fitting photometry of crowded fields, was conceived for one frame at a time on a single processor. \texttt{Source Extractor} (SExtractor) \citep{bertinSExtractorSoftwareSource1996} is widely used due to its speed and reliability, but its single-threaded architecture (in its classic version) limits its scalability on modern multi-core systems. While \texttt{SExtractor++} \citep{bertinSourceXtractorSoftware2020} aims to address this, it remains a CPU-bound solution. Python-based ecosystems like \texttt{Astropy} \citep{theastropycollaborationAstropyProjectSustaining2022}'s Photutils \citep{bradleyPhotutilsPhotometryTools2016} provide a broad set of photometric tools but operate on CPU and cannot exploit the data parallelism of modern GPUs for the element-wise array operations that dominate image processing at megapixel scales. Other tools like \texttt{Gnuastro} \citep{akhlaghiNoisebasedDetectionSegmentation2015} provide efficient command-line utilities but lack the direct integration required for complex, distributed Python pipelines.

To the best of our knowledge, few of these standard tools use GPUs for the core photometric operations, such as background estimation, convolution, and source detection. Recent efforts have shown the potential of the approach: \citet{niwanoGPUacceleratedImageReduction2021} obtained an order-of-magnitude speed-up of a full reduction pipeline by porting it to the GPU with CuPy, and \citet{caoImagePreprocessingFramework2025} assembled a GPU-accelerated preprocessing framework for the Ground-based Wide Angle Camera (GWAC) survey. These works either port existing CPU algorithms step by step or accelerate individual stages; what has been missing is a pipeline conceived so that every stage reduces to the same GPU-friendly primitives, deployed and evaluated as a system. In the context of time-domain astronomy, where latency is critical for transient follow-up, the processing time per image must be minimized. CPU-based pipelines often become the bottleneck when dealing with high-resolution images, such as 4k~$\times$~4k~pixels or larger, at high frame rates.

\subsection{The GPUPHOT Framework}
\gpuphot\ is designed for autonomous observatory pipelines where every image must be reduced, calibrated, and archived without operator intervention.

The main contributions of this work are:
\begin{enumerate}
    \item \textbf{A complete seven-stage pipeline deployed as a distributed system}: four stages run entirely on the GPU (background estimation, source detection, PSF modeling and aperture photometry), the catalog crossmatch selects its CPU or GPU backend by problem size (Section~\ref{sec:impl:adaptive_cuml}), and the zero-point and astrometric calibrations run on the CPU. Distributed workers consume images from a task queue and persist the calibrated catalogs to a spatially indexed database for sub-arcsecond queries (Section~\ref{sec:architecture}).  Benchmarked on eight NVIDIA GPUs spanning a 20$\times$ range of device memory (VRAM), the H100 reduces the densest 151.2\,MP field of the benchmark (134{,}206 sources) in 31.9\,s with one process per image, and GPU source detection on the A100 is 6--15$\times$ faster than the CPU library \texttt{sep} on the same background-subtracted frames (Section~\ref{sec:perf:cpu}).

    \item \textbf{Adaptive GPU memory management}: checkpoint-based VRAM monitoring with opt-in out-of-memory fallback via spatial image binning enables the same container image to operate across a 20$\times$ VRAM range, from 4\,GB laptop GPUs to 80\,GB datacenter accelerators, without container modifications.

    \item \textbf{Diagnosis of an allocator defect in the modern library stack}: in one-process-per-image benchmarks the newer stack (Python~3.12, CuPy~14, cuML) showed a median 38\% higher per-image latency and a 7\% lower peak VRAM than the older one (Python~3.8, CuPy~12).  Both trace to the RAPIDS allocator displacing the CuPy memory pool at import time; restoring the CuPy pool as the default allocator at the start of each task reduces the latency overhead to a median 11\% on datacenter GPUs and returns the peak to the older value, with identical catalogs and zero points.  The apparent memory benefit of the upgrade, and the concurrency it would have bought, are artifacts of the defect (Section~\ref{sec:perf:memory}).

    \item \textbf{Deterministic execution across GPUs and hosts}: three independent sources of run-to-run and host-to-host variation (the principal component analysis (PCA) solver, the seeds of the GPU clustering, and the host's basic linear algebra subprograms (BLAS) kernel) are identified, quantified and pinned, after which catalogs and zero points agree to the fourth decimal across the H100, A100 and L40S and across x86 hosts (Section~\ref{sec:impl_determinism}).  Pinned CUDA, CuPy and astrometric solver dependencies inside Docker \citep{merkelDockerLightweightLinux2014, morrisUseDockerDeployment2017}, and the benchmark scripts released with the code, make every result of this paper reproducible.

\end{enumerate}

\subsection{Scope of this Work}
The detailed scientific performance and full-scale astronomical validation, including completeness, contamination, and depth analysis, of the catalogs produced by \gpuphot\ are beyond the scope of this technical description and are presented in a companion paper \citep{alarconGPUPHOT2026}, which also gives the statistical derivation of the algorithms. This work focuses specifically on the computational architecture, GPU acceleration strategies, and system scalability. We evaluate the pipeline across the eight NVIDIA GPU platforms listed in
Table~\ref{tab:gpu_hardware}, covering image sizes from 4.2 to 151.2\,MP.

The remainder of this paper is organized as follows.
Section~\ref{sec:architecture} describes the high-level system architecture.
Section~\ref{sec:implementation} details the implementation of the four GPU-accelerated pipeline stages, the adaptive memory management system, the environment-aware execution model, and the measures that make execution deterministic.
Section~\ref{sec:deployment} covers deployment strategies, scalability, software availability, and reproducibility.
Section~\ref{sec:validation} presents functional validation of the pipeline.
Section~\ref{sec:performance} presents end-to-end performance benchmarks across eight GPUs.
Section~\ref{sec:discussion} discusses comparisons with existing tools, deployment recommendations, and limitations.
Section~\ref{sec:conclusion} summarizes contributions.

\section{System Architecture}
\label{sec:architecture}

\begin{figure*}[h!]
    \centering
    \includegraphics[width=\textwidth]{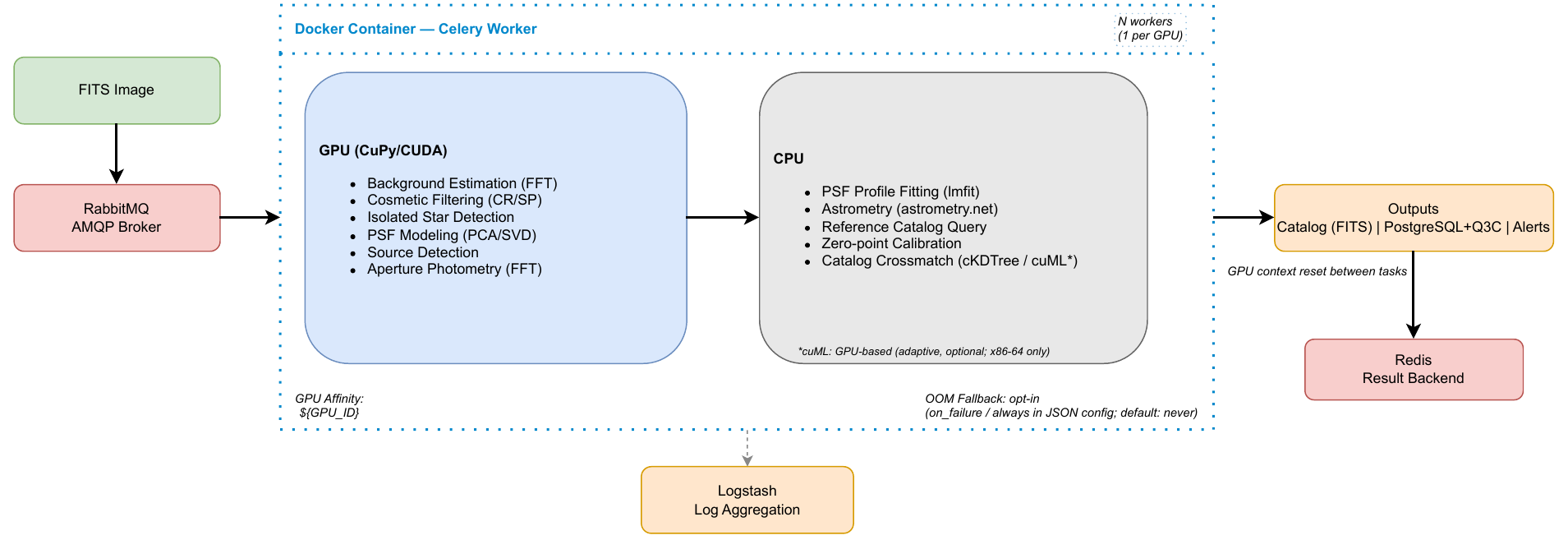}
    \caption{High-level architecture of the \gpuphot\ system. Incoming FITS images are ingested through the message broker and processed by containerized GPU workers, with calibrated catalogs and metadata persisted to a PostgreSQL/Q3C database.}
    \label{fig:architecture_diagram}
\end{figure*}

\gpuphot\ separates the reduction from the infrastructure that runs it. The engine that processes an image imports no task queue and no database driver, the worker service imports the engine and never the other way round, and the two carry separate requirements, so the same reduction code runs as a library inside an observatory control system and as a containerized service behind a task queue. Figure \ref{fig:architecture_diagram} shows the components and the path an image takes through them.

\subsection{Technology Stack}
The core of \gpuphot\ is built upon the following Python technology stack:

\begin{itemize}
    \item \textbf{Python 3.8 and later}: the two stacks tested in this paper are Python~3.8 with CuPy~12 and Python~3.12 with CuPy~14 (Section~\ref{sec:perf:setup}).
    \item \textbf{CuPy}: A NumPy \citep{harrisArrayProgrammingNumPy2020}/SciPy \citep{virtanenSciPy10Fundamental2020}-compatible array library for GPU-accelerated computing with NVIDIA CUDA \citep{okutaCuPyNumPycompatibleLibrary2017}. The pipeline is written against its NumPy-compatible API and contains no hand-written CUDA kernels.
    \item \textbf{Celery}: A distributed task queue that manages the execution of asynchronous image processing tasks. It allows \gpuphot\ to scale from a single machine to a cluster of worker nodes.
    \item \textbf{RabbitMQ}: The message broker between clients and Celery workers; its queues are persisted to disk, so queued images survive a broker restart.
    \item \textbf{Redis}: Used as the result backend for Celery, storing the state and return values of completed tasks.
    \item \textbf{Logstash}: Receives structured pipeline logs forwarded asynchronously from Celery workers via \texttt{logstash-async} (TCP) and optionally routes them to Elasticsearch for long-term storage and operational monitoring of the pipeline.
    \item \textbf{PostgreSQL with Q3C}: An open-source relational database used for persisting photometric catalogs and image metadata. We utilize the Q3C extension \citep{koposovQ3CQuadTree2006}, which implements a Quad Tree Cube indexing scheme, to enable efficient sub-arcsecond spatial crossmatching and cone searches on celestial coordinates. This capability is critical for managing the high density of sources detected in wide-field images, such as those from the 151~MP FERVOR-L camera \citep{alarconScientificCMOSSensors2023}. Database access is mediated through SQLAlchemy for transactional catalog insertions and \texttt{psycopg2} for direct analytical queries, providing both object-relational-mapping safety and low-level performance where needed. This persistence layer belongs to the worker package (\texttt{gpuphot\_worker}, below), not to the processing engine.
    \item \textbf{Pandas} \citep{mckinneyDataStructuresStatistical2010}: Tabular data throughout the pipeline, including detected source catalogs, photometric measurements, and crossmatch results, are represented as \texttt{DataFrame} objects. Column-wise operations, filtering and the serialization between stages and the database are done on these frames.
    \item \textbf{AstroQuery}: Remote catalog access, including queries to the VizieR service \citep{ochsenbeinVizieRDatabaseAstronomical2000} and the Astrometry.net web API, is handled through the \texttt{astroquery} library \citep{ginsburgAstroqueryAstronomicalWebquerying2019}. This provides a unified, Pythonic interface for retrieving reference photometric and astrometric catalogs during the calibration stage.
    \item \textbf{Docker \& Docker Compose}: The entire system is containerized \citep{merkelDockerLightweightLinux2014, morrisUseDockerDeployment2017}, allowing for reproducible builds and simplified deployment across different environments. Docker Compose is used to orchestrate the multi-container application.
\end{itemize}

\subsection{Deployment Modalities}
\gpuphot\ supports two deployment modes: as a standalone Python library for integration into existing observatory control systems, and as a containerized microservices stack (Celery, RabbitMQ, PostgreSQL/Q3C) for unattended queue-driven operation at robotic facilities. Both modes share the same GPU pipeline and memory management mechanisms. Full configuration recipes and scaling strategies are detailed in Section~\ref{sec:deployment}.

\subsection{Core Components}

\subsubsection{GPU Processing Engine (\texttt{gpuphot})}
This is the core component of the framework, containing the scientific code for image processing. It is organized into several sub-packages:
\begin{itemize}
    \item \texttt{phot}: Contains the core photometry algorithms, including background estimation, source detection, PSF modeling, and aperture photometry. These modules are optimized using CuPy.
    \item \texttt{utils}: Provides helper functions for GPU memory management, external catalog queries, and FITS header manipulation, as well as astrometric calibration helpers that integrate with local Astrometry.net solvers \citep{langAstrometrynetBlindAstrometric2010}. Ephemeris calculations for observational context, such as lunar distance and solar altitude at the time of observation, are performed via \texttt{PyEphem} \citep{rhodesPyEphemAstronomicalEphemeris2011}.
    \item \texttt{stats}: The sigma-clipped stacking used by the photometry stage (\path{stack_sigmaclip}), together with sub-pixel image-registration routines by phase cross-correlation, plain and masked, available in the package.
    \item \texttt{logger}: Hierarchical structured logging, the source of the records forwarded to Logstash.
\end{itemize}
The engine imports no infrastructure: no task queue and no database driver. Local catalog support is injected as a callable (\path{custom_vizier_search_func}), so PostgreSQL never enters the engine.
The complexity of the underlying GPU operations is abstracted behind a high-level Python API. Users interact with the system through a unified \texttt{ImageProcessor} interface, which handles the orchestration of calibration, detection, and photometry with a single method call. A caller supplies the instrument name, the image and its header and receives the catalog and the updated header (Section~\ref{sec:deployment:standalone}); the GPU is not visible in the interface.

\subsubsection{Distributed Worker System (\texttt{gpuphot\_worker})}
This package contains the infrastructure for distributed processing.
\begin{itemize}
    \item \texttt{tasks.py}: Defines the Celery tasks that can be executed remotely, such as \path{process_image_task} and \path{process_directory_task}. It orchestrates the calls to the core \texttt{gpuphot} engine.
    \item \texttt{worker\_app.py}: Configures and initializes the Celery application.
    \item \texttt{celeryconfig.py}: Dynamically loads Celery settings from environment variables, allowing for runtime configuration.
    \item \texttt{utils.py}: Contains worker-specific utilities, including the robust image opening function and the image reduction logic used for handling memory errors.
    \item \texttt{database\_insert\_utils.py}: Writes the calibrated header to the \texttt{imastats} table and the source catalog to the \texttt{imaphot} table of the PostgreSQL/Q3C database, each write in a single transaction.
    \item \texttt{database\_search\_utils.py}: Read-only queries on those tables, by date range, file name or transient, including the Q3C cone search.
    \item \texttt{celery\_exceptions.py}: The task base class and the exceptions serialized to the result backend.
    \item \texttt{header\_descriptions.py}: Human-readable descriptions of the FITS keywords stored with each image.
    \item \texttt{generators/}: Seven notebook generators (system overview, setup, instrument configuration, task execution, database query, multi-instrument detection, cuML configuration).
\end{itemize}
The dependency runs in one direction only: the worker imports the engine, the engine never imports the worker, and the two carry separate requirements.

\subsubsection{Multi-Instrument Configuration}
Adaptability to different instrument setups is achieved through a system of JSON configuration files. Each file maps instrument-specific FITS header keywords to a set of canonical internal keywords used by the pipeline, such as mapping \texttt{EXPTIME} or \texttt{EXPOSURE} to the internal \texttt{exposure\_time}. The same file carries the camera parameters (gain, read noise, pixel size, frame size) and the processing settings of Listing~\ref{lst:instrument_config}; eight such files describe the cameras of the TTT, the TST and ATLAS-Teide.

\subsubsection{Operational Autonomy and Local Catalog Support}
To ensure uninterrupted operation in isolated observatory environments where internet connectivity may be intermittent or restricted, \gpuphot\ features a modular catalog interface designed for operational autonomy. While the system defaults to querying remote services such as VizieR via \texttt{astroquery} \citep{ginsburgAstroqueryAstronomicalWebquerying2019}, it explicitly supports the integration of local, custom catalogs. By injecting a user-defined search function into the pipeline, observatories can query internal astrometric and photometric databases, such as a local copy of Gaia DR3 or a proprietary survey catalog stored in PostgreSQL. Section~\ref{sec:catalogue_backend} quantifies the effect: a local replica removes the network jitter of the remote query and reduces end-to-end latency by up to $1.8\times$ on dense fields.

\subsection{Data Flow}
The typical data flow through the system is as follows:
\begin{enumerate}
    \item A user or an automated script submits a task to the Celery queue via a Python call, such as \path{process_directory_task.delay(...)}.
    \item RabbitMQ receives the task and holds it until a worker is available.
    \item An idle Celery worker process fetches the task from the queue.
    \item The worker executes the task, which involves loading the image data, transferring it to the GPU, and running the \gpuphot\ processing pipeline.
    \item During processing, structured pipeline logs are forwarded asynchronously to a Logstash instance via \texttt{logstash-async} (TCP); Logstash can optionally route log records to Elasticsearch for long-term storage and indexing.
    \item Upon completion, the final photometric catalog and image metadata are inserted into the PostgreSQL database.
    \item The task's return value, such as a summary dictionary, is stored in the Redis result backend.
    \item The user can query the task's status and retrieve the result at any time.
\end{enumerate}
Catalog insertion is one transaction per image, so a failed task leaves no partial catalog.

\subsection{Fault Tolerance}
Out-of-memory (OOM) events trigger a configurable fallback: when \texttt{apply\_reduction} is set in the instrument JSON, the worker rebins or crops the image and retries locally rather than propagating a failure; unrecoverable errors are serialized and returned to the result backend without corrupting the PostgreSQL catalog. The implementation of the memory management and OOM fallback mechanisms is described in detail in Section~\ref{sec:impl_memory}.
\section{GPU Implementation}
\label{sec:implementation}

\gpuphot\ implements the photometric algorithms of the companion paper \citep{alarconGPUPHOT2026}, whose design principle is that every stage of the reduction is either a convolution with a fixed kernel or an element-wise operation on full-frame maps, so that no stage iterates over individual sources. That principle is what makes the pipeline map onto a GPU: the work of each stage is a small number of fast Fourier transforms (FFTs) and array operations whose cost depends on the pixel count of the frame and not on the number of sources. We use the CuPy library \citep{okutaCuPyNumPycompatibleLibrary2017} to run these operations on NVIDIA GPUs, minimizing data transfer between host and device. This section summarizes what each stage computes and how it is implemented, then describes the memory management, the environment-aware execution and the measures that make the execution deterministic; the statistical derivation, the defaults and the scientific validation of the algorithms belong to the companion paper.

\subsection{Image Preprocessing and Background Estimation}
\label{sec:impl_preprocessing}
The frame is loaded into GPU memory as a \texttt{float32} array. Two optional cosmetic corrections then run on the GPU, each enabled per camera: a cosmic-ray filter, conceptually related to L.A.Cosmic \citep{vandokkumCosmicrayRejectionLaplacian2001}, that flags sharp edges with a Laplacian and reconstructs the flagged pixels by FFT-based interpolation, and a salt-and-pepper filter for the random telegraph noise of scientific CMOS (sCMOS) sensors \citep{alarconScientificCMOSSensors2023} that replaces isolated outliers of a median-filtered residual. The background is a local estimate obtained by convolution, and convolutions of a 151.2\,MP frame are only practical through the FFT convolution theorem,
\begin{equation}
    I_{conv} = \mathcal{F}^{-1} \{ \mathcal{F}(I) \cdot \mathcal{F}(K) \},
\end{equation}
which scales as $O(N^2 \log N)$ instead of the $O(N^2 K^2)$ of a spatial convolution with a kernel of size $K$; \texttt{cupyx.scipy.fft} provides the transforms. Frames are padded by the kernel size with reflected borders, and invalid pixels, including those masked as compact sources, are filled from their valid neighbors before transforming. The stage also produces a per-pixel noise map, combining the sky, readout and dark-current variances, that the detection stages use as their significance normalization instead of a single global $\sigma_{\rm sky}$.

\subsection{Source Detection}
\label{sec:impl_detection}
Detection is a matched filter: the background-subtracted frame is cross-correlated with the PSF, the result is divided by the noise map, and sources are the local maxima above a significance threshold ($5\sigma$ by default), refined to sub-pixel positions with a first-moment centroid. A first pass, run before any PSF is known, uses a fixed Gaussian kernel and yields the isolated, unsaturated stars from which the PSF model of Section~\ref{sec:impl_psf} is built; its isolation test runs on the CPU with a $k$-d tree, which is faster than a GPU search at these sizes, and fewer than five such stars raise an \texttt{InsufficientStarsError}. The second pass uses the spatially varying PSF. Because that PSF is expressed as a fixed reference $P_0$ plus $K$ fixed eigenmodes $e_k$ with position-dependent coefficient maps $c_k(\mu)$, the position-dependent matched filter decomposes into $K+1$ genuine convolutions,
\begin{equation}
    S(\mu) = (M * \bar{P}_0)_\mu + \sum_{k=1}^{K} c_k(\mu)\,(M * \bar{e}_k)_\mu ,
\end{equation}
where $M$ is the background-subtracted frame and the bar denotes the reflected kernel; each term is evaluated by FFT and weighted by a precomputed map. The whole stage is therefore FFT convolutions, element-wise operations and a maximum filter, and no connected-component labeling is involved. The derivation of the statistic is given in the companion paper.

\subsection{PSF Modeling with Eigen-PSFs}
\label{sec:impl_psf}
The PSF of a wide-field frame varies across the field, and the companion paper models it as a reference PSF $P_0$ plus a small number of fixed deviation patterns, the eigenmodes or Eigen-PSFs, whose amplitudes vary with position; we refer to the step that builds them as the Eigen-PSF stage. \gpuphot\ builds this model in three steps, all on the GPU except the eigendecomposition itself. First, a square cutout of half-side $\max(\lfloor 10''/s \rfloor, 6)$ pixels, with $s$ the plate scale, is extracted around each isolated star (up to 1{,}000 per frame), recentered to sub-pixel accuracy and normalized to unit flux; stamps whose brightest pixel exceeds 30\% of the flux are rejected as poorly sampled. Second, the stamps of highest signal-to-noise ratio (S/N) in the central region of the frame (\texttt{center\_factor}, 0.7 by default) are stacked with iterative sigma clipping into $P_0$, at most \texttt{max\_stars\_ref} of them (15 by default). Third, a PCA of the deviations of all stamps from $P_0$, each normalized by its own scatter, yields $K = 5$ eigenmodes, and the coefficient of each star on each mode is interpolated into a full-frame map on square tiles of a per-instrument size (1{,}500 to 4{,}000\,px in the benchmark configurations). An analytical Moffat profile \citep{moffatTheoreticalInvestigationFocal1969} is fitted to $P_0$ with \texttt{lmfit} \citep{newvilleLMFITNonlinearLeastsquare2014a}; its full width at half maximum (FWHM) sets the separation and centroid window of the detection stage and the aperture grid of the photometry stage, and is recorded in the header together with the FWHM of the PSF reconstructed at four field corners.

The eigendecomposition operates on a matrix of at most 1{,}000 stamps by the stamp pixel count, which ranges from about $10^{3}$ pixels at the coarsest plate scale of the benchmark to $2 \times 10^{4}$ at the finest, and is delegated to \texttt{cuml.decomposition.PCA} when RAPIDS cuML is available, with \texttt{scikit-learn} \citep{pedregosaScikitlearnMachineLearning2012} as the fallback. Both paths use the full singular value decomposition (SVD) solver for the reasons of Section~\ref{sec:impl_determinism}, and Section~\ref{sec:perf:memory} shows that the cost of the cuML solver is governed by the stamp size. The stage can be disabled per instrument (\texttt{pca\_method}); two of the five benchmark cameras run without it (Section~\ref{sec:validation}).

\subsection{Adaptive Aperture Photometry}
\label{sec:impl_photometry}
Fluxes are measured in circular apertures of integer radii from $\lceil 0.75\,\mathrm{FWHM} \rceil$ to $\lceil 5\,\mathrm{FWHM} \rceil$. For each radius the aperture sums of all sources are one cross-correlation of the frame with a top-hat kernel, evaluated by FFT, so the growth curve of every detection costs $N_r$ full-frame convolutions instead of a loop over sources; the variance maps of the spatially varying noise terms are propagated through the same kernel, and the sums are accumulated in \texttt{float64}. The correction for the flux outside the aperture is measured rather than integrated from the PSF model: the growth curves of high-S/N isolated stars are grouped into spatial clusters by $k$-means on their pixel coordinates (backend in Section~\ref{sec:impl_environment}), each source adopts the mean curve of its nearest cluster, and when the configured tile exceeds the frame, as for the two smallest cameras of the benchmark, a single field-wide curve results. The radius that maximizes the S/N is searched only for those calibrators; a power law fitted to them, $\log_{10} r_{\rm opt} = a \log_{10} \nu + b$ with $\nu$ the detection significance, assigns the radius of every other source without recomputing its noise model.

\subsection{Photometric Zero-Point Calibration}
\label{sec:impl_zeropoint}
The zero point ties the instrumental magnitudes to a reference catalog, Pan-STARRS\,DR1, Gaia\,DR3 or SkyMapper depending on filter and declination (Section~\ref{sec:validation}), retrieved from VizieR \citep{ochsenbeinVizieRDatabaseAstronomical2000} or from a local database. Matched stars are filtered by color, S/N and position, and the zero point is the sigma-clipped mean of their catalog-minus-instrumental offsets at fixed unit slope, a choice the companion paper motivates by its frame-to-frame stability. The stage runs on the CPU, writes the zero point, its standard error and the number of calibrators to the header (\texttt{ZP}, \texttt{EZP}, \texttt{CATNSTAR}), and costs a negligible fraction of the pipeline at every tested source density (Table~\ref{tab:stage_breakdown}).

\subsection{GPU Memory Management}
\label{sec:impl_memory}
Processing high-resolution astronomical images, such as the 151.2~MP frames of the FERVOR-L camera on the TST, requires strict memory discipline on GPUs with limited VRAM. We organize the memory management strategy around five mechanisms, ordered from lowest to highest level of intervention.

\subsubsection{CuPy Memory Pools}
\gpuphot\ uses CuPy's memory pool facility (\path{cupy.cuda.MemoryPool}) to mitigate the overhead of frequent \texttt{cudaMalloc}/\texttt{cudaFree} calls during array-intensive pipeline stages. The pool retains previously freed GPU allocations and recycles them for subsequent requests of compatible size, amortizing allocation latency over the lifetime of a task. A \texttt{PinnedMemoryPool} is additionally used for host-side buffers, enabling asynchronous and higher-throughput PCIe transfers between host (CPU) and device (GPU). Importing RAPIDS cuML replaces CuPy's global allocator with the RAPIDS memory manager in unpooled mode, so \gpuphot\ re-seats the default CuPy pool, that is, installs it again as the default allocator, at the start of each task, inside the worker process; Section~\ref{sec:perf:memory} quantifies what not doing so costs. At the end of each image the pools are released and replaced, the FFT plan cache is cleared and the garbage collector is run. On CuPy~14.1 and later the replaced pool is not reclaimed (CuPy issue 9437): in a container test a long-lived worker without the per-task re-seat had grown to 24\,GiB of device memory by its fifth image and failed with an out-of-memory error at the sixth, whereas with it the footprint stayed flat over twelve; the benchmark stack, CuPy~14.0.1, is not affected.

\subsubsection{Checkpoint-Based Device-Memory Monitoring}
The \texttt{utils.gpu} module implements an \path{adaptive_memory_management()} function that is called at pipeline checkpoints: after background subtraction, inside isolated-star detection, and three times inside the optimal-photometry stage, whose batched aperture sums are the largest transient allocations. At each checkpoint the function evaluates an effective occupancy ratio, the larger of the device occupancy (used over total device memory) and the pool occupancy (used over reserved bytes in the CuPy pool), and applies a graduated response:
\begin{itemize}
    \item At 75\% occupancy, a debug-level memory report is logged.
    \item At 85\% occupancy, the system executes \path{mempool.free_all_blocks()} to release all cached allocations back to the CUDA driver, triggers Python garbage collection via \path{gc.collect()}, and synchronizes all active CUDA streams to ensure pending operations have completed before reclaiming memory.
\end{itemize}
The device term is the one that carries the safety guarantee, and it is the one that fired before the memory-exhaustion events discussed in Section~\ref{sec:perf:jetson}. The pool term is conservative by construction, because a healthy pool is nearly fully used most of the time. Under the CuPy~12 stack it triggers the release at nearly every checkpoint even at 2\% device occupancy, at a measured cost of about 0.2\,s per image. Under the CuPy~14 stack with cuML it never fired until the pool was re-seated, because the pool it inspects had been displaced by the RAPIDS allocator (Section~\ref{sec:limitations}).

\subsubsection{Out-of-Memory Fallback at the Worker Level}
The OOM fallback behavior is opt-in and controlled by the \texttt{apply\_reduction} key in the per-instrument JSON configuration, which supports three modes:
\begin{itemize}
    \item \texttt{never} (default): no automatic retry; a \texttt{MemoryError} propagates as a task failure.
    \item \texttt{on\_failure}: if a full-frame processing attempt raises a \texttt{MemoryError} or CUDA out-of-memory exception, the worker retries the task using a spatially binned and/or cropped version of the image.
    \item \texttt{always}: the reduced image is processed unconditionally, regardless of whether OOM occurred.
\end{itemize}
The spatial rebinning is performed via \texttt{scikit-image}'s \texttt{block\_reduce} function \citep{vanderwaltScikitimageImageProcessing2014}, which applies a configurable aggregation kernel (typically $2\times2$ mean) to reduce the array dimensions while preserving photometric flux. The \texttt{on\_failure} mode trades spatial resolution for successful completion, ensuring that no observation is silently dropped from the reduction queue.

\subsubsection{Process Recycling}
The configuration of the released worker package sets \path{worker_max_tasks_per_child=1}, which under the prefork execution pool, in which each slot is an operating-system process, would recycle that process after each image, destroying the CUDA context and all its allocations so that fragmentation cannot accumulate across heterogeneous image dimensions. Neither deployment in use applies it: the reference Compose file runs the workers on the \texttt{gevent} pool, to which Celery does not apply this limit, and the production deployment runs the prefork pool without it (Section~\ref{sec:deployment}). In both, the worker process persists across tasks, and the per-task pool re-seating and end-of-image release described above bound the memory carried from one image to the next. Section~\ref{sec:perf:memory} contrasts the one-process-per-image regime with a long-lived process.

\subsubsection{Infrastructure Provisions}
The \texttt{utils.gpu} module additionally exposes \path{offload_to_cpu()} and \path{load_to_gpu()} functions for reactive CPU$\leftrightarrow$GPU data migration under memory pressure, as well as \path{check_memory_availability()}, which verifies that sufficient headroom (a 10\% safety margin of total VRAM) exists before committing to large allocations such as the FFT workspace for full-frame convolution. These are provisions for future adaptive scheduling strategies; in the current pipeline, the checkpoint-based and worker-level mechanisms described above provide the primary memory safety guarantees.

\subsection{Environment-Aware Execution}
\label{sec:impl_environment}
\gpuphot\ employs a dynamic dispatch architecture designed to maximize performance across heterogeneous computing environments, ranging from embedded edge devices such as NVIDIA Jetson platforms to high-performance datacenter nodes.

\subsubsection{Dynamic Library Detection}
At import time, the framework checks for the presence of RAPIDS \texttt{cuML} \citep{rapidsdevelopmentteamCuMLGPUacceleratedMachine2024}. When present, three operations can run on the GPU: nearest-neighbor catalog crossmatching (\texttt{cuml.neighbors.NearestNeighbors}), the $k$-means clustering of the aperture-correction calibrators (\texttt{cuml.cluster.KMeans}), and the PCA of the Eigen-PSF stage (\texttt{cuml.decomposition.PCA}). When the library is absent, or when a GPU call fails at runtime, the system falls back to \texttt{scipy.spatial.cKDTree}, \texttt{sklearn.cluster.KMeans} and \texttt{sklearn.decomposition.PCA} respectively, requiring no monolithic dependency stack. The clustering and PCA problems are small, of order $10^3$ calibrators and a few hundred stamps; the seeded GPU clustering is 2.4 times faster than the CPU path at that size once the process is warm, whereas the cost of the GPU PCA is set by the stamp size and exceeds the CPU path on fine plate scales (Section~\ref{sec:cuml_eval}). The crossmatch is the only one of the three whose backend is selected by problem size (Section~\ref{sec:impl:adaptive_cuml}).

\subsubsection{Deployment Contexts}
We maintain two tested deployment configurations:
\begin{itemize}
    \item \textbf{Python 3.8} (CuPy 12, NumPy 1.23, no cuML): the lowest per-image latency in one-process-per-image operation: the Python~3.12 stack is slower by a median of 11\% on the datacenter GPUs with the allocator correction (Section~\ref{sec:perf:memory}), so this configuration remains a viable choice where strict per-image latency requirements dominate. Clustering, PCA and crossmatching run on the CPU.
    \item \textbf{Python 3.12\,+\,adaptive cuML} (CuPy 14, NumPy 2.0, cuML 25.2): \textbf{recommended for production deployments}. Its results are reproducible across GPU models and hosts without further configuration (Section~\ref{sec:impl_determinism}), and it runs on current library releases.  With the allocator correction of Section~\ref{sec:impl_memory} its memory footprint equals the Python~3.8 stack's and its per-image latency overhead is a median 11\% on datacenter GPUs in one-process-per-image operation, near parity in long-lived workers (Section~\ref{sec:perf:memory}).
\end{itemize}
The lower peak of the Python~3.12 stack without the pool is not a property of the libraries: with the pool re-seated, the peak returns to the Python~3.8 value within 3.3\% at all five benchmark sizes (Section~\ref{sec:perf:memory}).

\subsubsection{Crossmatch Backend Selection}
The catalog crossmatch stage matches detected source positions against astrometric and photometric reference catalogs using 2D nearest-neighbor queries.
Two backends are supported: \texttt{scipy.spatial.cKDTree} ($O(N \log N)$) and
\texttt{cuML.NearestNeighbors} ($O(N^2)$ brute-force).
An empirical evaluation across source counts ranging from 228 to 134{,}206 is presented in Section~\ref{sec:cuml_eval}; the outcome determines the production default and motivates the adaptive strategy described below.

\subsubsection{Jetson and ARM Considerations}
On aarch64 platforms such as the NVIDIA Jetson Orin, RAPIDS \texttt{cuML} is unavailable; both Python versions therefore use CPU-based fallbacks for clustering, PCA, and crossmatching. As a result, Jetson benchmarks reflect a different execution path from datacenter measurements. Deployment constraints and sensor compatibility guidance are discussed in Section~\ref{sec:deployment:jetson}.

\subsubsection{Adaptive cuML Crossmatch Strategy}
\label{sec:impl:adaptive_cuml}

Following the ablation study, we implemented an \emph{adaptive cuML crossmatch}
strategy that selects the crossmatch backend based on the number of detected
sources rather than applying a single backend unconditionally.
For each GPU model, two thresholds are configured: a minimum source count
(MIN) and a maximum source count (MAX).
When the detected source count falls within [MIN, MAX], cuML is activated;
outside this range, \texttt{cKDTree} is used.
This preserves cuML's algorithmic advantage in the synthetic crossover window
(Figure~\ref{fig:cuml_crossover}) while avoiding the per-call
floor of the GPU path, about 1\,ms per call, that dominates at low source counts.
Per-GPU activation thresholds, derived from the evaluation in
Section~\ref{sec:cuml_eval}, are listed in Table~\ref{tab:cuml_thresholds}.

The strategy is configured per deployment through two environment variables, \path{GPUPHOT_CUML_MIN_SOURCES} and \path{GPUPHOT_CUML_MAX_SOURCES}, holding the MIN and MAX source counts; a third, \path{GPUPHOT_USE_CUML_CROSSMATCH=1}, forces cuML for every crossmatch and exists for ablation only. With none of them set, \texttt{cKDTree} is used throughout, which is the default.

\subsection{Deterministic Execution}
\label{sec:impl_determinism}
Reproducibility of a GPU pipeline is not automatic: without the three provisions below, the same source code produces different catalogs on different GPU models and different zero points on different host CPUs. Each was identified in the benchmark and is pinned by the released code.
\begin{enumerate}[label=(\roman*)]
    \item \textbf{PCA solver.} The randomized SVD solver that \texttt{scikit-learn} selects by default for these matrix shapes is seeded from the global random state and therefore varies between processes; the difference propagates through the matched filter into $\pm 10{,}250$ detected objects on a 131{,}000-source field. Both PCA backends therefore use the full solver, which is exact and deterministic; flux, noise and S/N are bit-for-bit identical between the two solvers on a 19{,}565-source test field, so the choice affects reproducibility, not the science.
    \item \textbf{GPU clustering seeds.} The $k$-means implementation of cuML draws its initial centroids from a random-number generator whose stream advances by a multiple of the number of streaming multiprocessors of the device, so that two GPU models converge to different local optima from the same seed. The aperture-correction clusters, and through them the zero point, shifted by up to 0.047\,mag between the H100, A100 and L40S. The clustering is therefore seeded with explicit initial centroids computed on the CPU (ten deterministic $k$-means++ draws, keeping the one of lowest inertia), which removes the GPU generator from the computation; the CPU path pins its own random state.
    \item \textbf{BLAS kernel.} On the CPU path, OpenBLAS selects its kernel from the host microarchitecture and the kernels round differently, so the stages that run on the CPU can settle in different optima on different hosts.  On the host where OpenBLAS selects the Cooperlake kernel, an Intel Sapphire Rapids, a controlled comparison moved the zero point of one field by 0.0053\,mag, four times its formal error, and cut its calibrators from 186 to 112, while a second field failed to calibrate at all.  The SkylakeX and Zen kernels return the values of the rest of the fleet on the same images.  The three kernels detect the same number of objects on the divergent field, so the difference is not in the detection; the comparison does not separate the grouping from the principal component analysis, which runs on the CPU under this stack as well, and its Cooperlake arm rests on one observation per field. \gpuphot\ therefore pins \texttt{OPENBLAS\_CORETYPE} at import time from the CPU flags (SkylakeX on AVX-512 hosts, Haswell on AVX2 hosts, a pair verified to give identical results), unless the operator has set the variable explicitly, and warns once if the CPU grouping is entered with the variable unset.
\end{enumerate}
Section~\ref{sec:validation} reports the resulting agreement of catalogs and zero points across GPUs, hosts and library stacks.

\subsection{CPU Boundary: Astrometric Calibration}
\label{sec:impl_astrometry}
Astrometric calibration is the one stage that remains entirely on the CPU, because no GPU-native blind plate solver exists. The 500 most significant detections are passed to a local Astrometry.net solver \citep{langAstrometrynetBlindAstrometric2010}, seeded with the nominal plate scale and pointing, with a blind local solve and the online service as fallbacks; the World Coordinate System (WCS) solution, including a third-order distortion polynomial in the Simple Imaging Polynomial (SIP) convention, is written to the header and used to transform the catalog to celestial coordinates. The solver is held in an in-process singleton, so that the index files load once per worker process, and each attempt is bounded by a timeout implemented with \texttt{SIGALRM} (\texttt{GPUPHOT\_ASTROMETRY\_TIMEOUT}, 60\,s by default, overridable per instrument). The stage does not contend for VRAM, but it can dominate the latency of small frames and of hard fields: Section~\ref{sec:perf:jetson} shows a 4.2\,MP field whose single solve attempt takes two minutes on an ARM CPU, which is why the benchmarks use a 300\,s timeout. GPU-native astrometric calibration is identified as future work (Section~\ref{sec:future_directions}).

\section{Deployment, Availability, and Reproducibility}
\label{sec:deployment}

\gpuphot\ is deployed as a set of containers orchestrated by Docker Compose \citep{merkelDockerLightweightLinux2014, morrisUseDockerDeployment2017}, so that the CUDA runtime, the Python libraries and the astrometric index files travel with the image. The six x86 hosts of Section~\ref{sec:perf:setup}, from a laptop to datacenter servers, ran the same container image; the two ARM modules run the same code on a Jetson base image.

\gpuphot\ can also be used as a plain Python library without any Docker, broker, or database infrastructure (Section~\ref{sec:deployment:standalone}); the full Compose stack is only required for unattended queue-driven operation, additionally providing PostgreSQL/Q3C catalog persistence and horizontal multi-node scaling via Celery. Users evaluating \gpuphot\ on a new instrument are encouraged to start with standalone mode before deploying the distributed stack.

\subsection{Containerized Microservices}
The system is decomposed into several Docker services, defined in a \texttt{docker-compose.yml} file:
\begin{itemize}
    \item \textbf{\texttt{gpuphot\_worker}}: The core processing unit. This container includes the \gpuphot\ library, the Celery worker application, and the necessary CUDA runtime libraries. It is configured to access the host's GPU resources via the NVIDIA Container Toolkit.
    \item \textbf{\texttt{rabbitmq}}: The message broker service, based on the official RabbitMQ image. It handles task distribution and ensures persistence of the task queue.
    \item \textbf{\texttt{redis}}: The result backend service, providing fast in-memory storage for task results and status updates.
    \item \textbf{\texttt{postgres}}: The database service, utilizing a custom image that extends the official PostgreSQL image with the Q3C plugin for astronomical spatial indexing.
    \item \textbf{\texttt{flower}}: A web-based monitoring tool for Celery, allowing real-time inspection of worker status and task progress.
    \item \textbf{\texttt{lab}}: A JupyterLab service with GPU access, pre-configured with the \gpuphot\ environment for interactive analysis and notebook execution.
\end{itemize}

In the reference Compose file the Celery workers run on the \texttt{gevent} execution pool, which provides cooperative multitasking for the I/O-bound operations (database writes, message acknowledgements) without OS-level threading. The production deployment at the observatory runs instead the default prefork pool with long-lived worker processes, four per worker container on the A100 and H100 hosts and two on the L40S host, more than one container sharing each GPU, and no per-task recycling (deployed worker commands). In both configurations a worker process persists across images, which is the long-lived regime of Section~\ref{sec:perf:memory}. Periodic scheduling of maintenance tasks, such as catalog pruning or health checks, is managed by Celery Beat backed by the \texttt{celery-redbeat} scheduler, which persists the schedule state in Redis to ensure resilience across worker restarts.

\subsection{Scalability Strategies}
\gpuphot\ supports two primary scaling dimensions: vertical (multi-GPU) and horizontal (multi-node).

\subsubsection{Vertical Scaling (Multi-GPU)}
On a single machine with multiple GPUs, \gpuphot\ can launch multiple worker containers, each dedicated to a specific GPU device. We provide a helper script, \path{launch_gpuphot.sh}, which automatically detects the number of available NVIDIA GPUs and scales the \texttt{gpuphot\_worker} service accordingly. Each worker is assigned a unique \texttt{GPU\_ID} environment variable, ensuring that it exclusively uses its assigned device, preventing resource contention.

\subsubsection{Horizontal Scaling (Multi-Node)}
For larger workloads that exceed the capacity of a single server, \gpuphot\ supports distributed deployment across a cluster of machines. In this configuration, one node acts as the "Master," hosting the RabbitMQ, Redis, and PostgreSQL services. Other machines act as "Worker Nodes."

We provide a \texttt{launch\_external\_worker.sh} script for worker nodes. This script configures the local worker containers to connect to the message broker and result backend on the Master node via its IP address. Worker nodes can join or leave while tasks are running, because the queue lives in the broker and not in the workers.

\subsection{Configuration Management}
At runtime, GPU resource utilization is monitored through \texttt{GPUtil}, which provides programmatic access to per-device metrics such as VRAM occupancy, GPU load, and temperature. These readings are integrated into the hierarchical logging subsystem, enabling operators to correlate processing anomalies with hardware saturation events.

To manage the configuration across different environments, \gpuphot\ uses a centralized \texttt{.env} file. This file defines environment variables for file paths (such as image directories, calibration files), database credentials, and tuning parameters. This keeps credentials and site-specific paths out of the code and out of the container images.

\subsection{Edge Deployment: Jetson Orin}
\label{sec:deployment:jetson}

The containerized deployment described above targets x86\_64 servers with discrete NVIDIA GPUs.  NVIDIA Jetson Orin modules present two additional constraints that affect the deployment configuration.

\paragraph{Software stack limitations.}
RAPIDS \texttt{cuML} is unavailable on ARM (\texttt{aarch64}): no official wheels or conda packages are provided for this architecture.  All crossmatch and PCA operations therefore fall back to CPU-based \texttt{scipy} and \texttt{scikit-learn} on Jetson, regardless of the configured backend.  Additionally, NVIDIA Nsight Systems cannot profile GPU memory allocations on the Tegra platform; the unified memory architecture, in which CPU and GPU share the same physical LPDDR5 pool, makes discrete VRAM measurements architecturally ambiguous.

\paragraph{Sensor compatibility and sizing.}
The Jetson Orin modules are suitable for small-format sensors, including guiding cameras and calibration imagers, producing frames of about 4\,MP. The $2048 \times 2048$ COLORS frame (4.2\,MP) is the largest verified to complete on the 8\,GB unified memory pool with the released code. The 6.8\,MP FERVOR-M frame binned $3 \times 3$ ($3191 \times 2129$) exhausts the pool in the FFT stage on both modules, and on the JetPack~5 Orin~Nano the exhaustion hangs the host rather than raising an error (Section~\ref{sec:perf:jetson}).  Because that failure is a hang and not an error, an unattended Orin~Nano node should reject frames above the verified ceiling before dispatch; whether a memory limit on the container would convert the hang into a clean failure was not tested.  The pool is far from sufficient for the FERVOR-L survey cameras of the TTT1 and the TST, whose 37.8--151.2\,MP frames require more memory than is available for image data, FFT workspace, and operating system overhead simultaneously.  Two operational settings follow from the measurements of Section~\ref{sec:perf:jetson}: the astrometric timeout (\texttt{GPUPHOT\_ASTROMETRY\_TIMEOUT}) should be raised from its 60\,s default to 300\,s, because a hard field can need two minutes per plate-solve attempt on these CPUs, and the reference catalog should be reachable over a fast route, since a slow one dominated the latency of the Orin~Super by a factor of three in our tests.  Benchmark latencies for the Orin Super and Orin~Nano~8\,GB are reported in Section~\ref{sec:perf:jetson}.

\subsection{Standalone Library Integration}
\label{sec:deployment:standalone}

The full Docker Compose stack is designed for unattended queue-driven operation at a robotic facility.  For interactive analysis, instrument commissioning, or integration into an existing reduction workflow, \gpuphot\ can be used as a plain Python library without any broker, database, or container infrastructure.

In this mode, a caller constructs an \texttt{ImageProcessor} for the target instrument and calls \texttt{process\_image} on the FITS data and header:

\begin{lstlisting}[language=Python, basicstyle=\small\ttfamily]
from astropy.io import fits
from gpuphot.image_processor import create_processor

proc = create_processor('FERVOR-M')

with fits.open('/data/image.fits') as hdul:
    image_data = hdul[0].data
    fits_header = hdul[0].header

catalog, header = proc.process_image(image_data, fits_header)
\end{lstlisting}

\texttt{process\_image} returns a pandas \texttt{DataFrame} containing the source catalog and the translated FITS header with the photometric zero point (\texttt{ZP}, \texttt{EZP}, \texttt{CATNSTAR}) and the WCS solution appended; the WCS includes a third-order SIP distortion polynomial, which downstream consumers must apply to reproduce the catalog coordinates from pixel positions.  All memory management mechanisms described in Section~\ref{sec:impl_memory}, including the CuPy memory pool, checkpoint-based VRAM monitoring, and OOM fallback, remain active in this mode.  The only omitted components are the Celery task wrapper, RabbitMQ message routing, and PostgreSQL persistence; these are purely infrastructure concerns that do not affect the photometric algorithms.

The instrument identity passed to \texttt{create\_processor} resolves a JSON configuration file that maps FITS header keywords to pipeline parameters.  The repository ships only a default file; each observatory writes one such file per instrument, so the name in the example is not a file of the distribution.  Listing~\ref{lst:instrument_config} shows a representative configuration for the FERVOR-M camera on the TTT2, in the binned $2 \times 2$ mode of the 15.3\,MP benchmark frames.

\begin{lstlisting}[language=json, caption={Representative instrument configuration file for the FERVOR-M camera on the TTT2, binned $2 \times 2$ (4787$\times$3193\,px, 7.52\,\micron\ effective pixel).  The full parameter reference is in the repository documentation.}, label={lst:instrument_config}, basicstyle=\small\ttfamily]
{
  "header_keywords": {
    "gain":          "GAIN",
    "rdnoise":       "RDNOISE",
    "exposure_time": "EXPTIME",
    "filter":        "FILTER",
    "target_ra":     "OBJCTRA",
    "target_dec":    "OBJCTDEC"
  },
  "camera_specs": {
    "gain":    0.56,
    "rdnoise": 1.5,
    "pxsize":  7.52,
    "naxis1":  4787,
    "naxis2":  3193
  },
  "processing_params": {
    "center_factor":  0.7,
    "max_stars_ref":  15,
    "border":         50,
    "CR_filt":        false,
    "astrom":         true
  },
  "filter_map": {
    "Lum": "Lum",   "SDSSg": "SDSSg",
    "SDSSr": "SDSSr", "Ha": "SDSSr"
  }
}
\end{lstlisting}

Standalone mode is the recommended starting point on a new instrument: it runs the same pipeline without the broker, the database or the containers.

\subsection{Codebase, Licensing, and Versioning}
\gpuphot\ is developed as an open-source Python package hosted on a public Git repository (\url{https://github.com/Light-Bridges/GPUPhot}) under the MIT license. The framework follows a ``configuration-as-code'' philosophy: instrument definitions, deployment settings, and benchmark scripts are kept under version control alongside the core library and worker service. The software version accompanying this paper is release \texttt{v1.0.0}, which consolidates the framework with the CuPy memory pool restoration (Section~\ref{sec:impl_memory}) and the automatic selection of the host BLAS kernel (Section~\ref{sec:impl_determinism}). The runs behind the latency tables were executed with that allocator correction, verified inside each container by the content of the file that implements it, and under the same per-host BLAS kernel assignment that \texttt{v1.0.0} now selects by itself, so the measured deployment carries both provisions of the released version; the allocator ablation and the memory table also report, by design, the runs without the correction, which is what they measure. The reference-catalog comparison of Table~\ref{tab:speedup_local_vizier} predates this consolidation and is reported as measured. In addition, the empirical benchmark telemetry, screening scripts, and raw execution logs supporting the results in Section~\ref{sec:performance} are preserved under the \texttt{benchmarks/} directory. A persistent archive containing the exact release and benchmark datasets will be deposited in Zenodo upon acceptance for long-term independent verification.

\subsection{Supported Environments and Dependencies}
The framework is designed to operate across a range of observatory environments, from legacy control systems to modern GPU clusters. The standalone library mode supports the older Python versions (e.g., 3.8) still found in on-site telescope software, while the containerized microservices stack targets newer Python releases (e.g., 3.12) with recent CUDA toolkits. GPU acceleration currently relies on NVIDIA CUDA via CuPy; additional libraries such as RAPIDS \texttt{cuML} are detected at runtime, but \texttt{cKDTree} remains the default crossmatch backend because forcing cuML changes end-to-end latency by $-5$\% to $+5$\% at the source counts typical in real observations (Section~\ref{sec:cuml_eval}); the adaptive selection of Section~\ref{sec:impl:adaptive_cuml} enables cuML only inside its per-GPU window. When no GPU-capable library is available, the pipeline falls back to NumPy/SciPy-based implementations.

\subsection{Testing and Quality Assurance}
The codebase includes unit and integration tests targeting the core photometry pipeline, header translation, and worker orchestration. The test suite is built on the \texttt{pytest} framework to detect regressions across the supported Python environments. In addition, the project provides Jupyter notebooks under version control that exercise complete end-to-end workflows, serving both as executable documentation and as validation tests for typical instrument configurations.

\subsection{Benchmark Reproducibility}
Performance results in this work are derived from the per-run events that the workers ship to Elasticsearch, exported to CSV files kept under \texttt{benchmarks/data/}, with the raw per-run exports and their provenance notes under \texttt{benchmarks/data/raw/}. The chain is reproducible end to end: \path{benchmarks/build_unified_csv.py} rebuilds the benchmark CSV byte for byte from the raw export, applying the screening rules of Section~\ref{sec:perf:setup} with each rule documented in the code; \path{benchmarks/build_cell_status.py} builds the per-cell ledger of failed runs that labels the empty cells of Figure~\ref{fig:heatmap}; and \path{benchmarks/generate_manuscript_tables.py} and \path{benchmarks/generate_manuscript_figures.py} regenerate all benchmark figures and performance tables of Section~\ref{sec:performance} from those files (ten of the thirteen tables; the remaining three summarize hardware specifications, software versions, and configured cuML thresholds). The benchmark runner launches one process per repetition inside the profiler containers, with the warm-up, the catalog backend, the astrometric timeout and the per-host kernel selection fixed in the runner script (\path{benchmarks/results_collected/latency_campaign/run_latency_campaign.sh}). We encourage users to re-run these analyses on their own deployments and to regenerate the performance figures and tables by invoking the same scripts on locally collected benchmark data.

A persistent archive of the software and benchmark data will be deposited in Zenodo upon acceptance; the DOI will be included in the final version.
\section{Functional Validation with Real-World Data}
\label{sec:validation}

While computational acceleration is the primary goal of \gpuphot, it is necessary to verify that the photometric and astrometric outputs are scientifically valid. A full astronomical characterization, including completeness, contamination, depth analysis, and detailed comparison with established CPU-based photometry tools, is reserved for the companion paper \citep{alarconGPUPHOT2026}. In this section, we present a functional validation demonstrating that the pipeline operates correctly, and reproducibly, across the full range of instrumental setups and hardware platforms.

\subsection{Multi-Scale Dataset Description}
We utilized a validation dataset consisting exclusively of real scientific observations obtained from the TTT and TST at the Teide Observatory (Canary Islands, Spain). These robotic facilities provide a heterogeneous data environment, ranging from standard CCD frames to ultra-high-resolution sCMOS mosaics, which exercises the pipeline's throughput and memory management across a wide operating range.

The dataset covers five image sizes from five cameras in six configurations, as detailed in Table~\ref{tab:datasets}. These sizes are the ones the facility produces rather than the nominal sensor formats. In the results database of the observatory, seven read formats account for over 99\% of the frames: the full frames of 151.2\,MP (FERVOR-L on the TTT and the TST), 61\,MP (FERVOR-M and the ATLAS-Teide cameras) and 4.2\,MP (COLORS), the binned modes of the FERVOR-L ($2 \times 2$, 37.8\,MP) and of the FERVOR-M ($2 \times 2$, 15.3\,MP, and $3 \times 3$, 6.8\,MP), and an occasional 100\,MP window of the TST. Five of the seven are size classes of the benchmark, and the binned modes are not marginal: in 2026 the FERVOR-L of the TTT1 delivered 200\,210 frames binned $2 \times 2$ against 22\,552 at full resolution, and the FERVOR-M of the TTT2 180\,384 binned $2 \times 2$ against 9\,539 full frames. For each image size, between two and six images with different source densities were selected, nineteen in total, so that the effect of source count on processing time can be separated from that of pixel count. At 151.2\,MP the cataloged source count ranges from 808 to 134{,}206, and the end-to-end latency between these two extremes differs by a factor of 1.6 on the H100 under the CuPy~12 stack, 19.0 to 30.2\,s (Section~\ref{sec:perf:latency}). That effect would be masked if only one image per size were tested. Source counts quoted in this paper are the numbers of objects in the final catalog produced by the configuration of each camera in production use for this benchmark. A unification of the camera configurations across hosts changed these counts with respect to earlier measurements, for instance from 428 to 808 objects on the sparse 151.2\,MP field and from 131{,}397 to 134{,}206 on the densest one; the table that necessarily relies on those earlier measurements (Table~\ref{tab:speedup_local_vizier}) carries the earlier counts and says so.

Two of the five cameras, the COLORS (4.2\,MP) and the FERVOR-M on the TTT3 in its 6.8\,MP binned configuration, run with the Eigen-PSF stage disabled (\texttt{pca\_method = false}) in their production configuration. They contribute four of the nineteen benchmark images and, because small frames run on more hosts, 33\% of the repetitions of the benchmark. For these frames the detection kernel is the stacked reference PSF alone, and the aperture correction is field-wide because the configured tile size exceeds the frame. The benchmark uses each configuration unchanged, so the 4.2 and 6.8\,MP rows of the performance tables exercise a shorter pipeline than the larger frames; this matters when interpreting the cost of the GPU PCA backend (Section~\ref{sec:perf:memory}), which cannot contribute to those rows.

\begin{table*}[t]
\centering
\caption{Observational datasets used for validation and performance benchmarking, one row per camera configuration, all from real observing conditions at the Teide Observatory. The pipeline derives the plate scale from the pixel size and focal length of each header and sizes the PSF stamp from it (Section~\ref{sec:impl_psf}), so the stamp column follows from the two preceding ones. The sparse 151.2\,MP benchmark field is the unbinned FERVOR-L@TTT1 frame; the five dense 151.2\,MP fields are FERVOR-L@TST frames.}
\label{tab:datasets}
\footnotesize
\setlength{\tabcolsep}{4pt}
\begin{tabular}{l c c l c c c c r c c c}
\toprule
\textbf{Telescope} & \textbf{\shortstack{Aperture\\(m)}} & \textbf{\shortstack{Focal\\(m)}} & \textbf{Instrument} & \textbf{Bin} & \textbf{\shortstack{Pixel\\($\mu$m)}} & \textbf{\shortstack{Scale\\($''$/px)}} & \textbf{\shortstack{Frame\\(px)}} & \textbf{MP} & \textbf{\shortstack{Stamp\\(px)}} & \textbf{\shortstack{Eigen-\\PSF}} & \textbf{Images} \\
\midrule
TTT1 & 0.8 & 5.48 & FERVOR-L & $1 \times 1$ & 3.76  & 0.1415 & $14200 \times 10650$ & 151.2 & 141 & on  & 1 \\
TTT1 & 0.8 & 5.48 & FERVOR-L & $2 \times 2$ & 7.52  & 0.2831 & $7100 \times 5325$   & 37.8  & 71  & on  & 4 \\
TST  & 1.0 & 1.30 & FERVOR-L & $1 \times 1$ & 3.76  & 0.5966 & $14200 \times 10650$ & 151.2 & 33  & on  & 5 \\
TTT2 & 0.8 & 5.48 & FERVOR-M & $2 \times 2$ & 7.52  & 0.2831 & $4787 \times 3193$   & 15.3  & 71  & on  & 5 \\
TTT3 & 2.0 & 12.0 & FERVOR-M & $3 \times 3$ & 11.28 & 0.1939 & $3191 \times 2129$   & 6.8   & 103 & off & 2 \\
TTT3 & 2.0 & 12.0 & COLORS   & $1 \times 1$ & 13.50 & 0.2320 & $2048 \times 2048$   & 4.2   & 87  & off & 2 \\
\bottomrule
\end{tabular}
\begin{flushleft}
\footnotesize \textit{Columns:} Telescope, unit of the TTT or the TST that produced the frames, with its aperture and focal length; Instrument, camera; Bin, on-chip binning; Pixel, effective pixel size after binning, as read from the header; Scale, plate scale derived by the pipeline from the two preceding columns; Frame, pixel dimensions of the frame as processed; MP, pixel count in megapixels, the size label used throughout the paper; Stamp, side in pixels of the square PSF stamp that the rule of Section~\ref{sec:impl_psf} assigns to that plate scale; Eigen-PSF, whether the production configuration of the camera enables the principal component analysis (PCA) stage of the PSF model (Section~\ref{sec:validation}); Images, number of the nineteen benchmark images taken in that configuration.
\end{flushleft}
\end{table*}

\subsection{High-Resolution Stress Test}
The processing of full-resolution images from the TST, with a native resolution of 151.2\,MP, serves as a benchmark for the system. Handling such large arrays requires allocating about 605\,MB (577\,MiB) of contiguous memory for a single \texttt{float32} image buffer, and substantially more for intermediate processing steps such as FFT convolution and background estimation: the measured peak working set of a 151.2\,MP frame is 28.9\,GiB (Section~\ref{sec:perf:memory}). All benchmark images were processed successfully on GPUs with 48~GB or more VRAM (H100, A100, L40S) without manual intervention, confirming that the checkpoint-based memory management strategy described in Section~\ref{sec:impl_memory} is effective for the largest frames produced by the TTT/TST facility, while the 24\,GB RTX~3090 cannot hold the working set.

\subsection{Pipeline Correctness}
For each processed image, we verified the following functional outputs:

\begin{enumerate}
    \item \textbf{Source detection and reproducibility}: the same input image yields the same catalog on different GPU models. For all 22 exposures of the benchmark collection (the nineteen fields of Section~\ref{sec:performance}, two of which are represented by two or three exposures), the Python~3.12 stack produced identical source counts and identical zero points, to the fourth decimal, on the H100, A100 and L40S across all 38 combinations of image and configuration (adaptive and forced cuML crossmatch modes agreeing exactly; Section~\ref{sec:cuml_eval}); across all 57 combinations spanning both Python stacks, 55 returned identical values (7\,131 runs), the two exceptions arising under Python~3.8 in labels that pool several exposures where a different exposure was measured. Measured again with the allocator correction of Section~\ref{sec:perf:memory}, 167 of the 171 cells common to the two arms kept identical source counts and 157 returned a single zero point; the exceptions are all cells of the two labels that pool several exposures. The Python~3.8 stack agrees across hosts once the BLAS kernel is pinned (Section~\ref{sec:impl_determinism}); before that, one host returned zero points that differed by up to 0.0053\,mag on part of the images, and failed to calibrate one of them. The ARM Jetson Orin, on which no kernel pinning is applied, reproduces the x86 zero point of the 4.2\,MP field exactly.
    \item \textbf{Astrometric solution}: the Astrometry.net solver produces a WCS solution, including a third-order SIP distortion polynomial, for all images where sufficient reference stars are available ($\geq 5$ isolated stars with S/N $> 10$, as required by the pipeline); fields that fail this guard raise an \texttt{InsufficientStarsError} and are logged for manual review. Each solve attempt is bounded by a timeout (60\,s by default, 300\,s in all our benchmarks). The value matters: one of the 4.2\,MP benchmark fields, the comet C/2025~A6, needs about two minutes per attempt on the ARM CPU of the Jetson Orin~Nano and fails three consecutive attempts under the 60\,s default, while it solves reliably under the 300\,s setting (Section~\ref{sec:perf:jetson}).
    \item \textbf{Photometric calibration}: zero-point magnitudes are computed as the sigma-clipped mean offset at unit slope (Section~\ref{sec:impl_zeropoint}) against Pan-STARRS \citep{chambersPanSTARRS1Surveys2016}, Gaia~DR3 \citep{gaiacollaborationGaiaDataRelease2023}, or SkyMapper \citep{onkenSkyMapperSouthernSurvey2024} reference catalogs, depending on the sky position and filter. The zero point, its standard error (\texttt{EZP}) and the number of retained calibrators (\texttt{CATNSTAR}) are recorded in the structured catalog headers; a systematic characterization of zero-point residuals across instruments and filters is presented in the companion paper \citep{alarconGPUPHOT2026}.
    \item \textbf{Catalog output}: structured photometric catalogs are written to PostgreSQL with Q3C spatial indexing, enabling sub-arcsecond cone searches on the resulting database.
\end{enumerate}

\subsection{PSF Modeling Configuration}
The PSF model of Section~\ref{sec:impl_psf} runs with the production configuration of each camera. The reference PSF is stacked from the central region of the frame, selected per axis by \texttt{center\_factor} (0.5 to 0.9 across the benchmark cameras; the default 0.7 is the central 49\% of the area), from at most \texttt{max\_stars\_ref} = 15 stars, the default in all five cameras; the PCA uses all isolated stars of the frame so that the coefficient maps cover the field. Fewer than five isolated stars raise an exception and the image is logged for manual review. The FWHM and $\beta$ of the Moffat fits to the reference PSF and to the PSF reconstructed at four field corners are recorded as diagnostic header metadata.

\subsection{Scientific Validation}
Quantitative assessment of photometric accuracy (magnitude residuals vs.\ reference catalogs), astrometric precision (positional root-mean-square residuals against Gaia~DR3), and PSF modeling fidelity (residual flux after subtraction) are beyond the scope of this software-focused manuscript. The companion paper \citep{alarconGPUPHOT2026} validates the same pipeline through a controlled injection--recovery experiment on four instruments spanning an order of magnitude in plate scale, reporting 50\% completeness limits between $\mathrm{S/N} \simeq 5$ and 9, sub-pixel astrometric residuals of 0.11--0.38\,arcsec per axis, and a brightness-independent photometric offset of about 0.05\,mag that is absorbed by the zero point, together with the comparison against external catalogs.

\section{Performance Evaluation}
\label{sec:performance}

We evaluated the end-to-end processing performance of \gpuphot\ across eight
GPU configurations spanning datacenter, consumer, laptop, and edge hardware.
The evaluation covers latency, memory consumption, concurrency capacity,
and comparison with CPU-only pipelines.  Unless stated otherwise, the numbers
in this section are medians over repeated one-process-per-image runs of the
released code, with the CuPy pool re-seated per task; where a cell was
measured in more than one session, its sessions are pooled after the
exclusions of Section~\ref{sec:perf:setup} and its upper tail is trimmed by
the rule stated there, recorded with the data.  The reference-catalog
comparison of Section~\ref{sec:catalogue_backend} comes from an earlier set
of runs of the same benchmark, and Section~\ref{sec:perf:memory} analyzes
the unpooled allocator that importing cuML installs as an ablation.  The two
Jetson modules run without cuML, so their allocator is the pool in every
session.  Each table states the configuration of its data.

\subsection{Experimental Setup}
\label{sec:perf:setup}

\subsubsection{Hardware}

Table~\ref{tab:gpu_hardware} lists the eight GPU platforms used in this study.
The set includes two 80\,GB datacenter accelerators (H100 PCIe and
A100-SXM4-80GB), one 48\,GB datacenter GPU (L40S), two consumer-grade desktop
GPUs (RTX~3090 and RTX~3060), a laptop GPU (RTX~3050~Ti), and two NVIDIA
Jetson Orin modules with shared CPU/GPU unified memory.  Each GPU resides in a
different host, so host-level factors (CPU, driver, memory;
Table~\ref{tab:gpu_hardware}) are not separable from the GPU itself in
cross-platform comparisons.  The two Jetson modules in
particular are two units of the same module part number, p3767-0005 (Orin~Nano 8\,GB, where an Orin~NX would be p3767-0000 or p3767-0001), of different board revisions, running two software generations: the Orin~Nano runs JetPack~5 (CUDA~11.4, Python~3.8 only) and the Orin~Super runs
JetPack~6 (CUDA~12), on which both stacks were measured.  The Nano and Super
columns therefore compare two boards; only the Orin~Super compares the two
interpreters on one board.

\begin{table*}[t]
\centering
\caption{GPU hardware used for benchmarking and the host of each card.
VRAM figures denote total device memory; on Jetson Orin modules the memory
is shared between CPU and GPU.  The host columns were read from the hosts themselves; a dash marks a
value that could not be read.}
\label{tab:gpu_hardware}
\footnotesize
\setlength{\tabcolsep}{3pt}
\begin{tabular}{lrllrrll}
\toprule
\textbf{GPU} & \textbf{VRAM} & \textbf{Type} & \textbf{Host CPU} & \textbf{Cores/thr.} & \textbf{RAM} & \textbf{Driver} & \textbf{PCIe} \\
\midrule
H100 PCIe           & 80  & Datacenter     & AMD EPYC 9124                      & 16/32  &      62.5 & 565.57.01  & Gen5 $\times$16 \\
A100-SXM4-80GB      & 80  & Datacenter     & 2$\times$ Intel Xeon Gold 6354     & 36/72  & 1\,007.5 & 535.247.01 & SXM4 \\
L40S                & 48  & Datacenter     & 2$\times$ Intel Xeon Platinum 8468 & 96/192 &     251.6 & 535.309.01 & Gen4 $\times$16 \\
RTX 3090            & 24  & Consumer       & Intel Core i9-10940X               & 14/28  &     125.5 & 575.57.08  & Gen3 $\times$16 \\
RTX 3060            & 12  & Consumer       & Intel Core i5-10400F               & 6/12   &      62.7 & 570.133.07 & Gen3 $\times$16 \\
RTX 3050 Ti Laptop  &  4  & Laptop         & Intel Core i7-12700H               & 14/20  &      31.0 & 580.173.02 & Gen1 $\times$16$^a$ \\
Orin Super          &  8  & Edge (Tegra)   & Arm Cortex-A78AE                   & ---/6  &       7.4 & 540.4.0    & Integrated \\
Orin Nano 8\,GB     &  8  & Edge (Tegra)   & Arm v8                             & 6/6    &       7.3 & ---$^b$    & Integrated \\
\bottomrule
\end{tabular}
\begin{flushleft}
\footnotesize \textit{Columns:} GPU, device model; VRAM, nominal device memory in GB; Type, deployment class of the device; Host CPU, processor of the host, two sockets where marked; Cores/thr., physical cores and hardware threads of the host; RAM, host memory in GiB (\texttt{MemTotal}); Driver, NVIDIA driver version at the time of reading; PCIe, generation and width of the link reported by the driver, SXM4 for the A100, integrated for the Jetson modules.  $^a$Link state as read on the laptop; the generation the driver reports at rest is not necessarily the capability of the link.  $^b$JetPack~5 host without \texttt{nvidia-smi}; the driver version is not readable there.  The two Jetson entries report the same module part number, p3767-0005 (Orin~Nano 8\,GB), in the same field of the device tree of each host, attested as read from the module EEPROM on one of the two; their board revisions and serial numbers differ, so they are two units.  Their two RAM figures are the memory each JetPack release leaves to the system on that module, not different memory.  The JetPack~5 host carries the label \texttt{Orin NX 8GB} in the benchmark data files, assigned before the module was read.\par
\textit{Note:} The A100-SXM4-80\,GB was operated at a 300\,W power limit (500\,W default) and the H100~PCIe at 220\,W, the highest its chassis sustains, in every measurement of the tables and figures of this section, except 37 of its 2\,790 repetitions at 200\,W during thermal steps of the host.  The runs without the CuPy pool of Table~\ref{tab:allocator_ablation} predate that setting and are the exception: there the card ran mostly at 350\,W, and the composition of each of its cells is given with that table. Neither limit binds within one image: a single 151.2\,MP reduction draws at most 282\,W on the A100 and 180\,W on the H100 (0.5\,s sampling), with median draws of 73\,W and 94\,W; over the benchmark run windows the A100 median draw was 78\,W, with 1.5\,\% of the samples above 290\,W, transients that a limit enforced on a moving average allows. All other tested GPUs operated at their hardware-default power limits.
\end{flushleft}
\end{table*}

\subsubsection{Software}

All benchmarks were executed under two Python environments, hereafter py3.8
(Python~3.8 with CuPy~12 and no cuML) and py3.12 (Python~3.12 with CuPy~14 and
adaptive cuML), to quantify the impact of the newer CuPy~14 / NumPy~2.0 stack.  Table~\ref{tab:software_versions}
summarizes the key library versions.

\begin{table*}[t]
\centering
\caption{Software stack versions of the two benchmark environments.
The py3.12 stack ships CuPy~14, NumPy~2.0 and cuML~25.2; its memory and
latency differences from py3.8 trace to the allocator that cuML installs,
not to these libraries (Section~\ref{sec:perf:memory}).  Under py3.12, cuML
is activated only when the source count falls within per-GPU thresholds
(Section~\ref{sec:impl:adaptive_cuml}).  CUDA denotes the container base
image.  Every py3.12 measurement in this section has the CuPy pool
re-seated per task, the released configuration, except the reference-catalog
comparison of Table~\ref{tab:speedup_local_vizier}, measured with the
unpooled allocator that importing cuML installs;
Table~\ref{tab:allocator_ablation} quantifies the difference.}
\label{tab:software_versions}
\begin{tabular}{lll}
\toprule
\textbf{Component} & \textbf{py3.8} & \textbf{py3.12} \\
\midrule
Python        & 3.8.10  & 3.12.3  \\
CuPy          & 12.3    & 14.0.1  \\
NumPy         & 1.23    & 2.0.2   \\
SciPy         & 1.9     & 1.15.2  \\
scikit-learn  & 1.2     & 1.8.0   \\
CUDA (image)  & 12.2    & 12.6    \\
cuML          & ---     & 25.2.1 (adaptive) \\
\bottomrule
\end{tabular}
\begin{flushleft}
\footnotesize \textit{Columns:} Component, library or runtime; py3.8 and py3.12, the version installed in each of the two container images used in the benchmarks.  The CUDA runtime of each image is the same on the six x86 hosts (12.2.2 and 12.6.3); on the Orin~Super both images run CUDA~12.2.12 from JetPack~6, and the py3.8 image of the Orin~Nano, the only one that host runs, CUDA~11.4.
\end{flushleft}
\end{table*}

\subsubsection{Benchmark Images}

The benchmark collection comprises nineteen images at the five image sizes
produced by the cameras of the TTT and TST at the Teide Observatory (see
Table~\ref{tab:datasets} for full instrument details).  The images range from
4.2\,MP ($2048\times2048$, COLORS) to 151.2\,MP ($14200\times10650$, FERVOR-L
unbinned), spanning a $36\times$ factor in pixel count, and from 228 to 134{,}206
cataloged sources.  Each size class contains between two and six images of
different source density.  Two of the nineteen entries pool exposures of the same field taken
minutes apart, one two exposures (3\,977 and 4\,754 sources) and one three
(3\,515 to 3\,718 sources), 22 files in total; the reported time is the
median over all their repetitions and the Sources column lists the count of
one exposure.  Within a cell the medians of the pooled exposures differ by
2\% or less in median and by up to 9\% and 32\% in single cells.  As noted in
Section~\ref{sec:validation}, the two smallest cameras run without the
Eigen-PSF stage in their production configuration, and the benchmark uses
the production configuration of every camera unchanged.

\subsubsection{Methodology}

For each (GPU, image, configuration) cell, repetitions were run in sessions
of five warm-up and twenty timed repetitions.  The sampling depth differs
between stacks and hosts: the 57 datacenter cells of py3.12\,+\,adaptive
cuML were all measured in two or more sessions, a median
of 74 raw repetitions per cell (62 to 258); the 57 datacenter cells of
py3.8 come from the runs with the CuPy pool re-seated
(Section~\ref{sec:perf:memory}) and each rests on a single clean block, a
median of 12 raw repetitions (12 to 138); the 62 cells of the consumer and
Jetson GPUs have a median of 22 to 25 raw repetitions and, in eight of
them, two sessions.  Over the 176 cells of both tables the median is 40
raw repetitions, and 85 cells reach two sessions.  Each repetition is an
independent Python process launched inside the
benchmark container, so every timed run includes the per-process costs that
are paid inside the pipeline (CUDA context and library initialization,
loading of compiled kernels from cache, catalog connection); the warm-up
repetitions populate the on-disk caches (compiled kernels, astrometric index
files, database buffers) and are discarded.  The consequences of this
one-process-per-image regime are discussed in Section~\ref{sec:perf:memory}.

The timed quantity is the wall-clock time of the core \path{process_image}
function, recorded by the pipeline's own logging decorator and shipped to
Elasticsearch by the worker; it excludes FITS input and header translation.
Comparing this value with the Elasticsearch document timestamps on 177 runs
gives a median bias of $+0.039$\,s (0.26\% of the benchmark median), so the
decorator adds nothing measurable.  Warm-up removal is applied session by
session; upper outliers are removed with a criterion of three median
absolute deviations above the median of the cell, on the upper side only,
because the wall-clock time has a floor, the work, and no ceiling, so a
slow repetition can be an artifact of the host and a fast one cannot; and
the median of
the remaining clean repetitions (a median of 20 per cell: 53 in the
datacenter cells of py3.12 and 10 in those of py3.8) is reported.
The relative inter-quartile range (IQR) of the clean repetitions has a median
of 5.0\% and stays below 13\% for 90\% of the cells; the largest values, up to
43\%, occur in the 134\,206-source cells, where the query to the shared
catalog server is the largest variable component of the wall-clock time
(Section~\ref{sec:perf:latency}).

The hosts are shared machines, not an isolated testbed.  With the
production worker stopped on the GPU under test, each host still runs
production on its other GPUs, jobs of other users and, on the H100 host, a
numerical weather model on the CPU, and the three datacenter hosts query
one catalog server.  A test on the H100 host measured what
the sharing costs on two 151.2\,MP frames: production on the GPU under
test adds 14 to 19\% to the latency, the weather model alone about 27\%
relative to a cold reference, and the two together about 45\%; a session
under the model, excluded below, ran 48\% slower than a session of the
same host taken outside it.  The load is not constant within a block either,
0.14 to 0.89 per core across the 25 repetitions of an acceptance block,
and some repetitions run 25\% or more above their block
median, 6 of 36 in one session, with the host load, temperature, power and
memory identical inside and outside the excursion.  In two separate
observations on the H100 host, a witness frame read together with the
load of the catalog server ran at 58 and 66\,s against a clean value of
24 to 28\,s
while the GPU was idle, cool and at its normal power limit and the catalog
server stood at 0.83 to 0.94 load per core, an ordinary load for it (the
28th and 48th percentiles of its load during the two long runs below):
two observations, not a curve, which rule out that the mechanism needs
the GPU busy, and which have the size of the episodes described below,
2.1 and 2.7 times the clean value against 1.4 to 3.1 for the episodes, on
the same image and host.  On the A100 host one block was not measured because the
host, with its GPU idle and the production worker stopped, stood at 0.50
load per core from jobs of other users, one process at 561\% of a CPU,
and the py3.8 stack runs its clustering and PCA on the CPU.  Two
good-faith attempts at the same cell on the same host, five minutes apart
and under the same protocol, gave 25.3 and 28.5\,s, 13\% apart, one
failing the acceptance criterion and the other passing it; that is the
dispersion of the measurement taken directly, it exceeds several of the
effects discussed below, and it stands behind every single-cell figure of
the tables, which print one decimal by convention, so differences of that
order between cells are not differences.  The variability has temporal
structure: slow repetitions come in runs.  Marking as slow a repetition
above a fixed multiple of the first quartile of its block, and shuffling
each block 300 times, which keeps its distribution and destroys only its
order, the longest run of slow repetitions across the blocks of the
benchmark with at least 20 repetitions is 4.7 to 4.9 observed against 2.2
to 2.4 shuffled, at three thresholds ($z$ between 25 and 34), and at the
middle threshold 46 blocks hold a run of four or more against 14 expected
by chance.  Two continuous runs of 59 and 100 repetitions on one cell of
the H100 host each contained six episodes, the longest of 12 repetitions
near 60\,s and 8 near 76\,s over a base near 34\,s.  The episodes are
recurrent, their share varies widely between runs, they are separate from
the cold start (a run that starts cold begins at 82\,s, one that starts
warm begins at its base level), and their cause is not identified:
contention on the catalog server, the host load, a busy GPU, power and
temperature are each ruled out by the data, and whether the mechanism is
external to the pipeline or inside it is not established.  A block of 25
repetitions falls inside or outside an episode, which is consistent with
the 13\% between the two attempts above.  Four features
of the design contain this.  The two paired comparisons
(Figures~\ref{fig:py38_vs_adaptive} and~\ref{fig:always_vs_adaptive}) put
both arms of a cell inside the same session, so a host-level slowdown
enters numerator and denominator alike.  The ranking of the GPUs is read
on the medians over the 19 images, not on single cells.  Windows in which
a control shows the host, not the pipeline, at work are excluded by a
written criterion with citable evidence, five in all, listed next.  And
the signature of an intermittent interference was searched for in all 883
blocks of the benchmark and found in 10, six of them one heavy-tailed
image (below).

Whole sessions are excluded when a control inside the benchmark shows the
host, not the pipeline, at work; the exclusion list is part of the table
generator, each entry carries its evidence, and five windows are on it.
The session of the H100 host during which a numerical
weather model occupied the host CPU: its cells ran a median 48\% slower
than the session of the same host outside that window (interquartile range 44--52\%), and
the same paired control on the other two hosts, which do not run the model,
is indistinguishable from zero.  One hour of py3.8 on the RTX~3090 host while the
production catalog server co-located with it held the CPU at full load:
py3.8, whose clustering runs on the CPU, came out 31--41\% slower than the
rest of the benchmark from its first repetition, while py3.12 in the same runs came out
6\% faster.  One hour of py3.8 on the RTX~3050~Ti laptop
in which the time stepped from 19 to about 42\,s after four minutes of
back-to-back repetitions and did not return, so the session mixes two
regimes.  And one block of 25 repetitions of the 18\,712-source 151.2\,MP
field under py3.12 on the H100, in one session of that host: the
block ran at a median 53.2\,s while the other 18 images of the same session
all ran faster than in any other session, 17 of them by 14--34\%, every
repetition returned
the same zero point, error, calibrator count and source count, the power
limit and the temperature did not move, and the last three repetitions
fell back to 24--37\,s; the cause is not established, and the two known
loads of that host, production on the GPU and the weather model, reach
45\% together (above) and do not reproduce it; the block has the level of
the episodes measured on that host (above), 1.6 times the base against
1.4 to 3.1 for those measured, though it would be about twice as long as
the longest one measured, 25 repetitions against 12, and it is read as
one of them, of unidentified cause.  That cell is
reported from its remaining session, which ran about a third
slower than the excluded one on every image the two share, so its value
in Table~\ref{tab:latency_py312}, 33.2\,s, is a clean measurement that is
not comparable with its column neighbors, which come predominantly from
that other session.  And one block of ten repetitions of the 10\,168-source
field under py3.8 on the A100 host, which ran at a median
46.6\,s with the same signature, 44 to 57\,s with the last two repetitions
at 27.7 and 28.5\,s, while the twenty other blocks of that session showed
none (median-to-tail ratios of 0.94 to 1.07 against 1.66).  That cell was
measured again in a block of 25 repetitions that passed
three conditions fixed before the measurement, a median-to-tail ratio in
the band of its session companions (0.96), a witness image within 1\% of
its value in the runs the block replaces, and a drift of 2\% between the halves
of the block; the new block replaces the old one in
Table~\ref{tab:latency_py38} (28.3\,s over 22 clean repetitions) rather
than being pooled with it, because the median of the two together,
29.1\,s, would represent neither.  The new value differs from the 25.1\,s
of the block it replaces, a shift of 14\% relative to the witness image
that is not explained.  None of the three excluded
blocks was a borderline case: each fails the median-to-tail condition by
1.6 to 2.0 times the edge of its band.  The paired comparison of
Figure~\ref{fig:py38_vs_adaptive} keeps the session of the H100
host under the weather model, 15 of its 18 pairs, because its two arms alternate inside that window
and a host-level slowdown enters both arms.  Two pairs are excluded from
the figure by a rule written in its generator with its evidence, because
the slowdown did not enter both arms: the 18\,712-source field on the
H100, whose arms slowed by $2.4\times$ and $1.9\times$ relative to the
fastest session of each, and the 10\,168-source field on the A100, whose
py3.8 arm is the replaced block ($1.64\times$) while the py3.12 arm that
ran right after it was not slowed ($0.94\times$); the same test applied to
the other pairs of the H100 host justifies no further exclusion.  The
signature of the excluded block, a median that
exceeds the median of its last two repetitions by a factor of 1.3 or more
while the other blocks of the same session show no such tail, was then
sought in all 883 blocks of the benchmark.  It flags ten.  Three are
the excluded blocks, the two of the H100 ranked first at 2.1 and the
replaced A100 block at 1.7; six are blocks of the 134\,206-source field on
three hosts and both cuML modes of py3.12, the heavy tail of that field
rather than an interference (6 of its 40 blocks, against 1.1\% of all
blocks), three of which feed the L40S entry of
Table~\ref{tab:latency_py312}; and one comes from a run that no published
cell uses.  An interference that lasts a whole block leaves no fast tail
and is invisible to this test.

The reference catalog for every cell was a local PostgreSQL/Q3C replica of
Pan-STARRS\,DR1 on the observatory network, a single server shared by
every benchmark host and running on the RTX~3090 host, which served the
catalog to the others while measuring its own cells
(Section~\ref{sec:catalogue_backend} compares this backend against
VizieR), so the cells of different hosts share the catalog server and are
not independent of one another.  The astrometric
timeout was set to
300\,s, the production workers were stopped on the GPU under test, and the
H100 host was measured in windows in which no other process shared the GPU.
Only runs that produced a photometric catalog count as measurements.  The
outcome of every failed run was recorded in a per-cell ledger, which
distinguishes memory exhaustion on the device, exhaustion of pinned
(page-locked) host memory, and a host that hung, so that no empty cell of the tables or of
Figure~\ref{fig:heatmap} is left without a cause.
PCA computations in the PSF stage use \texttt{svd\_solver='full'} for
determinism (Section~\ref{sec:impl_determinism}).  Peak GPU memory was
measured with NVIDIA Nsight Systems (\texttt{nsys}) in separate runs, whose
timings are not used.  Measured memory is reported in binary units (MiB, GiB); the nominal device memory of Table~\ref{tab:gpu_hardware} is the manufacturer's figure in GB.

\subsection{End-to-End Latency}
\label{sec:perf:latency}

\begin{figure*}[t]
    \centering
    \includegraphics[width=\textwidth]{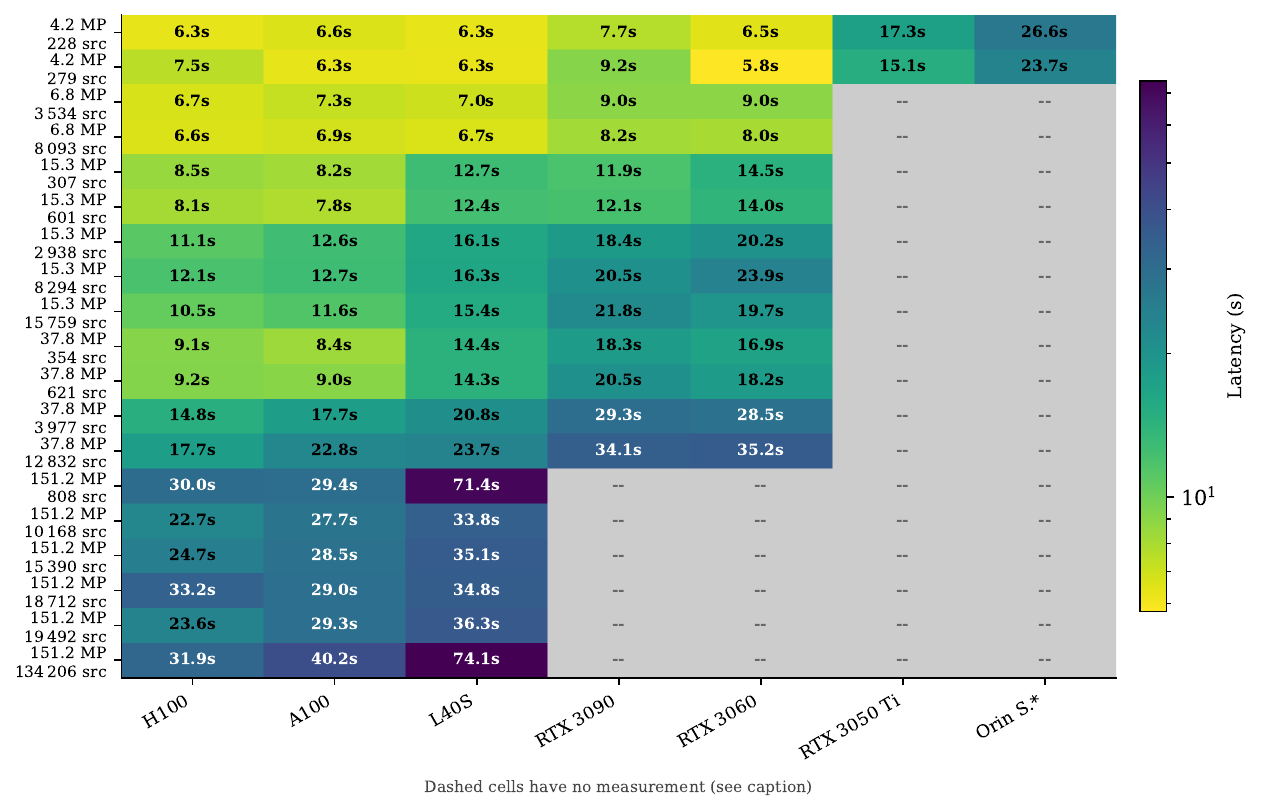}
    \caption{Median end-to-end latency (seconds) by GPU and benchmark image under py3.12\,+\,adaptive cuML, one process per image. Rows are the benchmark images, labeled by frame size and source count, and columns are the GPUs. Cell color encodes the median latency on a logarithmic scale, from bright (fast) to dark (slow), and the value in seconds is printed in each cell; gray cells marked with a dash have no measurement. Of the 46 empty cells, 19 were attempted and failed by memory exhaustion, 14 on the device and 5 in the pinned host memory of the RTX~3090 host, which the co-located catalog database starves (Section~\ref{sec:perf:latency}); the other 27, on the RTX~3060, RTX~3050~Ti and Orin~Super, were not launched because a frame of the same size or smaller had already exhausted the memory of that device. Orin~Super ($^*$) runs py3.12 on ARM without cuML; Orin~Nano is omitted (no py3.12 data; see Table~\ref{tab:latency_py38}).}
    \label{fig:heatmap}
\end{figure*}

Table~\ref{tab:latency_py312} presents the median end-to-end latency for
the py3.12 stack across seven GPU platforms and the nineteen benchmark images.
Each image size group contains two to six entries; 15.3\,MP has five entries
(307--15\,759 sources) and 151.2\,MP has six (808--134\,206 sources, of which
one is FERVOR-L@TTT1 and five are FERVOR-L@TST).  The Orin~Super module runs py3.12
on ARM without \texttt{cuML} (unavailable on \texttt{aarch64}); it is included
with an asterisk.  The Orin~Nano module has no py3.12 data and appears only in
Table~\ref{tab:latency_py38}: Python~3.12 is unavailable on that device
because the installed JetPack~5 does not provide the required CUDA~12
toolchain; upgrading requires a full module reflash to JetPack~6, which
was outside the scope of this study.
Both Jetson modules are discussed in detail in
Section~\ref{sec:perf:jetson}.  Dashes indicate cells without a measurement: either attempted and failed by
memory exhaustion, or not launched because a smaller frame had already
failed on that device (Figure~\ref{fig:heatmap} and the text below).

\begin{table*}[t]
\centering
\caption{Median end-to-end latency (seconds) for \textbf{py3.12\,+\,adaptive cuML},
the recommended configuration (Section~\ref{sec:discussion:deployment}), on
six x86 GPU platforms and the Orin~Super edge module ($^*$ARM; \texttt{cuML}
unavailable on \texttt{aarch64}), nineteen benchmark images, local catalog
backend.  The six x86 columns come from the runs on the released code, with the
pool re-seated (Section~\ref{sec:perf:memory}), the H100 column at its
220\,W power limit (Table~\ref{tab:gpu_hardware}); the Orin column comes
from runs that predate that correction, which does not reach it, because
cuML is unavailable on that module and its allocator is the pool in every
session.  Sessions of
five warm-up and twenty timed repetitions, each
repetition an independent process; every datacenter cell was measured in
two or more sessions on different days, a median of 74 raw and 53 clean
repetitions per cell, and the consumer and Jetson cells mostly in one
session, a median of 22 raw repetitions (Section~\ref{sec:perf:setup});
median of the clean repetitions over the sessions of the cell, after the
five exclusions listed there.  At 151.2\,MP the sparse 808-source field costs as much as the
dense fields on the H100 and A100 and is nearly as slow as the densest field on the L40S (71.4\,s vs.\ 74.1\,s), the
cost of the cuML PCA on its fine plate scale (Section~\ref{sec:perf:memory}).
The H100 entry at 18\,712 sources comes from a session about a third slower
than the one that produced most repetitions of its column
neighbors, after the exclusion of one anomalous block
(Section~\ref{sec:perf:setup}); the 134\,206-source cells
vary between sessions on the same host by more than the differences between
GPUs and are reported as measured
(Section~\ref{sec:perf:latency}).  The two 4.2\,MP
entries differ in filter (Lum and SDSSg); the Lum field contains the comet
C/2025~A6, a harder astrometric solve, and is the slower of the two on four
of the seven GPUs under py3.12 (and six of eight under py3.8; Table~\ref{tab:latency_py38}).  Dashes mark cells that did not complete (memory
exhaustion on the device or the host; Figure~\ref{fig:heatmap}).}
\label{tab:latency_py312}
\begin{tabular}{rr rrrrrrr}
\toprule
\textbf{MP} & \textbf{Sources}
            & \textbf{H100} & \textbf{A100} & \textbf{L40S} & \textbf{3090} & \textbf{3060} & \textbf{3050\,Ti} & \textbf{Orin Super}$^*$ \\
\midrule
4.2   &      228 & 6.3 & 6.6 & 6.3 & 7.7 & 6.5 & 17.3 & 26.6 \\
4.2   &      279 & 7.5 & 6.3 & 6.3 & 9.2 & 5.8 & 15.1 & 23.7 \\
6.8   &   3\,534 & 6.7 & 7.3 & 7.0 & 9.0 & 9.0 &  ---  &  ---  \\
6.8   &   8\,093 & 6.6 & 6.9 & 6.7 & 8.2 & 8.0 &  ---  &  ---  \\
15.3  &      307 & 8.5 & 8.2 & 12.7 & 11.9 & 14.5 &  ---  &  ---  \\
15.3  &      601 & 8.1 & 7.8 & 12.4 & 12.1 & 14.0 &  ---  &  ---  \\
15.3  &   2\,938 & 11.1 & 12.6 & 16.1 & 18.4 & 20.2 &  ---  &  ---  \\
15.3  &   8\,294 & 12.1 & 12.7 & 16.3 & 20.5 & 23.9 &  ---  &  ---  \\
15.3  &  15\,759 & 10.5 & 11.6 & 15.4 & 21.8 & 19.7 &  ---  &  ---  \\
37.8  &      354 & 9.1 & 8.4 & 14.4 & 18.3 & 16.9 &  ---  &  ---  \\
37.8  &      621 & 9.2 & 9.0 & 14.3 & 20.5 & 18.2 &  ---  &  ---  \\
37.8  &   3\,977 & 14.8 & 17.7 & 20.8 & 29.3 & 28.5 &  ---  &  ---  \\
37.8  &  12\,832 & 17.7 & 22.8 & 23.7 & 34.1 & 35.2 &  ---  &  ---  \\
151.2 &      808 & 30.0 & 29.4 & 71.4 &  ---  &  ---  &  ---  &  ---  \\
151.2 &  10\,168 & 22.7 & 27.7 & 33.8 &  ---  &  ---  &  ---  &  ---  \\
151.2 &  15\,390 & 24.7 & 28.5 & 35.1 &  ---  &  ---  &  ---  &  ---  \\
151.2 &  18\,712 & 33.2 & 29.0 & 34.8 &  ---  &  ---  &  ---  &  ---  \\
151.2 &  19\,492 & 23.6 & 29.3 & 36.3 &  ---  &  ---  &  ---  &  ---  \\
151.2 & 134\,206 & 31.9 & 40.2 & 74.1 &  ---  &  ---  &  ---  &  ---  \\
\bottomrule
\end{tabular}

\begin{flushleft}
\footnotesize \textit{Columns:} MP, frame size in megapixels; Sources, number of objects in the final catalog of the image; the remaining columns, median wall-clock time of the \path{process_image} function in seconds on each GPU (Section~\ref{sec:perf:setup}).
\end{flushleft}
\end{table*}

\begin{figure}[t]
    \centering
    \includegraphics[width=\columnwidth]{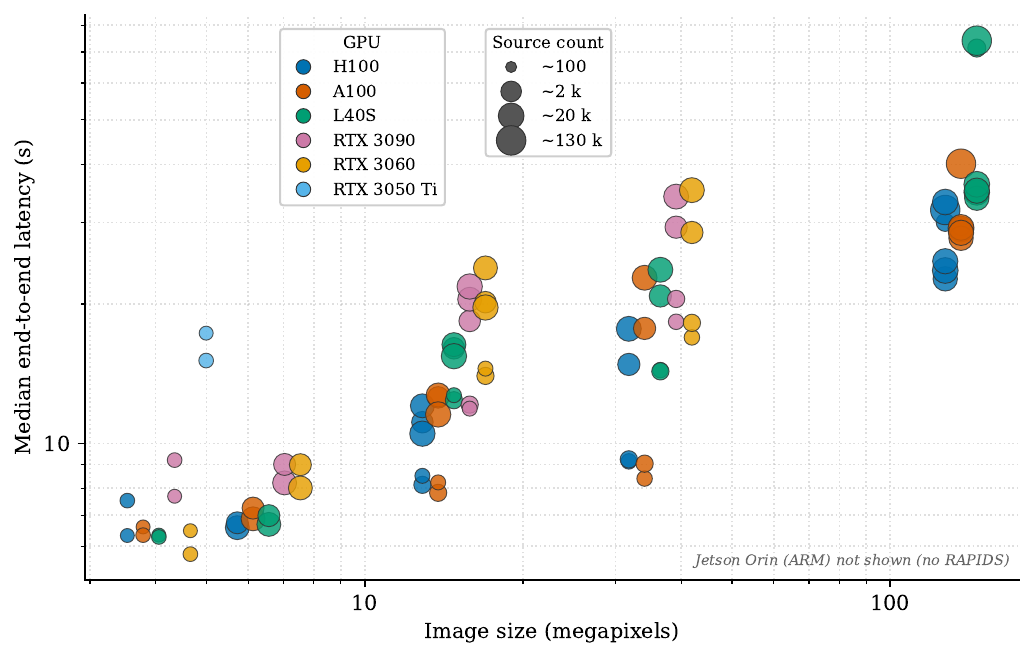}
    \caption{End-to-end latency vs.\ image size (py3.12\,+\,adaptive cuML, local catalog backend). Each marker is the median latency of one benchmark image on one GPU; markers are dodged horizontally within each image-size group. Color encodes the GPU and marker size the source count ($\propto \log_{10} N_{\rm src}$, 228 to 134\,206). The vertical spread within a group reflects source density and, at 151.2\,MP, the cost of the GPU PCA on the sparse unbinned FERVOR-L@TTT1 field (Section~\ref{sec:perf:memory}).}
    \label{fig:latency_vs_size}
\end{figure}

Figure~\ref{fig:latency_vs_size} shows the scaling behavior on a log--log
plot, and Figure~\ref{fig:heatmap} provides an alternative heatmap view that
also documents why each empty cell is empty.  Five observations emerge:

\begin{enumerate}
    \item \textbf{Source density affects latency within the same image size.}
    At 151.2\,MP, the H100 processes the sparse FERVOR-L@TTT1 field (808 sources)
    in 19.0\,s and the densest FERVOR-L@TST field (134\,206 sources) in 30.2\,s
    under py3.8, a factor of 1.6; the A100 spans 17.9 to 39.7\,s, a factor of
    2.2 (Table~\ref{tab:latency_py38}).  The growth is driven by the aperture
    photometry and catalog stages that scale with source count.
    Under py3.12\,+\,adaptive cuML the sparse field loses that advantage:
    30.0\,s on the H100, 29.4\,s on the A100 and 71.4\,s on the L40S, as much
    as or more than the dense fields.  This cost lives
    in the Eigen-PSF stage of the cuML stack and is set by the fine plate
    scale of this unbinned TTT1 frame, not by its source count
    (Section~\ref{sec:perf:memory}).
    \item \textbf{The datacenter ranking holds on the medians, and its origin
    is not established.}  Over the 19 images the medians under py3.12 are
    11.1\,s on the H100, 12.6\,s on the A100 and 16.1\,s on the L40S, and the
    ordering H100 $<$ A100 $<$ L40S holds in 9 of the 12 cells at 151.2\,MP;
    on the densest field the A100 is $1.26$--$1.31\times$ slower than the H100
    (both stacks) and the L40S $1.46\times$ under py3.8.  The three exceptions are of two kinds: one is a single cell,
    the H100 at 18\,712 sources under py3.12, reported from a session about
    a third slower than the one that produced most repetitions of
    its neighbors, after the exclusion of one anomalous block
    (Section~\ref{sec:perf:setup}); single cells are reported as measured and
    the ranking is read on the medians; the other two are the sparse
    FERVOR-L@TTT1 field under both stacks, where the H100 and the A100 lie
    within 1.1\,s of each other and the order is not distinguishable.  Nor can we
    attribute it to the accelerator: each GPU sits in a different host, and
    the one stage-resolved session on the densest field does not reproduce
    the ordering in any stage.  The ordering is reported as a reproducible
    observation without an established cause.
    \item \textbf{Small images are bound by fixed costs.}  At 4.2\,MP the
    three datacenter GPUs lie within 1.5\,s of each other (5.1--6.7\,s
    under py3.8, 6.3--7.5\,s under py3.12), the consumer cards within
    $1.3\times$ of the H100, and the RTX~3050~Ti laptop GPU within $3.4\times$.
    The fixed costs are the astrometric solve, the catalog query and the
    per-process initialization; for a 4.2\,MP frame the GPU stages account for
    43\% of the wall-clock time (Table~\ref{tab:stage_breakdown}), so faster GPUs have little left to accelerate.
    \item \textbf{At the highest density the catalog sets the variability.}
    On the 134\,206-source field the query to the shared catalog server
    takes 11--22\,s of a 30--74\,s wall clock and varies tenfold between repetitions on each host.  These cells carry the
    largest IQR of the benchmark (up to 43\%), and their sessions alternate
    between fast and slow modes.  Two facts tie the variability to the
    server: a probe of its active queries during the runs correlates with
    the repetition time ($r \approx 0.4$), and a second host measuring the
    same image at the same time moved with it.  A
    dedicated catalog replica per host would remove this dependence.
    \item \textbf{Cross-host residuals exist and are reported as such.}  Under
    py3.12 the L40S trails the RTX~3090 by 3--7\% in the two sparsest
    15.3\,MP fields and leads by 12--31\% in the other eleven cells; under
    py3.8 it leads in twelve of the thirteen cells by 3--24\% and trails by
    10\% on one 4.2\,MP field.  A
    stage-resolved decomposition of these cells could not be closed, because
    the RTX~3090 captures are few and their pairing by image is ambiguous, so
    the cause remains open.  Peak VRAM,
    in contrast, is a property of the image alone: across the six x86 GPUs the
    measured peak for the same image agrees within 2\% (Section~\ref{sec:perf:memory}).
\end{enumerate}

Every empty cell of Figure~\ref{fig:heatmap} that was attempted has its
cause in the ledger; across the three configurations of the benchmark it
records 40 cells that exhausted device memory, 16 that exhausted pinned host
memory, and 2 that hung the host, and the cells never launched are those of
devices that had already failed on a frame of the same size or smaller.  The device failures are
deterministic walls rather than probabilistic events: the RTX~3050~Ti at
6.8\,MP fails at the same allocation, byte for byte, in 45 of 45 attempts,
the requested working set exceeding the memory the device exposes by less
than 2\%, and the RTX~3060 at 151.2\,MP fails in the driver within three
seconds of the start.
The RTX~3090 is a different case.  Its host also runs the catalog database
used by the whole fleet, whose shared buffers leave 750--920\,MiB of free host
memory; the 151.2\,MP frame then dies during staging, with 264\,MiB of the
24\,GB of VRAM in use, because the pinned host allocation of 1.13\,GiB
(1\,210\,010\,624 bytes, identical across hosts and images of this size)
cannot be satisfied.  The card could not hold the 28.9\,GiB working set of this frame
in any case, but the wall it hits first is the host's, and a deployment that
co-locates the catalog server with a GPU worker inherits the same limit.  The
two hangs belong to the Orin~Nano and are discussed in Section~\ref{sec:perf:jetson}.

\subsection{Memory Analysis: Python 3.8 vs.\ 3.12}
\label{sec:perf:memory}

To isolate the impact of the library-stack upgrade, we repeated all
benchmarks under py3.8 (CuPy~12, NumPy~1.23).
Table~\ref{tab:latency_py38} presents the py3.8 median latencies.

\begin{table*}[t]
\centering
\caption{Median end-to-end latency (seconds) under py3.8 (CuPy~12), same
protocol and catalog backend as Table~\ref{tab:latency_py312}.  The H100, A100, L40S and RTX~3060
columns come from the runs with the pool re-seated
(Section~\ref{sec:perf:memory}), the H100 column at the 220\,W power limit
(Table~\ref{tab:gpu_hardware}); the other columns come from other sessions
of the same code, because py3.8 does not import cuML and the allocator
correction does not reach it, so for this stack the choice of session is a
choice of host conditions and not of code.  Each cell of these four
columns rests on a single clean block, a median of 10 clean repetitions (4 to 59), 32 of the 57
cells with ten or fewer; the other columns have a median of 20 clean
repetitions per cell (Section~\ref{sec:perf:setup}).  In the
one-process-per-image regime of the benchmark, py3.8 is faster than
py3.12\,+\,adaptive cuML in 80 of the 87 cells where both were measured, by
a median of 14\%; the seven exceptions are on the A100 (five), the L40S (one)
and the RTX~3050~Ti (one), and the residual is discussed in the text.
The A100 entry at 10\,168 sources is a block measured again after its
first block was excluded, which passed the acceptance conditions fixed
beforehand (Section~\ref{sec:perf:setup}); the runs it replaces measured
the same cell at 25.1\,s over 63 clean repetitions
(Section~\ref{sec:perf:memory}).  Dashes indicate cells that did not
complete.  $^*$Orin~Super 8\,GB under a Python~3.8 container on JetPack~6.
$^\dagger$Orin~Nano 8\,GB, Python~3.8 only; its 173.4\,s entry is
the C/2025~A6 field, whose single plate-solve attempt takes about two minutes
on this CPU (Section~\ref{sec:perf:jetson}).}
\label{tab:latency_py38}
\begin{tabular}{rr rrrrrrrr}
\toprule
\textbf{MP} & \textbf{Sources}
            & \textbf{H100} & \textbf{A100} & \textbf{L40S} & \textbf{3090} & \textbf{3060} & \textbf{3050\,Ti} & \textbf{Orin Nano}$^\dagger$ & \textbf{Orin Super}$^*$ \\
\midrule
4.2   &      228 & 5.1 & 5.8 & 5.8 & 6.6 & 5.9 & 17.3 & 173.4 & 26.1 \\
4.2   &      279 & 5.2 & 5.3 & 6.7 & 6.1 & 5.3 & 14.4 & 27.7 & 22.1 \\
6.8   &   3\,534 & 5.3 & 6.6 & 6.0 & 7.6 & 8.3 &  ---  &  ---  &  ---  \\
6.8   &   8\,093 & 4.7 & 5.7 & 5.7 & 7.0 & 7.2 &  ---  &  ---  &  ---  \\
15.3  &      307 & 6.2 & 8.1 & 8.0 & 10.2 & 11.1 &  ---  &  ---  &  ---  \\
15.3  &      601 & 5.7 & 7.1 & 8.1 & 9.5 & 10.6 &  ---  &  ---  &  ---  \\
15.3  &   2\,938 & 9.7 & 13.0 & 11.8 & 15.5 & 17.3 &  ---  &  ---  &  ---  \\
15.3  &   8\,294 & 10.7 & 12.0 & 12.1 & 14.4 & 20.0 &  ---  &  ---  &  ---  \\
15.3  &  15\,759 & 9.5 & 12.5 & 13.1 & 13.5 & 16.6 &  ---  &  ---  &  ---  \\
37.8  &      354 & 7.3 & 7.8 & 8.9 & 11.0 & 13.3 &  ---  &  ---  &  ---  \\
37.8  &      621 & 8.1 & 8.4 & 9.6 & 11.8 & 14.8 &  ---  &  ---  &  ---  \\
37.8  &   3\,977 & 14.1 & 19.4 & 19.1 & 23.2 & 26.4 &  ---  &  ---  &  ---  \\
37.8  &  12\,832 & 16.6 & 26.4 & 23.1 & 27.1 & 33.4 &  ---  &  ---  &  ---  \\
151.2 &      808 & 19.0 & 17.9 & 33.7 &  ---  &  ---  &  ---  &  ---  &  ---  \\
151.2 &  10\,168 & 21.2 & 28.3 & 33.8 &  ---  &  ---  &  ---  &  ---  &  ---  \\
151.2 &  15\,390 & 23.6 & 26.5 & 31.5 &  ---  &  ---  &  ---  &  ---  &  ---  \\
151.2 &  18\,712 & 23.5 & 27.5 & 32.1 &  ---  &  ---  &  ---  &  ---  &  ---  \\
151.2 &  19\,492 & 22.3 & 26.8 & 34.2 &  ---  &  ---  &  ---  &  ---  &  ---  \\
151.2 & 134\,206 & 30.2 & 39.7 & 44.1 &  ---  &  ---  &  ---  &  ---  &  ---  \\
\bottomrule
\end{tabular}

\begin{flushleft}
\footnotesize \textit{Columns:} MP, frame size in megapixels; Sources, number of objects in the final catalog of the image; the remaining columns, median wall-clock time of the \path{process_image} function in seconds on each GPU (Section~\ref{sec:perf:setup}).
\end{flushleft}
\end{table*}

\paragraph{Reference catalog backend: local replica against remote query.}\label{sec:catalogue_backend}
By default, \gpuphot\ queries VizieR for Pan-STARRS\,DR1 source
lists as part of the crossmatch stage.  Every cell of this paper was measured against a local
PostgreSQL/Q3C replica of the same catalog, and the comparison between the
two backends is the one set of runs in which both were measured on the same
code.  There, the six 151.2\,MP benchmark images were processed against VizieR and against the local PostgreSQL/Q3C replica on the observatory network, and the end-to-end wall-clock times were compared.  Those runs used a different camera configuration, whose catalogs hold 428 to 131\,397 sources for the same six images that
Table~\ref{tab:latency_py312} lists with 808 to 134\,206, so the backends are
compared on the same images and the same configuration as each other, and
the speedup is read against source count.

\begin{table*}[t]
\centering
\caption{Speedup of the local PostgreSQL/Q3C catalog backend relative to
VizieR (median VizieR time~$/$ median local time) on the six
151.2\,MP benchmark images; source counts are those of the camera
configuration in which this comparison was measured.  The speedup is $\approx 1.0$ on the sparse
428-source field and grows with source count as the catalog query becomes
dominant (Section~\ref{sec:catalogue_backend}).  The L40S/py3.12 cell at
$131\,397$ sources includes a five-repetition PostgreSQL buffer-cache
warm-up before measurement.  Median across 36 valid cells: $1.14\times$
(range $0.99$--$1.76\times$).}
\label{tab:speedup_local_vizier}
\begin{tabular}{lrcccccc}
\toprule
 & & \multicolumn{3}{c}{\textbf{py3.12\,+\,adaptive}} & \multicolumn{3}{c}{\textbf{py3.8}} \\
\cmidrule(lr){3-5}\cmidrule(lr){6-8}
\textbf{Image} & \textbf{Src} & \textbf{A100} & \textbf{H100} & \textbf{L40S} & \textbf{A100} & \textbf{H100} & \textbf{L40S} \\
\midrule
FERVOR-L@TTT1 Lum & 428 & 1.00 & 0.99 & 1.00 & 1.01 & 1.01 & 0.99 \\
FERVOR-L@TST SDSSg & 10\,005 & 1.24 & 1.18 & 1.12 & 1.28 & 1.20 & 1.17 \\
FERVOR-L@TST SDSSr$_{\rm 14k}$ & 14\,241 & 1.07 & 1.09 & 1.34 & 1.12 & 1.04 & 1.05 \\
FERVOR-L@TST Lum$_{\rm 18k}$ & 18\,888 & 1.14 & 1.14 & 1.19 & 1.17 & 1.16 & 1.09 \\
FERVOR-L@TST SDSSr$_{\rm 19k}$ & 19\,565 & 1.15 & 1.14 & 1.09 & 1.17 & 1.15 & 1.10 \\
FERVOR-L@TST Lum$_{\rm 131k}$ & 131\,397 & 1.65 & 1.66 & 1.51 & 1.76 & 1.64 & 1.53 \\
\midrule
Median (column) & & 1.14 & 1.14 & 1.15 & 1.17 & 1.15 & 1.10 \\
\multicolumn{8}{l}{\footnotesize Overall median (valid cells, $N=36$): 1.14$\times$;\quad range: 0.99--1.76$\times$}\\
\bottomrule
\end{tabular}

\begin{flushleft}
\footnotesize \textit{Columns:} Image, benchmark field named by camera and filter, with a subscript where two fields share both; Src, number of Pan-STARRS\,DR1 crossmatch sources in the configuration of this comparison; the six numeric columns, the speedup on each GPU under each stack, that is, the median end-to-end time against VizieR divided by the median with the local replica.
\end{flushleft}
\end{table*}

Table~\ref{tab:speedup_local_vizier} shows the speedup per image and
configuration.
The sparse control field (FERVOR-L@TTT1 Lum, 428~sources in that configuration)
yields speedup $\approx 1$ on all platforms: when the crossmatch radius
covers few sources the VizieR query latency is negligible relative
to the rest of the pipeline.
As source count grows the catalog query becomes the bottleneck.
At $131\,397$~sources the three datacenter GPUs reach py3.12 speedups of
$1.51\times$ (L40S), $1.65\times$ (A100), and $1.66\times$ (H100), and
py3.8 speedups of $1.76\times$ (A100), $1.64\times$ (H100), and $1.53\times$ (L40S).
In the intermediate range (10\,005--19\,565~sources) the speedup is
$1.04$--$1.34\times$; the speedup grows with source count because the
catalog query does (2.3\,s on the sparse and 13.4\,s on the densest
151.2\,MP field in Table~\ref{tab:stage_breakdown}) while the rest of the
pipeline does not.

The local replica also removes the variability of the remote query: the four cells whose IQR under VizieR exceeded 5\% (up to 26\% for the L40S on the 14\,241-source field) all fell below 3\% with the local backend, with no other change to the pipeline, stack or hardware, so the remote query and not the pipeline was the source of that spread.

For production deployments serving fields with source counts above
$10\,000$, a local catalog replica eliminates network-latency jitter and
reduces end-to-end latency by $1.1$--$1.8\times$ with no change to
catalog content or query semantics; the residual variability of the densest
cells of the benchmark (observation~4 of Section~\ref{sec:perf:latency})
comes from the load on the shared replica, not from the network.

\begin{figure}[t]
    \centering
    \includegraphics[width=\columnwidth]{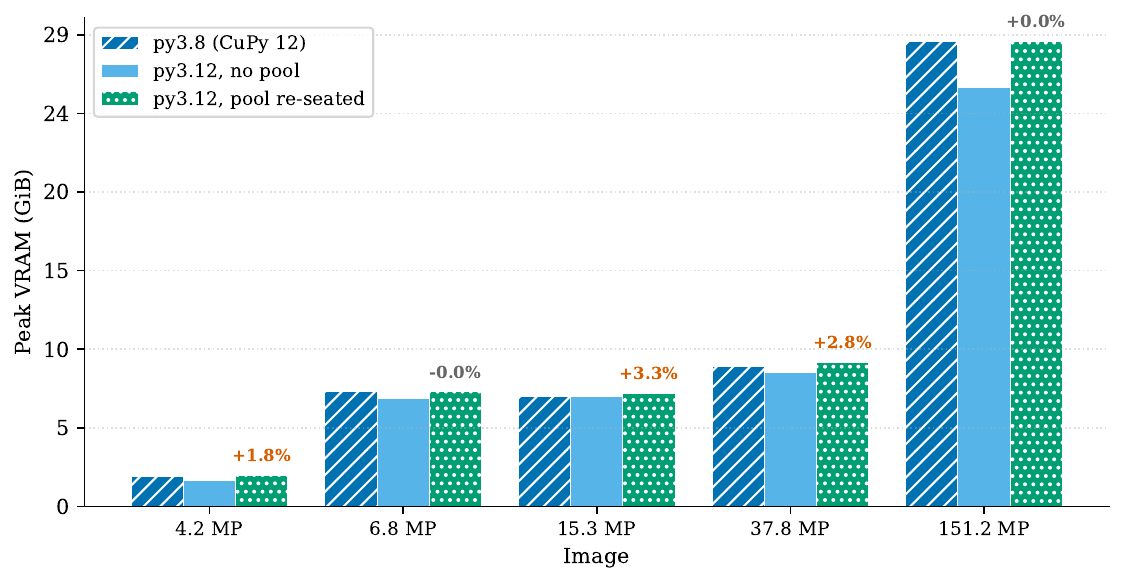}
    \caption{Peak GPU memory (A100) under py3.8 (CuPy~12, diagonal hatching), py3.12 without the CuPy pool (CuPy~14, solid) and py3.12 with the pool re-seated (dotted hatching) for five image sizes. The vertical axis is in GiB (1\,GiB = 1\,024\,MiB); Table~\ref{tab:vram_merged} lists the same peaks in MiB. The percentage above the third bar of each group is the difference of the pooled py3.12 peak from the py3.8 peak, blue where py3.12 peaks lower, orange where it peaks higher and gray where the two round to the same value; the difference of the unpooled state is read from the bar heights and listed in the table (Section~\ref{sec:perf:memory}).}
    \label{fig:memory_comparison}
\end{figure}

Table~\ref{tab:vram_merged} lists the peak VRAM per image size under py3.8
and under the two allocator states of the py3.12 stack, without the CuPy
pool (the state that importing cuML installs) and with the pool re-seated
per task (the released configuration); Figure~\ref{fig:memory_comparison}
shows the same three series.

\begin{table*}[t]
\centering
\caption{Peak GPU memory (MiB) on the A100-SXM4-80GB under py3.8 and under
the py3.12 stack without and with the CuPy pool, and the difference of each
py3.12 state from py3.8.  The A100 is the representative GPU; the peak for
the same image agrees within 2\% across the six tested x86 GPUs, with the
H100 56\,MiB higher in every cell at 15.3\,MP and above.  Without the pool
the py3.12 peak lies between 14.9\,\% below and 0.2\,\% above the py3.8
value (7.0\,\% below on average); with the pool it returns to the py3.8
value within 3.3\,\% at all five sizes.  The difference is produced by the
allocator, not by the libraries (Section~\ref{sec:perf:memory}).}
\label{tab:vram_merged}
\begin{tabular}{r rrr rr}
\toprule
\textbf{MP} & \textbf{\shortstack{py3.8\\(MiB)}} & \textbf{\shortstack{py3.12 no\\pool (MiB)}} & \textbf{\shortstack{py3.12 with\\pool (MiB)}} & \textbf{\shortstack{No pool vs\\py3.8 (\%)}} & \textbf{\shortstack{With pool vs\\py3.8 (\%)}} \\
\midrule
  4.2 & 1924 & 1637 & 1958 & $-14.9$ & $+1.8$ \\
  6.8 & 7285 & 6887 & 7284 & $-5.5$ & $-0.0$ \\
 15.3 & 6970 & 6982 & 7197 & $+0.2$ & $+3.3$ \\
 37.8 & 8914 & 8492 & 9165 & $-4.7$ & $+2.8$ \\
151.2 & 29569 & 26668 & 29570 & $-9.8$ & $+0.0$ \\
\bottomrule
\end{tabular}

\begin{flushleft}
\footnotesize \textit{Columns:} MP, frame size in megapixels; py3.8, py3.12 no pool and py3.12 with pool, peak device memory of the process in MiB under each stack and allocator state, measured with \texttt{nsys} in separate runs (Section~\ref{sec:perf:setup}); No pool vs py3.8 and With pool vs py3.8, relative difference of each py3.12 column from the py3.8 column, negative when py3.12 peaks lower.  The difference without the pool is an artifact of the allocator that importing cuML installs; with the pool re-seated the py3.12 peak returns to the py3.8 value within 3.3\,\% at all five sizes.
\end{flushleft}
\end{table*}

Three findings deserve emphasis:

\begin{enumerate}
    \item \textbf{The 7\,\% lower average peak of the py3.12 stack without the
    pool is the allocator, not the libraries.}
    Without the pool the py3.12 peak is 14.9\,\% lower than py3.8 at 4.2\,MP,
    5.5\,\% at 6.8\,MP, 4.7\,\% at 37.8\,MP and 9.8\,\% at 151.2\,MP, and
    12\,MiB higher at 15.3\,MP ($+0.2$\,\%; Table~\ref{tab:vram_merged}).  A stage-resolved memory
    timeline of the A100 captures, with a method that reproduces the
    tabulated peaks, shows the same largest allocations in both stacks
    (128\,MiB at 4.2\,MP, one padded $4096 \times 4096$ single-precision
    complex FFT plane; 11.8\,GiB at 151.2\,MP) and the peak in the same stage, the FFT cross-correlation of
    the batched aperture photometry.  What differs is the traffic: 190
    device allocations under py3.8 against 12\,301 under py3.12 at 4.2\,MP
    and 55\,674 at 151.2\,MP, the RAPIDS allocator of
    Section~\ref{sec:impl_memory} returning every temporary to the driver.
    Captures with the pool re-seated settle the origin.  The allocation
    count of py3.12 falls to 185 at 4.2\,MP and 633 at 151.2\,MP, and its
    peak returns to the py3.8 value within 3.3\,\% at all five sizes
    (Table~\ref{tab:vram_merged}), with the same largest allocation in both
    states.  The lower peak measured without the pool is therefore an effect of
    the unpooled allocator, the same defect that costs latency, and not a
    property of CuPy~14 or NumPy~2.0; the corrected stack has the memory
    footprint of the py3.8 stack.  How the pooled allocator raises the peak
    is only partly resolved: at the memory checkpoints of the py3.8 runs the
    pool holds 93--129\,MiB cached but free, at most 45\,\% of the 287\,MiB
    difference, and the whole pool is 36\,\% of the peak, so most of the
    difference lies outside the pool, in raw arrays, FFT planes and context,
    and is not decomposed.  The consequence for concurrency was verified on the RTX~3050~Ti: with the
    pool its 4.2\,MP peak rises from 1\,637 to 1\,923\,MiB and the device
    holds one image again, as under py3.8; Table~\ref{tab:concurrency} gives
    the count for the three states on every GPU
    (Section~\ref{sec:perf:concurrency}).

    \emph{The 6.8\,MP frame peaks above the 15.3\,MP frame under py3.8.}
    Peak memory is also not a monotonic function of pixel count under
    py3.8: the 6.8\,MP FERVOR-M@TTT3 frame peaks above the 15.3\,MP FERVOR-M@TTT2 frame
    (7\,285 against 6\,970\,MiB), while under the py3.12 stack of the benchmark
    the order is the normal one (6\,887 against 6\,982\,MiB).  A
    stage-resolved extraction of the benchmark captures gives the mechanism.
    The largest single allocation follows the padded frame area (656\,MiB at
    6.8\,MP against 1\,169\,MiB at 15.3\,MP, a ratio of 1.78 for an area ratio
    of 2.0) and is identical under both stacks, and in both frames the peak
    falls in the batched aperture photometry; what differs is how many
    buffers are alive at that instant.  Pairing every allocation with its
    release in a py3.12 capture shows ten buffers requested inside the
    cosmic-ray filter, 1\,277\,MiB in total, still alive when the 6.8\,MP
    frame reaches its peak, and none in the 15.3\,MP frame, whose camera
    runs without that filter.  The count was not repeated under py3.8; that
    the same buffers are retained there is inferred from the identity of
    every individual allocation compared between the two stacks so far.  On
    that inference the retention is a property of the filter, and whether it
    lifts the smaller frame above the larger one depends on the baseline of
    each stack, which is why the inversion appears under py3.8 and not under
    the uncorrected py3.12 stack.

    \emph{The 56\,MiB excess of the H100.}
    Across GPUs the peak is a property of the image: the measured
    values for the same frame agree within 2\%, and the H100 carries a constant excess. At 15.3\,MP and above, its set of per-run peaks is the A100 set shifted by exactly 56\,MiB, run for run, under both stacks; at 4.2 and 6.8\,MP the two sets coincide. The A100 and the L40S agree to the mebibyte at every size.  A stage-resolved capture under
    py3.12 localizes where the excess appears: every stage before the
    Eigen-PSF stage is identical between the two cards, the start-up
    baseline included (605\,MiB on both), and the divergence opens inside
    \path{get_eigen_psfs} as a 323\,MiB transient and a 59\,MiB residual
    that persists to the peak and is released when the process ends.  A
    solver workspace sized by the device would explain that capture, but not
    the same 56\,MiB under py3.8, whose PCA runs on the CPU, so the
    mechanism is not settled.  The two smallest frames are also the two
    whose cameras disable the Eigen-PSF stage, so size and configuration
    cannot be separated here.
    \item \textbf{The per-image cost of the py3.12 stack was largely an
    allocator defect.}
    Without the CuPy pool, py3.12\,+\,adaptive cuML was slower than py3.8 in
    every one of the 87 cells where both were measured: by a median of
    38\,\% (range $+2$ to $+240$\,\%), with per-GPU medians of $+35$\,\%
    (A100), $+40$\,\% (H100), $+67$\,\% (L40S), $+31$\,\% (RTX~3090),
    $+41$\,\% (RTX~3060), $+18$\,\% (RTX~3050~Ti) and $+5$\,\% (Orin~Super),
    and no GPU on which the pattern reversed; the 50 cells with more than
    2\,000 sources had a median of $+38$\,\% and a minimum of $+18$\,\%.
    With the pool re-seated, the released configuration of
    Tables~\ref{tab:latency_py312} and~\ref{tab:latency_py38}, the gap is a
    median of $+14$\,\% over the same 87 cells, py3.8 ahead in 80 of them,
    with per-GPU medians of $+14$\,\% (H100), $+7$\,\% (A100), $+16$\,\%
    (L40S), $+17$\,\% (RTX~3060), $+26$\,\% (RTX~3090), $+2$\,\%
    (RTX~3050~Ti) and $+5$\,\% (Orin~Super).
    Profiling the stages with NVIDIA Tools Extension (NVTX) ranges and the
    CUDA profiling interface showed that the affected stages are bound by the
    allocation overhead of their temporaries rather than by GPU work, and traced the excess to the allocator. Importing cuML replaces CuPy's pooled allocator with the RAPIDS memory manager in unpooled mode, so that every temporary array pays a driver allocation: an element-wise operation that allocates its result costs 17.6\,$\mu$s under the unpooled allocator against 8.5\,$\mu$s with the pool, and the median gap between consecutive kernels grows from 16 to 31\,$\mu$s, with identical GPU time on both stacks.  The batched aperture photometry, the stage that creates the most temporaries, carries most of the excess.  CuPy~14 itself is not slower:
    micro-benchmarks in the same containers launch kernels at least as fast
    as CuPy~12.
    Re-seating the default CuPy pool at the start of each task
    (Section~\ref{sec:impl_memory}) removes the penalty without changing the
    output: measured again on the three datacenter hosts with the pool in
    place (7\,131 runs in the 171 cells common to the two arms), 167 cells
    kept identical source counts and 157 returned a single zero point, the
    exceptions being all cells of the two labels that pool several
    exposures, where a different exposure was measured.  A stage-resolved
    comparison of the A100 captures with and without the pool places the
    whole gain in the two photometry stages, aperture and optimal, and leaves
    the PCA stage unchanged.  Table~\ref{tab:allocator_ablation} lists, per
    image and GPU, the median without the pool and the median of
    the clean sessions with it.  The raw change over the 19 images is
    $-4$\,\% on the H100, $-20$\,\% on the A100 and $-16$\,\% on the L40S;
    the two arms were measured in separate runs, so the raw figure mixes the
    allocator with everything else that differs between them, and on the H100
    that includes a lower power
    limit (220\,W against 350\,W for most repetitions;
    Table~\ref{tab:allocator_ablation}), although a single reduction draws at most
    180\,W on that card, so the limit does not bind within one image
    (Section~\ref{sec:perf:setup}).  The py3.8
    stack does not carry the correction, so its own change between the same
    two arms measures that difference alone; discounting it, the net effect of the
    pool is $-20$\,\% on the H100 (four cells with a clean py3.8
    measurement), $-22$\,\% on the A100 and $-19$\,\% on the L40S, three
    values within three points of each other.  Eleven of the 57 cells move
    against the trend, and eight of those differences are below one second.
    The largest is the L40S at 134\,206 sources ($+36$\,\%, 19.7\,s), the field whose
    catalog query varies most between repetitions and whose
    session-to-session variability exceeds the effect, flagged again in
    Figures~\ref{fig:py38_vs_adaptive} and~\ref{fig:always_vs_adaptive};
    the H100 at 18\,712 sources ($+10$\,\%, 3.0\,s) is reported from a session about
    a third slower than the excluded one on that host, after
    the exclusion of one anomalous block (Section~\ref{sec:perf:setup}); and
    the H100 at 279 sources ($+20$\,\%, 1.3\,s) is the only 4.2\,MP cell in which
    that card is the slowest of the three, which we do not attribute.
    Figure~\ref{fig:py38_vs_adaptive}, measured with the pool in place and
    the two stacks alternating in blocks inside the same session, gives the overhead that
    the released version carries: a median of $+11$\,\%
    over 50 cells, $+9$\,\% on the H100, $+5$\,\% on the A100 and $+38$\,\%
    on the L40S, $+6$\,\% in the 31 dense cells and $+18$\,\% in the 19
    sparse ones, with py3.12 ahead in 10 cells.

\begin{table*}[t]
\centering
\caption{Allocator ablation on the three datacenter GPUs: median end-to-end
latency (seconds) of py3.12\,+\,adaptive cuML per benchmark image without the
CuPy pool (the RAPIDS allocator that importing cuML installs) and with the
pool re-seated per task (the released configuration; clean sessions), one
process per image, local catalog.  The two arms were measured in separate
runs, which is what the py3.8 reference of the last row controls for.
Power limit of the
H100 (Section~\ref{sec:perf:setup}): the with-pool repetitions of that
card ran at 220\,W, 37 of 2\,790 at 200\,W during thermal steps of the
host; without the pool, the py3.8 reference ran at 350\,W in all 19
cells and py3.12 in 16 of them, while three py3.12 cells mix limits, with
34 to 42\% of their repetitions below 350\,W: the 18\,712- and
15\,390-source fields mostly at 220\,W and the 808-source field at 250 and
300\,W.}
\label{tab:allocator_ablation}
{\setlength{\tabcolsep}{4.5pt}\begin{tabular}{rr rrr rrr rrr}
\toprule
\textbf{MP} & \textbf{Sources} & \multicolumn{3}{c}{\textbf{H100}} & \multicolumn{3}{c}{\textbf{A100}} & \multicolumn{3}{c}{\textbf{L40S}} \\
\cmidrule(lr){3-5}\cmidrule(lr){6-8}\cmidrule(lr){9-11}
 & & \textbf{\shortstack{No\\pool (s)}} & \textbf{\shortstack{With\\pool (s)}} & \textbf{\shortstack{$\Delta$\\(\%)}} & \textbf{\shortstack{No\\pool (s)}} & \textbf{\shortstack{With\\pool (s)}} & \textbf{\shortstack{$\Delta$\\(\%)}} & \textbf{\shortstack{No\\pool (s)}} & \textbf{\shortstack{With\\pool (s)}} & \textbf{\shortstack{$\Delta$\\(\%)}} \\
\midrule
4.2   &      228 & 6.2 & 6.3 & $+2$ & 6.5 & 6.6 & $+2$ & 7.0 & 6.3 & $-9$ \\
4.2   &      279 & 6.2 & 7.5 & $+20$ & 6.2 & 6.3 & $+2$ & 6.0 & 6.3 & $+5$ \\
6.8   &   3\,534 & 6.8 & 6.7 & $-1$ & 9.1 & 7.2 & $-20$ & 7.6 & 7.0 & $-8$ \\
6.8   &   8\,093 & 6.0 & 6.6 & $+10$ & 6.5 & 6.9 & $+6$ & 7.4 & 6.7 & $-10$ \\
15.3  &      307 & 8.8 & 8.5 & $-4$ & 9.4 & 8.2 & $-12$ & 14.3 & 12.7 & $-11$ \\
15.3  &      601 & 8.1 & 8.1 & $+1$ & 8.2 & 7.8 & $-5$ & 13.1 & 12.4 & $-5$ \\
15.3  &   2\,938 & 14.7 & 11.1 & $-24$ & 17.2 & 12.6 & $-27$ & 21.6 & 16.1 & $-26$ \\
15.3  &   8\,294 & 13.8 & 12.1 & $-13$ & 14.8 & 12.7 & $-14$ & 19.6 & 16.3 & $-16$ \\
15.3  &  15\,759 & 13.0 & 10.5 & $-20$ & 14.2 & 11.6 & $-18$ & 17.9 & 15.4 & $-14$ \\
37.8  &      354 & 9.1 & 9.2 & $+1$ & 11.5 & 8.4 & $-27$ & 15.7 & 14.4 & $-8$ \\
37.8  &      621 & 9.5 & 9.2 & $-3$ & 11.4 & 9.0 & $-21$ & 18.1 & 14.3 & $-21$ \\
37.8  &   3\,977 & 21.7 & 14.8 & $-32$ & 28.6 & 17.7 & $-38$ & 29.9 & 20.8 & $-30$ \\
37.8  &  12\,832 & 27.4 & 17.7 & $-36$ & 34.1 & 22.8 & $-33$ & 35.9 & 23.7 & $-34$ \\
151.2 &      808 & 30.9 & 30.0 & $-3$ & 59.8 & 29.4 & $-51$ & 90.5 & 71.4 & $-21$ \\
151.2 &  10\,168 & 26.6 & 22.7 & $-15$ & 34.0 & 27.7 & $-18$ & 41.3 & 33.9 & $-18$ \\
151.2 &  15\,390 & 29.4 & 24.7 & $-16$ & 38.2 & 28.5 & $-25$ & 43.7 & 35.1 & $-20$ \\
151.2 &  18\,712 & 30.2 & 33.2 & $+10$ & 38.1 & 29.0 & $-24$ & 44.0 & 34.8 & $-21$ \\
151.2 &  19\,492 & 29.9 & 23.6 & $-21$ & 39.1 & 29.3 & $-25$ & 45.0 & 36.3 & $-19$ \\
151.2 & 134\,206 & 38.0 & 31.9 & $-16$ & 48.4 & 40.2 & $-17$ & 54.4 & 74.1 & $+36$ \\
\midrule
\textbf{Median $\Delta$} & &  &  & $-4$~{\scriptsize($n$=19)} &  &  & $-20$~{\scriptsize($n$=19)} &  &  & $-16$~{\scriptsize($n$=19)} \\
\textbf{Median net} & &  &  & $-20$~{\scriptsize($n$=4)} &  &  & $-22$~{\scriptsize($n$=19)} &  &  & $-19$~{\scriptsize($n$=19)} \\
\bottomrule
\end{tabular}
}
\begin{flushleft}
\footnotesize \textit{Columns:} MP, frame size in megapixels; Sources, number of objects in the final catalog of the image; for each GPU, No pool, median latency without the CuPy pool; With pool, median latency with the pool re-seated; $\Delta$, relative change from the former to the latter, negative when the pool is faster.  \textit{Rows:} Median $\Delta$, median of the $\Delta$ column over the 19 images, which mixes the allocator with everything else that differs between the two arms; Median net, the same median after dividing each cell by the change of the py3.8 stack on the same GPU and image between the same two arms, which isolates the allocator because py3.8 does not carry the correction ($n$, cells with a clean py3.8 measurement on that GPU).
\end{flushleft}
\end{table*}

    \emph{The residual: the Eigen-PSF cost on fine plate scales.}
    The residual is concentrated in one image class: the sparse 151.2\,MP
    FERVOR-L@TTT1 field, still $+60$\,\% (H100), $+64$\,\% (A100) and $+110$\,\%
    (L40S) slower than py3.8 with the pool in place
    (Figure~\ref{fig:py38_vs_adaptive}).  Its excess is the cost of
    \texttt{cuml.decomposition.PCA} in \texttt{get\_eigen\_psfs}, which NVTX
    profiling of this field measured at 10.2, 11.4 and 34.2\,s
    on the H100, A100 and L40S, against 0.23\,s for the same stage on the
    densest 151.2\,MP field and 0.14\,s under \texttt{scikit-learn}.  The
    cause is not the sparseness of the field but the size of the matrix that
    the PCA decomposes.  The PSF stamp is sized in arcseconds, with a
    half-side of $\max(\lfloor 10''/s \rfloor, 6)$ pixels for a plate scale
    $s$, so its pixel count is set by the focal length of the telescope.  The
    unbinned FERVOR-L@TTT1 field (focal length 5.48\,m, 0.1415\,arcsec per
    pixel) has a stamp of $141 \times 141$ pixels, and the PCA sees
    19\,881 features.  The FERVOR-L@TST fields (1.3\,m, 0.5966\,arcsec per
    pixel) have a stamp of $33 \times 33$ pixels and 1\,089 features, for
    the same sensor and frame size.  A synthetic sweep
    of \texttt{cuml.decomposition.PCA} (five components, full solver, on the
    L40S) shows that the number of stamp pixels alone governs the cost: with
    19\,881 columns the decomposition takes 9.8\,s for 50 rows and 10.8\,s
    for 25\,000, flat over a factor of 500 in the number of stars, whereas
    with 1\,089 columns it stays between 0.01 and 0.02\,s over the same
    range.  That signature is the one of the eigendecomposition of the
    $p \times p$ covariance matrix, of cost $O(p^{3})$ in the number of
    features $p$; had the formation of the covariance dominated, the cost
    would have grown with the number of rows, and it does not.  The
    discriminant is therefore the number of pixels of the stamp, not the
    number of sources and not the aspect ratio of the matrix, and every
    unbinned frame of the TTT1 falls in the ten-second regime by
    construction, because its plate scale fixes stamps of 19\,881 pixels.
    The benchmark contains one such frame.  By the same rule, applied to
    the pixel size and focal length that the pipeline reads from each
    header, the other images that run the stage have stamps of
    $71 \times 71$ pixels (5\,041 features: the four binned TTT1 images
    and the five TTT2 images) or $33 \times 33$ (1\,089: the five TST
    images), and the two configurations whose stamps come next in size,
    the FERVOR-M binned $3 \times 3$ ($103 \times 103$, 10\,609) and the
    COLORS ($87 \times 87$, 7\,569), both on the TTT3, run with the
    Eigen-PSF stage disabled.  With the pool in place, the 27 datacenter
    cells with 5\,041-feature stamps differ between the stacks by $-15$ to
    $+5$\,s, none of them near the 10--34\,s that the PCA alone costs on
    the unbinned frame (Figure~\ref{fig:py38_vs_adaptive}).  The regime
    therefore requires both a fine plate scale and the stage enabled, and
    one image of the benchmark meets both conditions.

    \emph{The CPU solver.}
    The CPU solver of the py3.8 and ARM paths behaves in the opposite way.
    Swept on the CPU of the same host (thread count and BLAS kernel left
    unpinned, so the absolute values are indicative), the cost of
    \texttt{scikit-learn} grows with the number of stars: from 0.10\,s at 50
    to 2.4\,s at 1\,000 with 19\,881 features, and from 0.008 to 0.20\,s with
    1\,089.  At the star counts of this frame the pipeline measured it at
    0.14\,s against 10.2\,s for cuML, two orders of magnitude in favor of
    the CPU path in the real regime.  The allocator correction does
    not touch this stage (11.4\,s before and after it on this frame), which is why the cell remains the worst of
    Figure~\ref{fig:py38_vs_adaptive}; routing the PCA to the CPU above a
    stamp-size threshold removes it, and the sweep gives that threshold a
    numerical basis (Section~\ref{sec:future_directions}).

    \emph{The execution regime.}
    The regime matters, and the mechanism says why.  The
    end-of-image cleanup of the pipeline installs a fresh CuPy pool as the
    allocator, so a process that reduces several images pays the unpooled
    allocator on its first image only; the benchmark, which launches a new
    process per image, pays it on every image, and a long-lived worker
    once.  Measured directly, in a long-lived process cycling through six
    different fields (96 runs on two hosts under the unpooled allocator, in which only the
    first image of each sequence paid the displaced allocator),
    py3.12 reached parity with py3.8: between $-20$ and $+55$\,\% per
    image, parity in aggregate, and ahead on the large dense fields.  In
    those sequences the first image, the only one still on the displaced
    allocator, paid an excess of 0 to 15\,s on the four frames that occupied
    that position, an excess that tracks the work of the frame rather than
    its size (the two 15.3\,MP frames, with 601 and 15\,759 sources, differ
    fifteenfold).  The latency tables therefore describe
    the one-process-per-image regime, which is what the prefork
    configuration of Section~\ref{sec:impl_memory} pays; the reference
    deployment runs its workers on the \texttt{gevent} pool, which persists
    across tasks, and the production workers of the observatory are
    long-lived as well, so both operate in the second regime.  With the
    correction, every image starts with the default pool in place and the
    two regimes differ only by the start-up costs of the process; a
    long-lived py3.12 worker with the pool in place, cycling through six
    fields of up to 37.8\,MP for 96 images on the A100, completed all of them
    without a memory failure and without drift, the per-size medians of the
    second half of the cycle within 0.7\,s of the first.
    
\end{enumerate}

\subsection{Concurrency Implications}
\label{sec:perf:concurrency}

In the \gpuphot\ production deployment, multiple Celery workers share a
single GPU.  The maximum number of concurrent images is bounded by
$\lfloor \text{VRAM}_{\text{usable}} / \text{VRAM}_{\text{peak}} \rfloor$,
where the usable VRAM is the total reported by the device (3\,771\,MiB on the
RTX~3050~Ti laptop GPU against a nominal 4\,096\,MiB) and the peak is the
measured working set of the image.
Table~\ref{tab:concurrency} applies this bound to the 4.2\,MP COLORS images,
the most frequent workload in the TTT/TST pipeline, for py3.8 and for the
two allocator states of py3.12.

\begin{table*}[t]
\centering
\caption{Maximum concurrent 4.2\,MP images per GPU under py3.8 and under
py3.12 without and with the CuPy pool, computed as
$\lfloor \text{VRAM} / \text{peak VRAM per image} \rfloor$ from the
device-reported total and the measured peak of each state on each GPU.  The
apparent gain of the unpooled state (up to $+100$\,\% on the RTX~3050~Ti) is
an artifact of the allocator; with the pool the count returns to the py3.8
value on every GPU except the A100, whose pooled peak (1\,958\,MiB) lies
just above the py3.8 peak (1\,924\,MiB) and costs one image
(Section~\ref{sec:perf:memory}).}
\label{tab:concurrency}
\begin{tabular}{lrrrr}
\toprule
\textbf{GPU} & \textbf{VRAM (GB)} & \textbf{py3.8} & \textbf{\shortstack{py3.12\\no pool}} & \textbf{\shortstack{py3.12\\with pool}} \\
\midrule
H100           & 80 & 41 & 49 & 41 \\
A100           & 80 & 42 & 49 & 41 \\
L40S           & 48 & 23 & 27 & 23 \\
RTX 3090       & 24 & 12 & 14 & 12 \\
RTX 3060       & 12 &  6 &  7 &  6 \\
RTX 3050 Ti    &  4 &  1 &  2 &  1 \\
\bottomrule
\end{tabular}

\begin{flushleft}
\footnotesize \textit{Columns:} GPU, device; VRAM, nominal device memory in GB; py3.8, py3.12 no pool and py3.12 with pool, number of 4.2\,MP images that fit concurrently in that memory under each stack and allocator state, computed as the integer part of the device-reported memory divided by the measured peak of one image on that GPU, a capacity bound obtained by arithmetic and not a measured concurrency; the measured latency and throughput of two and three concurrent 151.2\,MP reductions on the A100 are given in the text.
\end{flushleft}
\end{table*}

The largest apparent gain of the unpooled state is on the RTX~3050~Ti
laptop GPU, where two unpooled py3.12 frames (3\,274\,MiB) fit in the
3\,771\,MiB the device exposes while two py3.8 frames (3\,848\,MiB) do not;
with the pool re-seated the py3.12 peak on that device is 1\,923\,MiB and
one frame fits, as under py3.8.  The same holds at
the other end of the range.  A 151.2\,MP frame occupies 28.9\,GiB under
py3.8 and under the corrected py3.12 stack (26.0\,GiB only under the
defect), so an 80\,GB device holds two such reductions and the 48\,GB L40S
one, under either stack.  These are capacity bounds; the throughput they allow was measured on the
A100 with the pool in place.  Two concurrent 151.2\,MP reductions complete
in 37.4\,s each against 31.9\,s alone, 1.7 times the throughput of one
process (0.044 against 0.026 images per second); with three in flight the
per-image latency rises to 47.2\,s and the throughput to 2.1 times, but four
of the nine processes failed by memory exhaustion, so the bound of two is
also the practical limit on an 80\,GB device.

\paragraph{Deployment recommendation.}
On the basis of the results above, we recommend
\texttt{py3.12\,+\,adaptive cuML} for production deployments for its
reproducibility and its current library stack, at no memory cost and at a per-image latency cost of a median 11\,\% in
one-process-per-image operation, which the long-lived regime of production
reduces (Section~\ref{sec:perf:memory});
the rationale is given in Section~\ref{sec:discussion:deployment}.

\begin{figure*}[t]
    \centering
    \includegraphics[width=\textwidth]{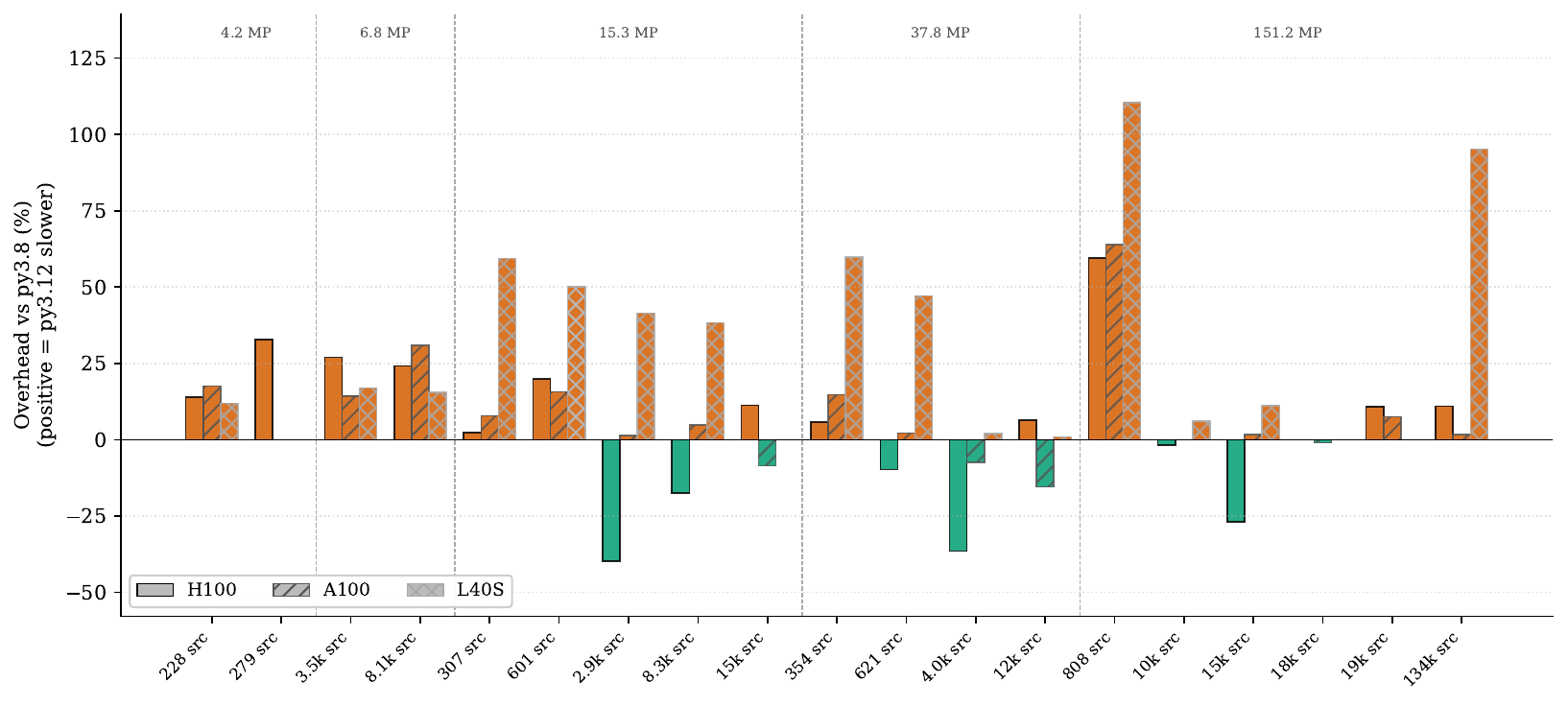}
    \caption{Per-image latency overhead of py3.12\,+\,adaptive cuML relative
    to py3.8 (positive\,=\,py3.12 slower) on the three datacenter GPUs,
    measured with the pool in place and the two stacks alternating in
    blocks of about twenty repetitions inside the same session on each host
    (50 cells).  Bars are grouped by frame size and labeled by source count; the
    hatching identifies the GPU and the bar color follows the sign, orange
    where py3.12 is slower and green where it is faster.  A missing bar is a
    cell whose two stacks were not measured within the same session on that
    host or, for the H100 at 18\,712 sources and the A100 at 10\,168, a pair
    excluded because the slowdown of its session did not enter both arms
    (Section~\ref{sec:perf:setup}).  The median overhead is $+11$\,\% ($+9$\,\% on the H100, $+5$\,\%
    on the A100, $+38$\,\% on the L40S), $+6$\,\% in the 31 cells with more
    than 2\,000 sources and $+18$\,\% in the 19 sparse cells; py3.12 is
    faster in 10 cells.  The three GPUs sit in different hosts.  The largest
    overhead, the 151.2\,MP FERVOR-L@TTT1 field ($+60$ to $+110$\,\%), is the cost
    of the cuML PCA on the 19\,881-pixel stamps of the unbinned TTT1 frame
    (Section~\ref{sec:perf:memory}).  The other bar above $+70$\,\%, the
    L40S at 134\,206 sources ($+95$\,\%), is the field whose catalog query
    varies most between repetitions (Section~\ref{sec:perf:latency}).  Without the pool the overhead has a median of $+38$\,\% and no negative
    cell (Section~\ref{sec:perf:memory}).}
    \label{fig:py38_vs_adaptive}
\end{figure*}

\subsection{cuML Crossmatch Evaluation}
\label{sec:cuml_eval}

GPU acceleration does not universally improve all pipeline stages.
We tested this for the catalog crossmatch stage, where source positions
are matched against astrometric and photometric reference catalogs
using nearest-neighbor queries in 2D sky coordinates.

\subsubsection{Experimental Setup}
\begin{figure*}[t]
    \centering
    \includegraphics[width=\textwidth]{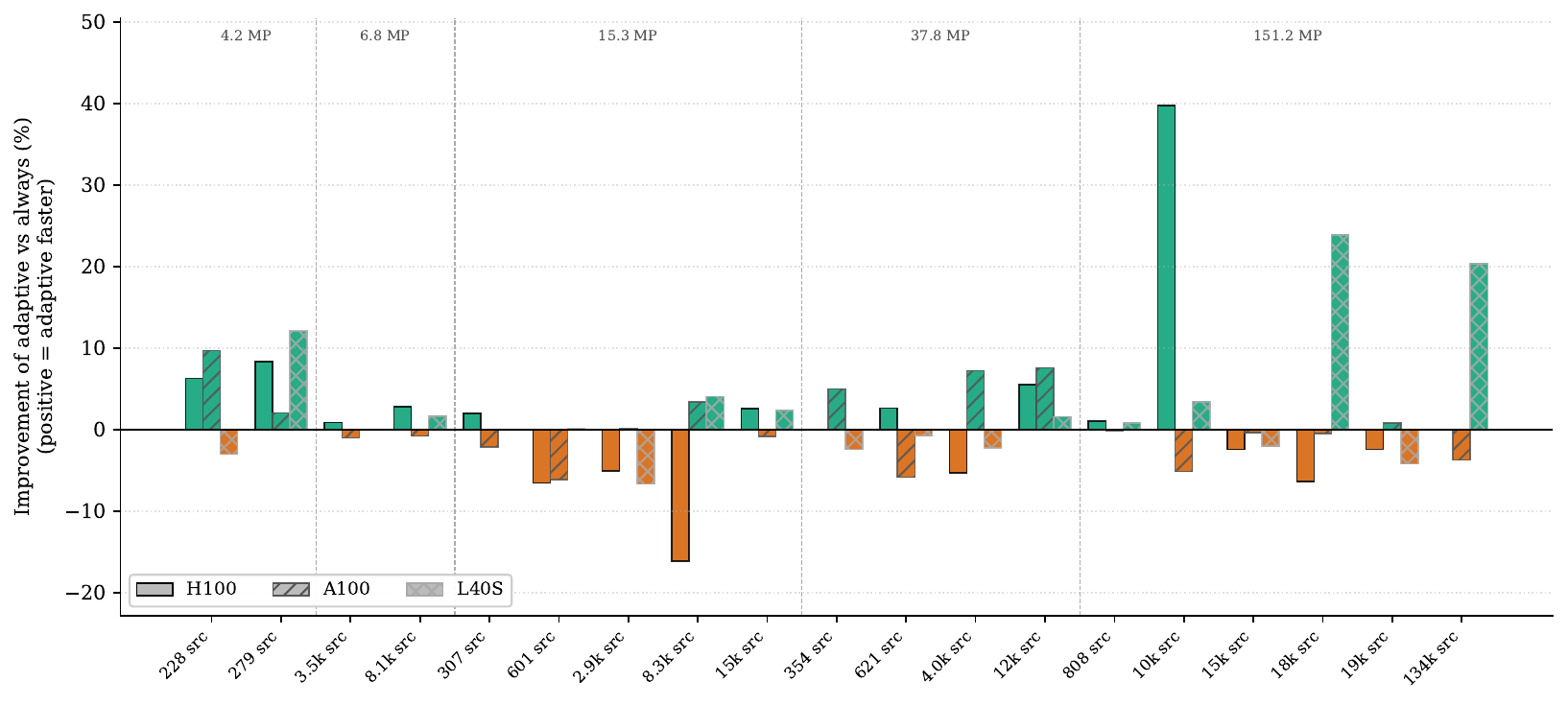}
    \caption{Improvement of the adaptive cuML crossmatch over forced cuML
    (positive\,=\,adaptive faster) per benchmark image and datacenter GPU,
    the two modes alternating in blocks of ten repetitions inside the same
    session on each host with the pool in place (56 cells).  Bars are grouped by
    frame size and labeled by source count; the hatching identifies the GPU
    and the bar color follows the sign, green where the adaptive mode is
    faster and orange where forced cuML is faster.  The one missing bar (H100
    at 134\,206 sources) is a cell whose two modes were not measured within
    the same session.  The median difference
    is 0.0\,\% ($+1.0$\,\% on the H100, $-0.4$\,\% on the A100, $+0.1$\,\% on
    the L40S) and 37 of the 56 cells lie within $\pm5$\,\%.  The four cells
    beyond $\pm15$\,\% (H100 at 10\,168 and 8\,294 sources, L40S at 18\,712
    and 134\,206) have no established cause; the two modes produced identical
    catalogs and zero points in every cell.}
    \label{fig:always_vs_adaptive}
\end{figure*}

We performed an ablation study on the A100-SXM4-80GB, which accommodates
the full benchmark collection including the 151.2\,MP frames, comparing
end-to-end pipeline latency with \path{cuML.NearestNeighbors} forced for
every crossmatch versus the default \path{scipy.spatial.cKDTree} forced, on
the nineteen benchmark images with their production configurations, five repetitions per cell, with the pool in place; two of the 190 runs failed and are excluded, which changes no median.
The comparison between the adaptive and the forced mode was measured
separately on the three datacenter GPUs, with the two modes interleaved
inside the same session (Figure~\ref{fig:always_vs_adaptive}).

\subsubsection{Results}

\begin{table}[t]
\centering
\caption{cuML ablation on the A100 (pool in place, production camera configurations): end-to-end latency with
\texttt{cuML.NearestNeighbors} forced for every crossmatch versus
\texttt{cKDTree} forced, medians of five repetitions per cell.  Penalty is
the relative change when cuML is forced (negative: cuML faster).  Across the
19 images the ratio of the two arms has a median of 1.01 and stays within
0.95--1.05.  Source counts are those of the current configuration; for the
two labels that pool several exposures a different exposure is listed than
in Table~\ref{tab:latency_py312}.}
\label{tab:cuml_ablation}
\begin{tabular}{rrrrr}
\toprule
\textbf{MP} & \textbf{Sources} & \textbf{\shortstack{With\\cuML (s)}} & \textbf{\shortstack{Without\\cuML (s)}} & \textbf{Penalty (\%)} \\
\midrule
4.2   & 228 & 8.2 & 8.1 & $+0$ \\
4.2   & 279 & 7.4 & 7.5 & $-2$ \\
6.8   & 3\,515 & 7.9 & 7.9 & $+0$ \\
6.8   & 8\,093 & 7.7 & 7.5 & $+3$ \\
15.3   & 307 & 9.5 & 9.1 & $+4$ \\
15.3   & 601 & 8.6 & 8.9 & $-4$ \\
15.3   & 2\,938 & 14.6 & 14.0 & $+5$ \\
15.3   & 8\,294 & 14.8 & 14.8 & $+0$ \\
15.3   & 15\,759 & 12.6 & 12.4 & $+2$ \\
37.8   & 354 & 10.2 & 10.0 & $+2$ \\
37.8   & 621 & 11.0 & 10.5 & $+5$ \\
37.8   & 4\,754 & 20.7 & 20.1 & $+3$ \\
37.8   & 12\,832 & 24.5 & 24.2 & $+1$ \\
151.2   & 808 & 30.8 & 30.2 & $+2$ \\
151.2   & 10\,168 & 27.5 & 28.9 & $-5$ \\
151.2   & 15\,390 & 29.6 & 31.0 & $-5$ \\
151.2   & 18\,712 & 30.2 & 29.9 & $+1$ \\
151.2   & 19\,492 & 32.8 & 32.5 & $+1$ \\
151.2   & 134\,206 & 42.5 & 44.4 & $-4$ \\
\bottomrule
\end{tabular}

\begin{flushleft}
\footnotesize \textit{Columns:} MP, frame size in megapixels; Sources, number of objects in the final catalog; With cuML and Without cuML, median end-to-end latency in seconds with the crossmatch forced to \texttt{cuML.NearestNeighbors} and to \texttt{cKDTree}; Penalty, relative change of the former with respect to the latter.
\end{flushleft}
\end{table}

\subsubsection{Analysis}
Across the 19 images, forcing \texttt{cuML.NearestNeighbors} for the
crossmatch changes end-to-end latency by $-5$\,\% to $+5$\,\% (median
$+1$\,\%), within the repeatability of the cells; even at 134\,206 sources,
where the crossmatch workload is heaviest, the forced mode is 4\,\% faster,
a difference of the size of the session-to-session variability of that
cell.  The crossmatch is not a measurable share of the pipeline at any
tested source density; the dominant stages are the photometry stages and,
at the highest densities, the catalog query
(Table~\ref{tab:stage_breakdown}).

The interleaved comparison of Figure~\ref{fig:always_vs_adaptive} extends
this conclusion to the recommended adaptive mode on three GPUs: its median
difference from the forced mode is 0.0\,\%, and the two modes are
indistinguishable in output, with identical catalogs and zero points in all
22 images of the collection.  The comparison also carries a methodological
lesson.  A first version of it, in which the two modes were measured in
separate sessions instead of interleaved, showed apparent effects of up to 15\,\% whose
size correlated with the separation between the two arms
($r = 0.50$): on a host whose catalog query varies with the production load,
comparing arms measured in separate sessions measures the catalog, not cuML.
The effects vanished when the arms were interleaved.  The py3.8 comparison
of Figure~\ref{fig:py38_vs_adaptive} was insensitive to the separation
($r = -0.05$), which is what a mechanical cause such as the launch penalty
predicts and a load artifact does not.

\begin{figure}[t]
    \centering
    \includegraphics[width=\columnwidth]{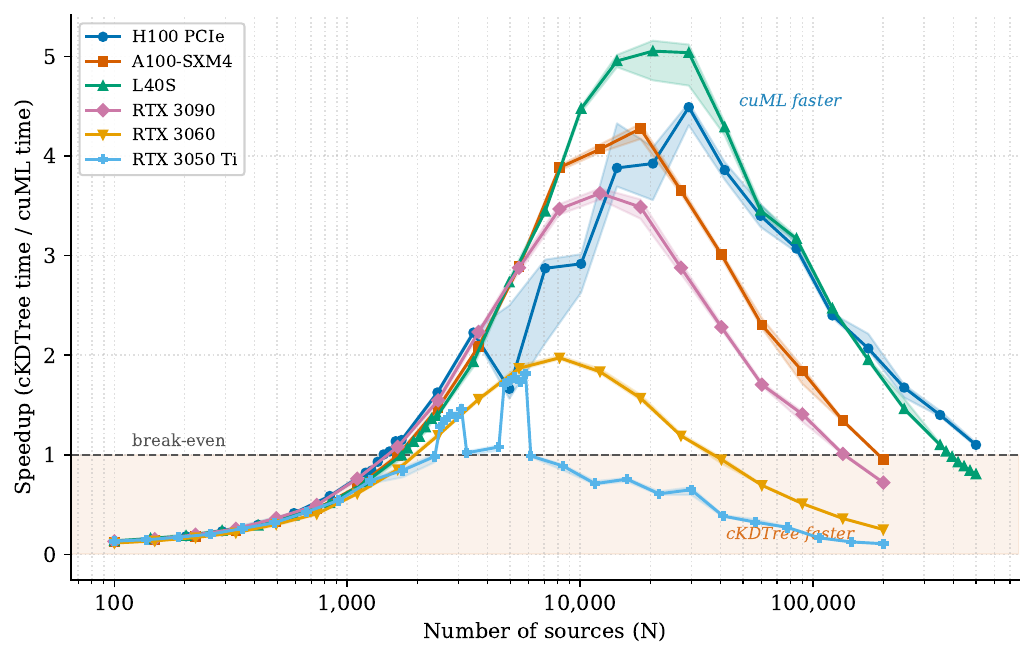}
    \caption{Synthetic crossmatch speedup (\texttt{cKDTree} time / \texttt{cuML}
    time) vs.\ number of sources for six GPUs.  Each curve is one GPU, the
    line joining the measured speedup at each source count and the shaded
    band the spread implied by the interquartile ranges of the two timings;
    the horizontal axis is logarithmic.  The dashed line marks break-even and
    the tinted region below it the regime where \texttt{cKDTree} is faster;
    values above~1.0 indicate \texttt{cuML} is faster.  The first crossover
    lies between 1{,}400 and 2{,}500 sources on every GPU; above it \texttt{cuML}
    leads up to 27{,}000 sources on the RTX~3060, 134{,}000 on the A100 and
    RTX~3090 and 372{,}000 on the L40S, and still leads at the largest count
    tested on the H100, whereas on the RTX~3050~Ti it leads only between
    2{,}500 and 5{,}800 sources.  The H100 and L40S were swept to 500{,}000
    sources, the other four to 200{,}000.  The isolated speedup
    ratio is diluted in end-to-end
    measurements by the multi-stage pipeline context; however, the crossover
    source count informs the per-GPU adaptive thresholds
    (Table~\ref{tab:cuml_thresholds}).}
    \label{fig:cuml_crossover}
\end{figure}

We additionally conducted synthetic benchmarks isolating the crossmatch
operation (Figure~\ref{fig:cuml_crossover}).
In controlled tests, \texttt{cuML} achieves lower latency than \texttt{cKDTree}
starting at 1{,}400--2{,}500 sources depending on the GPU.  The
synthetic timings show why: the GPU path has a floor of about 1\,ms per
call that does not move between 100 and 2\,500 sources (0.9\,ms on the
L40S, 1.1--1.2\,ms on the A100, H100, RTX~3090 and RTX~3060, 1.8\,ms on the
RTX~3050~Ti), whereas the \texttt{cKDTree} query grows from 0.1--0.2\,ms at 100 sources to 1.4--1.8\,ms at 2\,500. The crossover is where the growing tree-search cost meets the flat floor of the brute-force $O(N^2)$ GPU search.
The production thresholds (Table~\ref{tab:cuml_thresholds}) define the
source-count window within which the adaptive strategy activates cuML; the lower
bound is the crossover point identified in Figure~\ref{fig:cuml_crossover}.
The negligible end-to-end overhead across all ablation cells is consistent with
this picture: in the real pipeline the crossmatch is called multiple times per
image but each call operates on a moderate number of sources, placing most
observations outside the cuML-beneficial window.

\subsubsection{Decision}
The adaptive mode is retained as the recommended production configuration: configured via environment variables with the per-GPU thresholds of Table~\ref{tab:cuml_thresholds}, it is indistinguishable from the forced mode in latency and identical in output. Of the two other cuML backends, \texttt{cuml.decomposition.PCA} is negligible on the TST frames and dominant on the native-resolution TTT1 frame, whose PSF stamps are 18 times larger in pixels (Section~\ref{sec:perf:memory}); the seeded \texttt{cuml.cluster.KMeans}, with its ten initial-centroid draws matched to the ten restarts of the CPU path, takes 0.053\,s per call against 0.125\,s for \texttt{scikit-learn} on the calibrators of the densest field (1\,000 points, 24 clusters, A100, warm process), so the GPU clustering does pay at these sizes; the 3 to 6\,s that a new process spends before its first kernel on the datacenter hosts fall on its first call only. Both backends are selected by availability, not by size; the PCA would benefit from the per-stage selection already implemented for the crossmatch (Section~\ref{sec:future_directions}).
Table~\ref{tab:cuml_thresholds} lists the per-GPU source-count window within
which cuML is activated in the recommended adaptive configuration.  These thresholds follow the synthetic crossover curves (Figure~\ref{fig:cuml_crossover}), with the upper bound of the H100 at the largest count tested, where \texttt{cuML} still leads, and were checked against the end-to-end ablation measurements (Table~\ref{tab:cuml_ablation}, Figure~\ref{fig:always_vs_adaptive}).

\begin{table}[h]
\centering
\caption{Adaptive cuML crossmatch thresholds per GPU model.  cuML is
activated only when the detected source count falls within [MIN, MAX];
outside this range \texttt{cKDTree} is used.  The MIN values lie within two sweep steps of the first crossover of the synthetic benchmark (Figure~\ref{fig:cuml_crossover}) on the A100, RTX~3090, RTX~3060 and L40S, whose first crossovers fall between 1{,}645 and 2{,}454 sources, and well above it on the H100 (first crossover at 1{,}432 sources against a MIN of 5{,}000), a deliberately conservative margin; the MAX values lie below the last source count at which \texttt{cuML} still leads in that benchmark (134{,}058 on the A100 and RTX~3090, 27{,}061 on the RTX~3060, 371{,}992 on the L40S), except on the H100, where \texttt{cuML} still leads at the largest count tested, 500{,}000 sources.  Where a MAX lies below that last count, the value is the last point at which \texttt{cuML} still led in the coarser first sweep, whose steps were 2 to 2.5 times apart; the denser sweeps later placed the crossover higher, so each of those MAX values now sits below the stretch in which the speedup falls to unity (on the L40S, 1.46 at 245{,}880 sources and 1.03 at 371{,}992).  They come from the timing sweep, not from a memory limit.
These thresholds are configured per deployment via the \texttt{GPUPHOT\_CUML\_MIN\_SOURCES} and \texttt{GPUPHOT\_CUML\_MAX\_SOURCES} environment variables; values shown reflect recommended settings for each GPU model.}
\label{tab:cuml_thresholds}
\begin{tabular}{lrr}
\toprule
\textbf{GPU} & \textbf{MIN sources} & \textbf{MAX sources} \\
\midrule
H100 PCIe         & 5{,}000 & 500{,}000 \\
L40S              & 2{,}000 & 200{,}000 \\
A100-SXM4         & 2{,}000 & 100{,}000 \\
RTX 3090          & 2{,}000 & 100{,}000 \\
RTX 3060          & 2{,}000 &  20{,}000 \\
RTX 3050 Ti       & \multicolumn{2}{c}{---$^\dagger$} \\
\bottomrule
\end{tabular}
\begin{flushleft}
\footnotesize \textit{Columns:} GPU, device model; MIN and MAX sources, lower and upper source counts of the window within which the adaptive crossmatch selects cuML.\par
$^\dagger$ \texttt{cuML} leads only in a narrow window, 2{,}503 to 5{,}822 sources (peak $1.8\times$, two of its twelve points at parity), so \texttt{cKDTree} is used exclusively.
\end{flushleft}
\end{table}

\subsection{Edge Deployment: Jetson Orin}
\label{sec:perf:jetson}

We tested \gpuphot\ on two NVIDIA Jetson Orin modules to assess
the feasibility of edge deployment for autonomous observatories and
robotic telescopes.  Both modules use a unified memory architecture in
which 8\,GB of LPDDR5 is shared between the CPU and the integrated
NVIDIA Ampere GPU; as noted in Section~\ref{sec:perf:setup}, they are two
units of the same module part number on two software generations.

\paragraph{Orin Super.}
The Orin Super (JetPack~6) processed the two 4.2\,MP COLORS benchmark
images in 26.6\,s (Lum, 228 sources) and 23.7\,s (SDSSg, 279 sources) under
py3.12 (ARM; \texttt{cuML} unavailable), that is, $1.5$--$1.6\times$ slower
than the RTX~3050~Ti laptop GPU on the same images (17.3 and 15.1\,s) and
about $4\times$ slower than the datacenter GPUs.  Its zero points on the two
fields are identical to the x86 values under both stacks.

\paragraph{Orin Nano 8\,GB.}
The Orin~Nano (JetPack~5, Python~3.8 only) processed the same two images in
27.7\,s (SDSSg) and 173.4\,s (Lum).  The factor of six between two frames of
the same size and similar source count is not a property of the module: the
Lum field targets the comet C/2025~A6 and is astrometrically hard.  Its
single plate-solve attempt takes 119\,s on this CPU, against 16.7\,s for the
SDSSg field, and the solve time varies by a third between consecutive
attempts (160 and 119\,s in two probes), which is exactly the spread the cell
shows in the benchmark (96--199\,s over 53 repetitions).  The same field is
the more expensive of the two on six of the eight hosts under py3.8 (and four of seven under py3.12), and a
stage-resolved capture on an x86 host confirms that the difference sits in
the plate solve itself (0.43--0.47\,s against 0.38\,s under py3.8, 0.33--0.37 against 0.26--0.31\,s under py3.12).  Its magnitude does not
scale with the speed of the machine: a few hundredths of a second on x86 against
102\,s on the Orin~Nano, two orders of magnitude beyond what the ratio of CPU
speeds predicts.  With the default
astrometric timeout of 60\,s, the field fails three consecutive attempts and
loses three and a half minutes; the 300\,s timeout used in all our
benchmarks is therefore a measured requirement for slow CPUs, not a
convenience.

\paragraph{Memory ceiling.}
Neither module completes the 6.8\,MP frame with the released code.  Its FFT
working set overruns the module: a 224\,MiB cuFFT plan allocation fails with
5.71\,GiB already reserved, on a module that exposes 5.81\,GiB free at start,
under both stacks and from a freshly restarted container.  The two modules fail differently: the Orin~Super raises a clean
out-of-memory error, whereas the Orin~Nano hangs the host after the
critical memory-pressure checkpoint fires (effective ratio 0.984, its device term at 0.981) and requires
physical intervention; we observed this twice, on the 6.8 and 37.8\,MP
frames, and these are the two hang entries of the ledger in
Section~\ref{sec:perf:latency}.  The two units carry the same module part number but different board
revisions, and they run two software generations, so we do not
attribute the different failure mode to either alone.  A development build with a smaller working set had processed
the 6.8\,MP frame on the Orin~Super; the released code peaks higher, so the
verified ceiling is 4.2\,MP on both modules.

Deployment constraints and hardware-level restrictions for Jetson devices are discussed in Section~\ref{sec:deployment:jetson}.

\subsection{CPU Baseline Comparison}
\label{sec:perf:cpu}

To contextualize \gpuphot's absolute latency, we compare it against two
widely used CPU-only photometry tools: \texttt{sep}
\citep{barbarySEPSourceExtraction2016} and Photutils \citep{bradleyPhotutils2026}.
A critical scope mismatch must be stated before interpreting the results:
\texttt{sep} and Photutils perform source detection and circular aperture
photometry, corresponding to approximately two pipeline stages.  In
contrast, \gpuphot\ executes seven stages: FFT-based background
estimation, source detection, PSF modeling via Eigen-PSFs, aperture
photometry, multi-catalog crossmatch with \texttt{cKDTree}, astrometric
calibration via Astrometry.net (CPU-bound), and structured catalog output
to PostgreSQL.  Direct wall-clock comparison is therefore not
scope-equivalent; we present it for transparency.

Table~\ref{tab:cpu_baseline} reports the median latencies for all three
tools across ten representative images spanning the full resolution range.
The three columns were measured in the same py3.12 container
on the A100 host, so they share CPU, libraries and image files; the
\gpuphot\ column is the A100 entry of Table~\ref{tab:latency_py312}.

\begin{table*}[t]
\centering
\caption{CPU baseline comparison.  \texttt{sep} and Photutils execute
source detection and aperture photometry only (${\sim}2$ stages);
\gpuphot\ executes seven stages including PSF modeling, crossmatch,
and astrometric calibration.  All three tools were measured on the same
physical host as the A100, ensuring a fair hardware comparison.
\gpuphot\ times correspond to the A100 under py3.12.
\texttt{sep} timings at 151.2\,MP are dominated by image segmentation
and background estimation, which scale with pixel count rather than
source density, explaining the near-constant values across the three
source-density rows at that resolution.}
\label{tab:cpu_baseline}
\begin{tabular}{rrrrr}
\toprule
\textbf{MP} & \textbf{Sources} & \textbf{sep (s)} & \textbf{Photutils (s)} & \textbf{\gpuphot\ (s)} \\
\midrule
4.2   &      228 & 0.11 & 1.1 & 6.6 \\
4.2   &      279 & 0.10 & 1.1 & 6.3 \\
6.8   &   8\,093 & 0.17 & 2.0 & 6.9 \\
15.3  &      601 & 0.42 & 4.5 & 7.8 \\
15.3  &      307 & 0.49 & 4.2 & 8.2 \\
37.8  &      354 & 1.01 & 11.1 & 8.4 \\
37.8  &      621 & 1.02 & 11.0 & 9.0 \\
151.2 &      808 & 4.32 & 48.1 & 29.4 \\
151.2 &  15\,390 & 4.61 & 48.1 & 28.5 \\
151.2 &  18\,712 & 4.76 & 47.7 & 29.0 \\
\bottomrule
\end{tabular}

\begin{flushleft}
\footnotesize \textit{Columns:} MP, frame size in megapixels; Sources, number of objects in the final \gpuphot\ catalog; sep and Photutils, wall-clock time in seconds of detection and aperture photometry on the CPU of the A100 host; \gpuphot, median end-to-end latency of the seven-stage pipeline on the A100 under py3.12 (Table~\ref{tab:latency_py312}).
\end{flushleft}
\end{table*}

On this same-machine comparison the outcome depends on image size.
At 4.2\,MP, Photutils completes in 1.1\,s and \texttt{sep} in 0.1\,s, against
6.3--6.6\,s for \gpuphot: the five additional stages (background estimation,
PSF modeling, catalog query and \texttt{cKDTree} crossmatch, zero-point
calibration and Astrometry.net plate solving) and the per-process
initialization are fixed costs that a small frame cannot amortize.
At 37.8\,MP \gpuphot\ is already ahead: Photutils requires 11.0--11.1\,s and
\gpuphot\ 8.4--9.0\,s for its seven stages.
At 151.2\,MP, Photutils completes in 47.7--48.1\,s regardless of source
density, while \gpuphot\ reduces the
15\,390- and 18\,712-source FERVOR-L@TST fields in 28.5--29.0\,s, that is, seven
stages in 0.6 of the time of two, and the sparse 808-source FERVOR-L@TTT1 field
in 29.4\,s (0.61), where the cuML PCA of Section~\ref{sec:perf:memory} sets the cost.
This comparison remains scope-non-equivalent; \gpuphot's seven-stage
output (calibrated magnitudes on a spatially varying PSF model, astrometric
solution, structured catalog in PostgreSQL) is not equivalent to Photutils'
two-stage output.
The \texttt{sep} library, implemented in C with minimal Python overhead,
remains faster at all scales for its narrower scope of two stages
(4.3--4.8\,s at 151.2\,MP).
Two notes on provenance apply.  Photutils~3.0.0, the version installed for
this benchmark, is 24\% slower than the version measured earlier on the same
host, image by image, while \texttt{sep} varies by $-10$ to $+13$\% image by image (median $-3$\%) and detects
identical source lists; and the \gpuphot\ column reflects the camera
configuration of this benchmark, which catalogs between 0.7 and 3.9 times the objects of the earlier one, image by image,
whereas \texttt{sep} and Photutils are insensitive to that configuration.
The comparison is therefore between complete tools on the same files, not
between equal workloads.

To compare the detection stage alone, we captured the exact array that
\gpuphot\ passes to its matched-filter detection, that is, the
background-subtracted frame after the cosmetic filter of each camera, and
ran \texttt{sep} and Photutils on that same array in the same container and
process, with a $5\sigma$ threshold in all three (A100 host, pool in place).  The compared operation is the one
each tool performs from that input to its own source list: \gpuphot's
\path{detect_sources_psf}, \texttt{sep.extract}, and the complete Photutils
detection (segmentation, deblending and source catalog).  The three lists
differ, by up to a factor of three on the densest field, and no threshold
makes them coincide; the comparison is of the time each tool needs for its
own detection on the same pixels, not of identical outputs.
Table~\ref{tab:nvtx_detection} reports the medians.

\begin{table*}[t]
\centering
\caption{Source detection stage on the same input: \gpuphot's matched-filter
detection against \texttt{sep}~1.4.1 and Photutils~3.0.0, run in the same
container and process on the A100 host on the exact background-subtracted
array that the pipeline passes to its detection (pool in place), all with a $5\sigma$ threshold.  The source lists the three tools return
differ and no threshold equalizes them, so the compared quantity is the
time from the shared input to each tool's own list.  \gpuphot\ times are
medians of five synchronized calls, and the cold call includes the creation
of the cuFFT plan for the frame.  \texttt{sep} is the median of five (with
an enlarged pixel stack, which enables the dense fields without altering its
detection); Photutils the median of two for the complete detection
(segmentation, deblending and source catalog).  With the Eigen-PSF stage
enabled the detection applies $K+1 = 6$ convolutions, without it one, so
the two groups are not pooled.}
\label{tab:nvtx_detection}
\begin{tabular}{rlrrrrrrr}
\toprule
\textbf{MP} & \textbf{PCA} & \textbf{\shortstack{N\\\gpuphot}} & \textbf{\shortstack{N\\sep}} & \textbf{\shortstack{\gpuphot\\cold (s)}} & \textbf{\shortstack{\gpuphot\\warm (s)}} & \textbf{sep (s)} & \textbf{\shortstack{Photutils\\complete (s)}} & \textbf{Speedup} \\
\midrule
4.2 & off & 231 & 80 & 0.015 & 0.004 & 0.415 & 1.168 & 27.5$\times$ \\
4.2 & off & 285 & 104 & 0.018 & 0.003 & 0.283 & 0.190 & 15.4$\times$ \\
6.8 & off & 7\,782 & 6\,093 & 0.017 & 0.005 & 1.931 & 1.284 & 117.0$\times$ \\
15.3 & on & 294 & 180 & 0.106 & 0.097 & 0.652 & 0.638 & 6.1$\times$ \\
15.3 & on & 750 & 156 & 0.107 & 0.097 & 0.627 & 0.487 & 5.9$\times$ \\
37.8 & on & 448 & 180 & 0.108 & 0.101 & 1.264 & 1.214 & 11.7$\times$ \\
37.8 & on & 713 & 332 & 0.119 & 0.103 & 1.288 & 1.310 & 10.8$\times$ \\
151.2 & on & 893 & 303 & 0.449 & 0.435 & 4.539 & 6.158 & 10.1$\times$ \\
151.2 & on & 16\,518 & 7\,975 & 0.431 & 0.425 & 5.014 & 19.878 & 11.6$\times$ \\
151.2 & on & 19\,916 & 10\,554 & 0.442 & 0.427 & 6.611 & 15.830 & 15.0$\times$ \\
\bottomrule
\end{tabular}

\begin{flushleft}
\footnotesize \textit{Columns:} MP, frame size in megapixels; PCA, whether the Eigen-PSF stage (a principal component analysis of the PSF stamps) is enabled in the production configuration of the camera; $N$ \gpuphot\ and $N$ sep, number of sources each tool returns from the shared array; \gpuphot\ cold and warm, time in seconds of the first detection call in the process and of the subsequent calls; sep and Photutils complete, time of each tool's own detection; Speedup, \texttt{sep} time divided by the cold \gpuphot\ time.
\end{flushleft}
\end{table*}

On the seven frames whose configuration enables the Eigen-PSF stage,
where the detection applies six convolutions, \gpuphot\ is 5.9--15.0$\times$
faster than \texttt{sep} and 4.6--46$\times$ faster than the complete
Photutils detection.  The ratio grows with image size and with source
density because the two tools scale differently: the matched filter costs
$K+1$ FFT convolutions of the frame whatever its content, so its time is
0.43--0.45\,s on the three 151.2\,MP frames whether they hold 893 or
19\,916 sources, whereas the segmentation of \texttt{sep} grows with the
sources it finds, from 4.5 to 6.6\,s on the same three frames.  On the
three frames whose configuration disables the stage, the detection is a
single convolution and the ratios are larger, 15--117$\times$ against
\texttt{sep}; they measure a lighter operation and are reported separately
for that reason.  The cold \gpuphot\ times include the creation of the
cuFFT plan of this stage, whose padded size differs from the background stage's and is therefore not shared with it. The plan is paid once per process: on every image in the one-process-per-image benchmark, and once by a long-lived worker. It is 3\% of the stage at 151.2\,MP but four fifths of it at 4.2\,MP, where a warm call takes 3--4\,ms instead of 15--18.  The end-to-end gap of Table~\ref{tab:cpu_baseline} on small
frames is therefore not in the detection: it is in the five stages that
\gpuphot\ executes beyond it, and in the fixed costs of the process.

To quantify the contribution of each stage, including the CPU-bound
astrometric calibration (Astrometry.net), Table~\ref{tab:stage_breakdown}
reports the per-stage wall-clock breakdown measured with NVTX
instrumentation on the A100 for a small (4.2\,MP), a sparse (151.2\,MP)
and a dense (151.2\,MP)
image.

\begin{table*}[t]
\centering
\caption{Per-stage wall-clock breakdown on the A100 under NVIDIA Nsight
Systems with NVIDIA Tools Extension (NVTX) instrumentation (py3.12\,+\,adaptive cuML, pool in place).  The images are the COLORS@TTT3 Lum
frame (4.2\,MP, 228 sources), the unbinned FERVOR-L@TTT1 Lum frame
(151.2\,MP, 808 sources, sparse) and the FERVOR-L@TST Lum frame (151.2\,MP,
134\,206 sources, the densest of the benchmark); stage rows are medians of
five runs.  A dash means that the camera configuration does not run that
stage (the COLORS runs the cosmic-ray filter and not the Eigen-PSF
stage).
\emph{Not instrumented} is the wall clock outside every NVTX range (process
start, transfers, serialization).  The profiler inflates the wall clock by
factors of 0.99, 1.13 and about 1.2 with respect to the same runs without it
(the last from a single clean repetition), so the absolute times are not
those of Table~\ref{tab:latency_py312}; the shares are.}
\label{tab:stage_breakdown}
\begin{tabular}{llrrrrrr}
\toprule
\textbf{Stage} & \textbf{Type} & \textbf{\shortstack{4.2\,MP}} & \textbf{\%} & \textbf{\shortstack{151.2\,MP\\sparse}} & \textbf{\%} & \textbf{\shortstack{151.2\,MP\\dense}} & \textbf{\%} \\
\midrule
Optimal photometry & GPU & 1.773 & 26.9 & 6.216 & 18.7 & 17.692 & 35.7 \\
\quad of which Aperture photometry & GPU & 1.149 & 17.5 & 5.376 & 16.2 & 16.529 & 33.3 \\
\midrule
Astrometry.net solver & CPU & 1.207 & 18.3 & 1.697 & 5.1 & 1.747 & 3.5 \\
\midrule
Catalog query (network I/O) & I/O & 0.863 & 13.1 & 2.288 & 6.9 & 13.386 & 27.0 \\
\midrule
Background estimation (FFT) & GPU & 0.854 & 13.0 & 1.716 & 5.2 & 1.763 & 3.6 \\
\midrule
WCS fitting \& header update & CPU & 0.506 & 7.7 & 0.497 & 1.5 & 0.664 & 1.3 \\
\midrule
PSF cutout extraction & GPU & 0.125 & 1.9 & 0.559 & 1.7 & 2.728 & 5.5 \\
Star detection & GPU & 0.029 & 0.4 & 0.087 & 0.3 & 0.208 & 0.4 \\
FFT convolution & GPU & 0.018 & 0.3 & --- & --- & --- & --- \\
PSF source detection & GPU & 0.017 & 0.3 & 0.454 & 1.4 & 0.443 & 0.9 \\
\midrule
Moffat PSF fitting & CPU & 0.016 & 0.2 & 0.059 & 0.2 & 0.049 & 0.1 \\
Zero-point calibration & CPU & 0.005 & 0.1 & 0.009 & 0.0 & 0.170 & 0.3 \\
\midrule
Source crossmatch & GPU & 0.002 & 0.0 & 0.004 & 0.0 & 0.717 & 1.4 \\
Eigen-PSF (PCA) & GPU & --- & --- & 11.505 & 34.7 & 0.289 & 0.6 \\
PSF coefficient map & GPU & --- & --- & 4.780 & 14.4 & 3.382 & 6.8 \\
\quad of which Tile statistics & GPU & --- & --- & 2.863 & 8.6 & 2.118 & 4.3 \\
Star projection & GPU & --- & --- & 0.001 & 0.0 & 0.001 & 0.0 \\
\midrule
\textbf{Not instrumented} &  & 1.165 & 17.7 & 3.329 & 10.0 & 6.330 & 12.8 \\
\textbf{Wall clock (under nsys)} &  & 6.581 & 100.0 & 33.199 & 100.0 & 49.568 & 100.0 \\
\bottomrule
\end{tabular}

\begin{flushleft}
\footnotesize \textit{Columns:} Stage, pipeline stage, with indented rows for sub-stages contained in the row above and not added again; Type, where the stage executes (GPU via CuPy, CPU, or I/O for the catalog query; the crossmatch is classified as GPU because its cuML path executes there); the three pairs of columns, self time of the stage in seconds and its share of the wall clock under the profiler for the 4.2\,MP image, the sparse 151.2\,MP image and the dense 151.2\,MP image.
\end{flushleft}
\end{table*}

For the small 4.2\,MP image the GPU stages account for 43\,\% of the
wall clock and no single stage dominates: the optimal photometry, which
contains the batched aperture photometry, takes 27\,\%, the astrometric
solver and the WCS fit 26\,\%, the
background estimation 13\,\%, the catalog query 13\,\%, and 18\,\% falls
outside the instrumented stages, in process start-up, transfers and
serialization.  Faster GPUs therefore have little to accelerate on such
frames, which is what the flat 4.2\,MP rows of Table~\ref{tab:latency_py312}
show.  For the large frames the picture is set by stages that are not the
GPU photometry.  On the sparse 151.2\,MP frame the Eigen-PSF stage alone is
35\,\% of the wall clock, the cuML PCA cost of Section~\ref{sec:perf:memory},
and the coefficient maps another 14\,\%; the photometry stages are 19\,\%,
and the astrometric solver shrinks to 5\,\% because Astrometry.net uses
only the brightest 500 sources regardless of image size.  On the dense
frame the photometry stages rise to 36\,\% and the catalog query to
27\,\%, the two of them 63\,\% of the run, while the PCA is 0.6\,\%: at
this density the time is set by the photometry and by the catalog server,
not by detection or astrometry.  A stage-resolved comparison of the A100
captures without and with the pool places the whole gain of the pool in the
two photometry stages and leaves the PCA
unchanged, as expected from the launch-bound character of the former
(Section~\ref{sec:perf:memory}) and the solver-bound character of the latter.

\section{Discussion}
\label{sec:discussion}

\subsection{Comparison with Existing Tools}

Table~\ref{tab:tool_comparison} positions \gpuphot\ among established photometric software.

\begin{table*}
\centering
\caption{Comparison of \gpuphot\ with existing astronomical photometry tools in scope, execution platform, and deployment support.}
\label{tab:tool_comparison}
\begin{tabular}{lllllll}
\toprule
Tool & Language & CPU/GPU & Pipeline Stages & Distributed & Containerized & Active \\
\midrule
SExtractor & C & CPU & Detection, photometry & No & No & Yes (v2.x) \\
\texttt{sep} & Python/C & CPU & Detection, photometry & No & No & Yes \\
Photutils & Python & CPU & Detection, photometry & No & No & Yes \\
IRAF/PyRAF & IRAF/Python & CPU & Full reduction & No & No & Legacy \\
\gpuphot\ & Python & GPU+CPU & 7 stages (full) & Yes (Celery) & Yes (Docker) & Yes \\
\bottomrule
\end{tabular}
\begin{flushleft}
\footnotesize \textit{Columns:} Tool, software package; Language, implementation language or languages; CPU/GPU, processor on which the computation runs, GPU+CPU for \gpuphot\ because four of its seven stages run on the GPU, the crossmatch on either processor by problem size, and the calibrations on the CPU (Section~\ref{sec:implementation}); Pipeline stages, breadth of the operations covered natively by each tool, from source detection through catalog generation, ``7 stages (full)'' being the complete pipeline of Section~\ref{sec:implementation}; Distributed, built-in support for multi-node task distribution, with the mechanism in parentheses; Containerized, official Docker deployment recipes; Active, maintenance status at the time of writing: ``Yes'', actively maintained; ``Yes (v2.x)'', actively maintained in its 2.x release series; ``Legacy'', no longer under active development.
\textit{Note:} Wall-clock comparison between \gpuphot\ and detection-only tools is not scope-equivalent; see Table~\ref{tab:cpu_baseline} and the scope-equivalent per-stage comparison in Table~\ref{tab:nvtx_detection}.
\end{flushleft}
\end{table*}

We do not compare with the Vera C. Rubin Observatory pipeline \citep{juricLSSTDataManagement2017} or DRAGONS \citep{labrieDRAGONSDataReduction2019}, as these tools address complementary needs at different operational scales: the Rubin pipeline targets survey-scale data management across petabyte archives, while DRAGONS provides instrument-specific reduction for the Gemini Observatory. Neither is designed as a general-purpose, GPU-accelerated photometry framework for heterogeneous telescope networks.

Among the tools listed, \gpuphot\ is the only one combining GPU acceleration, distributed task processing via Celery, and containerized deployment via Docker. The \texttt{sep} library \citep{barbarySEPSourceExtraction2016} remains faster than \gpuphot\ for detection-only workloads at every image size (Section~\ref{sec:perf:cpu}), which is expected given its optimized C backend and narrower scope. However, \texttt{sep} does not provide PSF modeling, astrometric calibration, multi-catalog crossmatch, or distributed processing. Similarly, Photutils \citep{bradleyPhotutilsPhotometryTools2016} offers a broad Python API for photometric analysis but does not exploit GPU parallelism, limiting its throughput on high-resolution images. \gpuphot\ targets the operational niche where all seven pipeline stages must execute autonomously with low latency across image sizes ranging from 4.2 to 151.2\,MP. On the same host, its seven stages take longer than the two stages of Photutils on small frames, where fixed costs dominate, are ahead from 37.8\,MP, and complete in 0.6 of the Photutils time on the 151.2\,MP fields, sparse or dense (Table~\ref{tab:cpu_baseline}, Section~\ref{sec:perf:cpu}); restricted to the source detection stage alone, the A100 is 6--15$\times$ faster than \texttt{sep} on the same background-subtracted frames (Table~\ref{tab:nvtx_detection}).

\subsection{Deployment Configuration Recommendation}
\label{sec:discussion:deployment}

For production deployments we recommend Python~3.12 with adaptive cuML
crossmatching (Section~\ref{sec:impl:adaptive_cuml}), on two grounds.
First, reproducibility: the py3.12 stack produced identical catalogs and
zero points across the H100, A100 and L40S on all 22 benchmark exposures
and across every mode of the crossmatch (Section~\ref{sec:validation}).
The py3.8 stack, whose clustering runs on the CPU, depends on the host BLAS
kernel, which \gpuphot\ pins at import time; without the pinning one host returned a zero point
0.0053\,mag away from the rest on one field, with 112 calibrators instead
of 186, and failed to calibrate another, while py3.12 was
invariant in every combination tested (Section~\ref{sec:validation}).  Second, maintenance: it runs
on current releases of CuPy, NumPy and scikit-learn, whereas Python~3.8
reached end of life in October 2024, so it no longer receives security
fixes, and the CuPy~12 line is no longer maintained.  Memory is not a ground either way, since the two stacks have the same footprint, within 3.3\%, once the allocator is corrected, and latency is a cost rather than a benefit: a median 11\% in one-process-per-image operation on the datacenter GPUs, 6\% on fields with more than 2\,000 sources (Figure~\ref{fig:py38_vs_adaptive}), and near parity in a long-lived worker (Section~\ref{sec:perf:memory}).

Python~3.8 remains the correct choice for deployments that prioritize the
latency of each individual image in one-process-per-image operation,
provided the BLAS kernel is pinned, and it is the only option on JetPack~5
modules.  Concurrency is the same under both stacks to within one image: an 80\,GB device holds two 151.2\,MP reductions, the 48\,GB L40S one, and the 4\,GB RTX~3050~Ti one 4.2\,MP frame (Section~\ref{sec:perf:concurrency}).  One instrument
inverts the recommendation: on native-resolution frames of the TTT1, whose
fine plate scale gives the PSF stamps 19\,881 pixels, 18 times the TST
value, the cuML PCA made py3.8 faster by $1.6$--$2.1\times$ on the one such
frame of the benchmark, even with the pool in place
(Section~\ref{sec:perf:memory}), until the PCA backend can be selected per
stage (Section~\ref{sec:future_directions}).

\subsection{Limitations}
\label{sec:limitations}

We identify the following limitations of the current \gpuphot\ implementation and of its evaluation:

\begin{enumerate}[label=(\alph*)]
    \item \textbf{cuML crossmatch performance.}
    As demonstrated in Section~\ref{sec:cuml_eval}, forcing
    \texttt{cuML.NearestNeighbors} for every crossmatch changes end-to-end
    latency by $-5$\% to $+5$\% across 19 real benchmark images (median $+1$\%), with 12 of
    19 within $\pm$3\%, and the adaptive mode is indistinguishable from the
    forced one in interleaved measurements (median 0.0\%,
    Figure~\ref{fig:always_vs_adaptive}).
    Synthetic benchmarks show \texttt{cuML} is faster than \texttt{cKDTree} only
    above 1{,}400--2{,}500 sources, depending on the GPU: the GPU path has a measured floor of
    about 1\,ms per call, flat up to 2\,500 sources, that the $O(N \log N)$
    tree search undercuts at low counts;
    \texttt{cKDTree} is therefore the production default, with \texttt{cuML}
    activated adaptively for dense fields.

    \item \textbf{Python~3.12 per-image latency.}
    Without the CuPy pool, py3.12\,+\,adaptive cuML was slower than py3.8 in every one of the 87 cells where both were measured (median 38\%, up to $+240$\% on the sparse 151.2\,MP field; Table~\ref{tab:allocator_ablation}); with the pool re-seated, the released configuration, the gap is a median of 14\% over the same cells (Tables~\ref{tab:latency_py312} and~\ref{tab:latency_py38}) and 11\% in the same-session comparison of Figure~\ref{fig:py38_vs_adaptive}.  The residual is concentrated in the
    Eigen-PSF stage of the native-resolution TTT1 frame, where the full SVD
    solver of \texttt{cuml.decomposition.PCA} takes tens of seconds on a
    matrix of 19\,881 features against 0.23\,s on the 1\,089-feature TST
    frames and 0.14\,s under \texttt{scikit-learn}; a per-stage backend
    selection would remove it.
    The benchmark measures the one-process-per-image regime, which is what
    the prefork worker configuration pays; a long-lived worker on the
    \texttt{gevent} pool operates closer to parity, and, although a long-lived worker with the pool in place ran 96 images without drift or memory failure (Section~\ref{sec:perf:memory}), the throughput of that regime under production load was not measured.

    \item \textbf{Pool term of the memory-pressure checkpoint.} Of the two ratios the checkpoint of Section~\ref{sec:impl_memory} compares, only the device ratio measures pressure on the card; the pool ratio measures how compact the pool is and fires the release at nearly every checkpoint under py3.8 (about 0.2\,s per image at 2\% occupancy), and never under the uncorrected py3.12 stack. It should be replaced by a ratio to the device capacity.

    \item \textbf{Jetson Orin 8~GB memory ceiling.} With the released code the
    two Jetson Orin modules complete the 4.2\,MP frames only; the 6.8\,MP
    frame exceeds the unified memory in the FFT stage under both stacks, and
    on the Orin~Nano the exhaustion hangs the host instead of raising an
    out-of-memory error (Section~\ref{sec:perf:jetson}).  In addition, cuML
    is unavailable on the ARM architecture, and the devices are not suitable
    for large-format sCMOS workloads (15.3--151.2\,MP).

    \item \textbf{Astrometry on CPU.} The astrometric calibration step relies on Astrometry.net, which runs entirely on the CPU. No GPU-native blind plate solver is currently available in the community, and on slow CPUs a hard field can cost minutes per frame (Section~\ref{sec:perf:jetson}).

    \item \textbf{Single-GPU per image.} Intra-image multi-GPU parallelism is not implemented. The current scaling model relies on Celery worker replication, where each worker processes one image on one GPU. This is sufficient for the image sizes tested (up to 151.2\,MP on an 80\,GB A100) but may become a constraint for future mosaic sensors.

    \item \textbf{Shared catalog server.} At the highest source densities the
    end-to-end time is set by the query to the reference catalog rather than
    by the GPU, and the benchmark ran against a single replica, shared by
    every benchmark host and with the production pipeline and hosted on the
    RTX~3090 machine, so that host served the catalog to the others while
    measuring its own cells.  Three observations of
    Section~\ref{sec:performance} trace to this topology rather than to the
    pipeline: the episodic instability of the densest cells (IQR up to
    43\%, Section~\ref{sec:perf:latency}), the hour of py3.8 excluded on
    that host when the database held its CPU at full load
    (Section~\ref{sec:perf:setup}), and the exclusion of the RTX~3090 from
    the 151.2\,MP measurements: the co-located catalog database reserves
    32\,GiB of shared memory on that 125\,GiB host and leaves it with less
    than 1\,GiB free, and the pinned host memory that frame staging needs
    cannot come from reclaimable cache.

    \item \textbf{Scope of the per-stage measurements.} The detection
    comparison (Table~\ref{tab:nvtx_detection}) isolates the detection
    sub-stage only, and the stage breakdown (Table~\ref{tab:stage_breakdown})
    was recorded under the profiler, which inflates the wall clock by up to
    20\%; only its shares are used.  Two of the five cameras run without the
    Eigen-PSF stage, so the 4.2 and 6.8\,MP rows exercise a shorter pipeline
    (Section~\ref{sec:validation}).
    \item \textbf{Observations without an established cause.} Three
    observations of Section~\ref{sec:performance} are reproducible and are stated as observations only.  The H100~$<$~A100~$<$~L40S ordering holds on the medians over the 19 images and in 9 of the 12 cells at 151.2\,MP, the exceptions being one single cell reported from a slower session after the exclusion of an anomalous block (Section~\ref{sec:perf:setup}) and the sparse TTT1 field, on which the H100 and A100 lie within 1.1\,s of each other, and it is not the accelerator's.  The L40S trails
    the RTX~3090 by 3--7\% in two sparse cells under py3.12, and the stage decomposition
    of those cells could not be closed.  The 56\,MiB excess of the H100 is
    localized to the Eigen-PSF stage in the py3.12 stack but is present under
    py3.8 as well, and its mechanism is not known.  One attribution is
    settled by intervention and reported as such: the lower peak memory of
    the py3.12 stack, which the allocator correction removes, although how
    the pooled allocator raises the peak is only partly decomposed
    (Section~\ref{sec:perf:memory}).
\end{enumerate}

\subsection{Future Directions}
\label{sec:future_directions}

Building on the limitations identified above, we outline the following near-term and medium-term development priorities:

\begin{enumerate}[label=(\alph*)]
    \item \textbf{GPU-native astrometry} via parallelized quad-matching algorithms, eliminating the CPU bottleneck imposed by Astrometry.net.

    \item \textbf{Per-stage backend selection} extended from the crossmatch to
    the PCA: on frames with a fine plate scale the CPU implementation is
    faster and removes the only remaining large py3.12 overhead, the cuML
    PCA, whereas the GPU clustering, 2.4 times faster than the CPU path at
    the sizes of the benchmark, stays on the GPU; the sweep of
    Section~\ref{sec:perf:memory} gives the stamp-size threshold for that
    routing a numerical basis.

    \item \textbf{A device-referenced pressure metric} for the memory
    checkpoint, together with \textbf{per-function GPU memory monitoring}
    via decorator-based instrumentation, enabling fine-grained VRAM profiling
    across all pipeline stages.

    \item \textbf{Multi-GPU per image} for sensors exceeding 151\,MP, partitioning individual images across multiple GPUs with halo regions to handle the overlaps.

    \item \textbf{VRAM-adaptive FFT batching}, using the runtime memory monitoring already present in \texttt{gpu.py} to dynamically adjust batch sizes based on available VRAM.

    \item \textbf{Public benchmark suite} with ATLAS-scale images (61\,MP, $6 \times 10^3$--$1.6 \times 10^5$ sources) and a long-lived-worker mode, so that other groups can compare against it and reproduce the evaluation.

    \item \textbf{PSF-fitting photometry mode}, analogous to DAOPHOT ALLSTAR \citep{stetsonDAOPHOTComputerProgram1987}, reusing the existing Eigen-PSF infrastructure for iterative subtraction and re-fitting in crowded fields.
\end{enumerate}

\section{Conclusion}
\label{sec:conclusion}

We have presented \gpuphot, an open-source Python framework that accelerates the full photometric reduction pipeline for the TTT/TST robotic facility at Teide Observatory.

Three quantitative results characterize the system's performance envelope.
First, the NVIDIA H100 processes the densest 151.2\,MP sCMOS field of the benchmark, 134{,}206 cataloged sources, end-to-end, with one process per image, in 31.9\,s under the recommended py3.12\,+\,adaptive cuML configuration and in 30.2\,s under the CuPy~12 stack, through the full seven-stage pipeline; the sparsest field of the same size takes 19.0\,s under CuPy~12.
Second, in one-process-per-image benchmarks the CuPy~14 / cuML stack showed a 7\% lower peak VRAM and a median 38\% higher per-image latency than the CuPy~12 stack.  We traced both to the RAPIDS allocator displacing the CuPy memory pool at import.  Restoring the pool per task returns the peak to the CuPy~12 value and reduces the median latency overhead on the datacenter GPUs to 11\% (6\% on fields with more than 2{,}000 sources), with identical catalogs and zero points in every cell measured again, the cells of the two labels that pool several exposures aside.
Third, peak VRAM for the same input image agrees within 2\% across the six discrete GPUs tested and is the same under both stacks, within 3.3\%, once the allocator is corrected, so the memory footprint is set by the image and its processing configuration, 28.9\,GiB for a 151.2\,MP frame, and bounds concurrency at two such reductions on an 80\,GB device.

Because \gpuphot\ executes seven pipeline stages, a direct wall-clock
comparison with Photutils (two stages) is not scope-equivalent; on the same
physical host, \gpuphot\ is slower on frames up to 15.3\,MP, faster from 37.8\,MP, and completes the 151.2\,MP fields in 0.6 of the Photutils time (Section~\ref{sec:perf:cpu}).
When the comparison is restricted to the source detection stage alone,
the A100 detects sources 6--15$\times$ faster than \texttt{sep} on the same
background-subtracted frames (Table~\ref{tab:nvtx_detection}), demonstrating that GPU acceleration
is effective where the operations are genuinely parallel.

The recommended py3.12\,+\,adaptive cuML configuration is chosen for
reproducibility and for its current library stack, not for memory or speed:
it has the memory footprint of the py3.8 stack, a residual latency cost that
a long-lived worker removes, and one exception, the cuML PCA on frames with
a fine plate scale, where the number of pixels of the PSF stamp, and nothing
else, sets the cost of the full SVD solver of cuML.
The deployment range extends from the H100 datacenter accelerator to the
Jetson Orin edge modules (8\,GB shared memory), which remain viable for
4.2\,MP frames and for which the astrometric timeout must accommodate
plate solves of minutes on hard fields.

The crossmatch stage uses an adaptive backend: \texttt{cKDTree} by default, with \texttt{cuML.NearestNeighbors} activated only when source counts fall within the per-GPU window where GPU acceleration is beneficial (Section~\ref{sec:impl:adaptive_cuml}, Table~\ref{tab:cuml_thresholds}); forcing cuML unconditionally changes end-to-end latency by $-5$\% to $+5$\% across the 19-image benchmark, and the adaptive and forced modes are indistinguishable in interleaved measurements (Section~\ref{sec:cuml_eval}).

Reproducibility required identifying and pinning three sources of non-determinism, the PCA solver, the seeding of the GPU clustering and the host BLAS kernel; once pinned, the pipeline returns the same catalog and the same zero point, to the fourth decimal, on the H100, A100 and L40S, across x86 hosts, and on an ARM module for the frames it can hold (Sections~\ref{sec:impl_determinism} and~\ref{sec:validation}).

Known limitations and near-term development priorities are discussed in
Sections~\ref{sec:limitations} and~\ref{sec:future_directions}.

\printcredits

\section*{Declaration of competing interest}
\label{sec:doci}
The authors declare that they have no known competing financial interests or personal relationships that could have appeared to influence the work reported in this paper.

\section*{Declaration of generative AI and AI-assisted technologies in the writing process}
\label{sec:ai_declaration}
During the preparation of this work, the authors used Large Language Models and AI-assisted technologies for text validation and to verify consistency between the manuscript, the benchmark data, and the source code. After using these systems, the authors reviewed and edited the content as needed and take full responsibility for the content of the publication.

\section*{Software Availability}
\label{sec:software_availability}

The source code of \gpuphot, including the worker service and the benchmark scripts, is released under the MIT license at \url{https://github.com/Light-Bridges/GPUPhot}. Docker containers and benchmark scripts are provided to enable reproducible evaluation across the GPU platforms tested in this work. A persistent archive of the specific version used for the benchmarks presented in this paper will be deposited in Zenodo upon acceptance.

\section*{Data Availability}
\label{sec:data_availability}

The benchmark data and analysis scripts supporting the results presented in
this article are available in the \gpuphot\ GitHub repository at
\url{https://github.com/Light-Bridges/GPUPhot}, under the \texttt{benchmarks/}
directory. A complete, citable archive of the specific version used in this
paper will be deposited in Zenodo upon acceptance alongside the software.

\section*{Acknowledgements}
\label{sec:acknowledgements}

This work was supported by Light Bridges S.L. under an Industrial PhD agreement with the Universidad de La Laguna (ULL). S. Lemes-Perera acknowledges the funding and support provided by the company for this doctoral research.

The authors acknowledge support from the PID2022-138933OB-I00 ATQUE and 2023DIG28 IACTA research projects funded by MCIN\slash AEI\slash 10.13039\slash 501100011033\slash FEDER EU, and the CajaCanarias la Caixa Foundation, respectively. We also acknowledge the support provided by the CryptULL Research Group and the Cátedra de Ciberseguridad Binter-ULL for their contribution to the development of secure data management frameworks, as well as the institutional support provided by the Instituto Tecnológico y de Energías Renovables (ITER).

We express our gratitude to Light Bridges S.L. and ASTROPOC (\url{https://www.astropoc.com/}) for providing the high-performance GPU computing infrastructure, hardware resources, and astronomical datasets essential for the development, benchmarking, and performance evaluation of the \gpuphot\ library. The storage and computing capacity at ASTROPOC's EDGE computing center in Tenerife was made available in the form of Indefeasible Computer Rights (ICR), provided by Light Bridges in cooperation with Bechtle and LENOVO.

This article includes observations made with the Two-meter Twin Telescope (TTT) and the Transient Survey Telescope (TST), both located at the Teide Observatory of the Instituto de Astrofísica de Canarias and operated by Light Bridges on the island of Tenerife, Canary Islands (Spain), using observation time rights (DTO). Dr. Antonio Maudes’s insights in economics and law were instrumental in shaping the development of this work.

\bibliographystyle{cas-model2-names}
\bibliography{references_zotero}

\end{document}